\documentclass[twocolumn]{aastex62}
\usepackage{graphicx,natbib, amsmath}

\received{?, 2018}
\revised{?, 2018}
\accepted{?, 2018}
\submitjournal{ApJ}

\shorttitle{Stellar surface magneto-convection  as a source of astrophysical noise}
\shortauthors{Cegla et al.}
\begin{document}
\title{Stellar Surface Magneto-Convection as a Source of Astrophysical Noise\\ III. Sun-as-a-star Simulations and Optimal Noise Diagnostics}

\newcommand{\vdag}{(v)^\dagger}

\correspondingauthor{H.~M. Cegla}
\email{h.cegla@unige.ch}

\author{H.~M. Cegla}
\altaffiliation{CHEOPS Fellow, SNSF NCCR-PlanetS}
\affiliation{Observatoire de Gen\`eve, Universit\'{e} de Gen\`eve, Versoix 1290, Switzerland}
\affiliation{Astrophysics Research Centre, School of Mathematics \& Physics, Queen's University Belfast, Belfast BT7 1NN, UK}

\author{C.~A. Watson}
\affiliation{Astrophysics Research Centre, School of Mathematics \& Physics, Queen's University Belfast, Belfast BT7 1NN, UK}
\author{S. Shelyag}
\affiliation{School of Information Technology, Deakin University, Melbourne, Australia}
\author{M. Mathioudakis}
\affiliation{Astrophysics Research Centre, School of Mathematics \& Physics, Queen's University Belfast,  Belfast BT7 1NN, UK}
\author{S. Moutari}
\affiliation{Mathematical Science Research Centre, School of Mathematics \&  Physics, Queen's University Belfast, Belfast BT7 1NN, UK}

\begin{abstract}
Stellar surface magnetoconvection (granulation) creates asymmetries in the observed stellar absorption lines that can subsequently manifest themselves as spurious radial velocities shifts. In turn, this can then mask the Doppler-reflex motion induced by orbiting planets on their host stars, and represents a particular challenge for determining the masses of low-mass, long-period planets. Herein, we study this impact by creating Sun-as-a-star observations that encapsulate the granulation variability expected from 3D magnetohydrodynamic simulations. These Sun-as-a-star model observations are in good agreement with empirical observations of the Sun, but may underestimate the total variability relative to the quiet Sun due to the increased magnetic field strength in our models. We find numerous line profile characteristics linearly correlate with the disc-integrated convection-induced velocities. Removing the various correlations with the line bisector, equivalent width, and the $V_{asy}$ indicator may reduce $\sim$50-60\% of the granulation noise in the measured velocities. We also find that simultaneous photometry may be a key diagnostic, as our proxy for photometric brightness also allowed us to remove $\sim$50\% of the granulation-induced radial velocity noise. These correlations and granulation-noise mitigations breakdown in the presence of low instrumental resolution and/or increased stellar rotation, as both act to smooth the observed line profile asymmetries. 
\end{abstract}


\defcitealias{cegla13}{I}
\defcitealias{cegla18a}{II}

\section{Introduction}
\label{sec:intro}
In this series of papers, we aim to disentangle the impact of stellar surface magneto-convection on the radial velocity (RV) confirmation of low-mass, long-period planets. Sun-like stars have an outer convective envelope, wherein hot bubbles of plasma, known as granules, rise to the surface, cool, and eventually sink back down into the surrounding regions known as intergranular lanes. Naturally, these up- and down-flows induce blue- and redshifts accordingly. Since the granules are brighter and cover more surface area than the intergranular lanes, there is a net convective blueshift and the observed line profiles have an asymmetric shape; for the Sun, this asymmetry produces a `C-shaped' line bisector \citep{gray05}. Individual plasma flows on the Sun move with 1-4~km~s$^{-1}$ velocities, but much of the up- and down-flows cancel out over the $\sim$10$^6$ visible granules on the solar disc, such that the net root-mean-square (rms) is several 10s of cm~s$^{-1}$. This is particularly troublesome for exoplanet hunters as these stellar-induced velocity variations can mask the minute Doppler reflex motion of the host star produced by low-mass, long-period planetary companions. For example, the Earth only induces a mere 9~cm~s$^{-1}$ Doppler wobble on the Sun; hence granulation variability has the potential to completely mask an Earth-analogue signal. In the past, the detection of such small amplitude planetary signals was precluded foremost by instrumental precision; however, this is no longer the case thanks to the next generation of spectrographs, e.g. ESPRESSO and EXPRES. Consequently, photospheric magneto-convection/granulation is a source of astrophysical `noise' that must be overcome if we are to push the planet confirmation barriers to the level of habitable worlds around Sun-like stars.

The aim of this series is to identify the spectral fingerprints of stellar surface magnetoconvection and to use them to diagnose and disentangle the granulation-induced velocity variations in exoplanet confirmation and characterisation. Note, this distinction makes this work fundamentally different from past attempts to characterise the level of stellar variability using a so-called `jitter' term \citep[e.g.][]{santos00, saar03, wright05}, which not only combined various stellar phenomena into a single, unphysical term, but also treated them as `independent, identically distributed Gaussian noise process[es]' \citep{aigrain12}. Such a formalism may be useful for a first-order prediction of the stellar-induced RV variability; however, as \cite{aigrain12} point out, stellar variability manifests itself as red and/or pink noise and therefore the impact on RV measurements will often be significantly larger than that predicted by a random jitter term with the same mean amplitude. Accordingly, our aim is not to predict the bulk amplitude of the granulation-induced variations, but rather to identify key signatures that allow us to mitigate its impact on RV measurements. For such predictions, we refer the readers to \cite{cegla14a} and \cite{bastien14}, which show how photometry may be used to predict the amplitude of RV variability for magnetically quiet stars, and \cite{oshagh17}, who show how this may potentially be extended to more active stars. For predictions of the stellar oscillation-induced RV amplitudes (excited by convection), see \cite{yu18}. For techniques to mitigate the impact of stellar oscillations on precision RV measurements, see \cite{medina18} and \cite{chaplin19}.

The present methods to mitigate the impact of granulation in exoplanet observations revolve around optimising observing strategies to bin down this noise source \citep{dumusque11a, meunier15}. However, such approaches are time and cost intensive, and may reach a fundamental noise floor. Moreover, even though the lifetimes of individual granules are only $\sim$5-10 minutes, because the granulation signal is correlated it can take more than an entire night to bin to the 10~cm~s$^{-1}$ level \citep{meunier15}. Instead, we propose to use the information available in the stellar spectra to disentangle the granulation impact. The premise for this work is that because the RV shifts originate from changing stellar line profile asymmetries (from the granulation evolution), we can use our knowledge of these asymmetries to predict and mitigate the convection-induced RV variability. To understand how granulation impacts stellar absorption lines, we turn to three-dimensional (3D) magnetohydrodynamic (MHD) solar simulations, coupled with one-dimensional radiative transport. Our aim is to use the realistic line profiles output by such simulations to tile a stellar grid and mimic Sun-as-a-star observations, which ultimately can be used to search for correlations between the line profile asymmetries and the measured RVs. Since the MHD and corresponding line synthesis are very computationally heavy, it is not feasible to populate a stellar grid wherein each tile is independent. To overcome this aspect, we parameterised the absorption line profiles at disc centre in \citet[][hereafter Paper~I]{cegla13} and extended this across the stellar limb in \citet[][hereafter Paper~II]{cegla18a}. In this paper, we use this granulation parameterisation to create the aforementioned Sun-as-a-star observations and investigate the usefulness of a variety of stellar line characteristics as granulation noise mitigation tools. 

In Section~\ref{sec:creat_mod}, we detail how the Sun-as-a-star model observations are constructed and compare the outputs to solar observations. We use these model observations to search for correlations between convectively-induced line profile asymmetries and RVs in Section~\ref{sec:corr}, and investigate the impact of stellar rotation and instrumental resolution in Section~\ref{sec:add_fact}. Each potential granulation noise diagnostic is evaluated in Section~\ref{sec:reduct}, and we conclude in Section~\ref{sec:conc}.

\section{Creating Sun-as-a-star Observations}
\label{sec:creat_mod}
In Papers~\citetalias{cegla13} and \citetalias{cegla18a}, we demonstrated how we can parameterise solar surface granulation across the stellar disc by breaking it down into four components, split by magnetic field and photospheric continuum intensity: (bright, non-magnetic) granules, (dark) non-magnetic intergranular lanes, (dark) magnetic intergranular lanes, and magnetic bright points (MBPs). The accuracy of this parameterisation was validated against its ability to recreate the Fe~I~6302~$\AA$ line profile (synthesised using NICOLE; \citealt{NICOLE1,NICOLE2})\footnote{1D radiative transport is performed, which neglects spatial coupling; however, the impact on the disc-integrated line profile asymmetries is likely small and full 3D radiative transport calculations would require significantly more computational power.} from a time-series of 3D MHD solar simulation snapshots, at limb angles 0-80$^{\rm{o}}$ (in 2$^{\rm{o}}$ steps), generated with the MURaM code \citep{vogler05} for a net magnetic field of 200~G and corresponding to a physical size of 12$\times$12~Mm$^2$. As noted in Paper~\citetalias{cegla13}, this magnetic strength was chosen to ensure we were able to capture the magnetic components of granulation that are present even in the quiet Sun (which is likely closer to $\sim$130~G; \citealt{Trujillo04}), and the Fe~I~6302~$\AA$ line was chosen as it is widely used in both solar observations and simulations as a diagnostic tool for magnetic field and temperature.  

Filling factors for each component were determined for each snapshot using the same magnetic field and continuum intensity cuts used to create the parameterisation. These filling factors were used to add the average four component line profiles together in the correct proportions to reconstruct a line profile representative of the granulation pattern for each particular snapshot. Doing so allowed us to confirm our parameterisation could produce line profiles with the same net shifts and shapes as the more computationally heavy radiative, 3D MHD simulations. We refer the readers to Papers~\citetalias{cegla13} and \citetalias{cegla18a} for further details. 

Herein, we use probability distributions derived from this time-series of 3D MHD snapshots to select new filling factors and use them in conjunction with the four average (limb-dependent) component line profiles to generate new line profiles for 12$\times$12~Mm$^2$ patches with the same fundamental convection characteristics. This then allows us to generate enough independent line profiles to tile an entire stellar grid and create numerous Sun-as-a-star model observations of photospheric magnetoconvection. 
\begin{figure*}[t!]
\centering
\includegraphics[width = 8.5 cm]{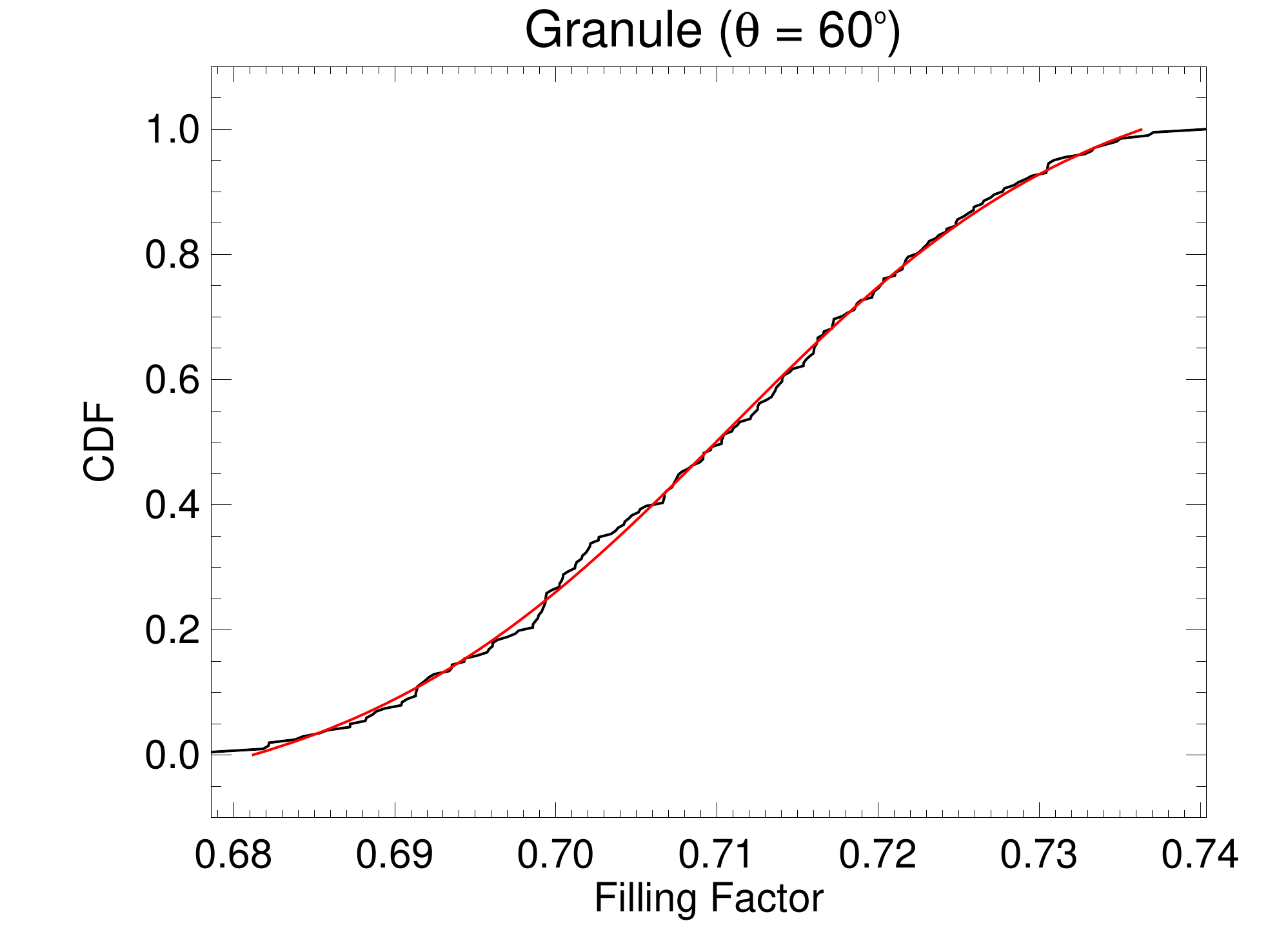}
\includegraphics[width = 8.5 cm]{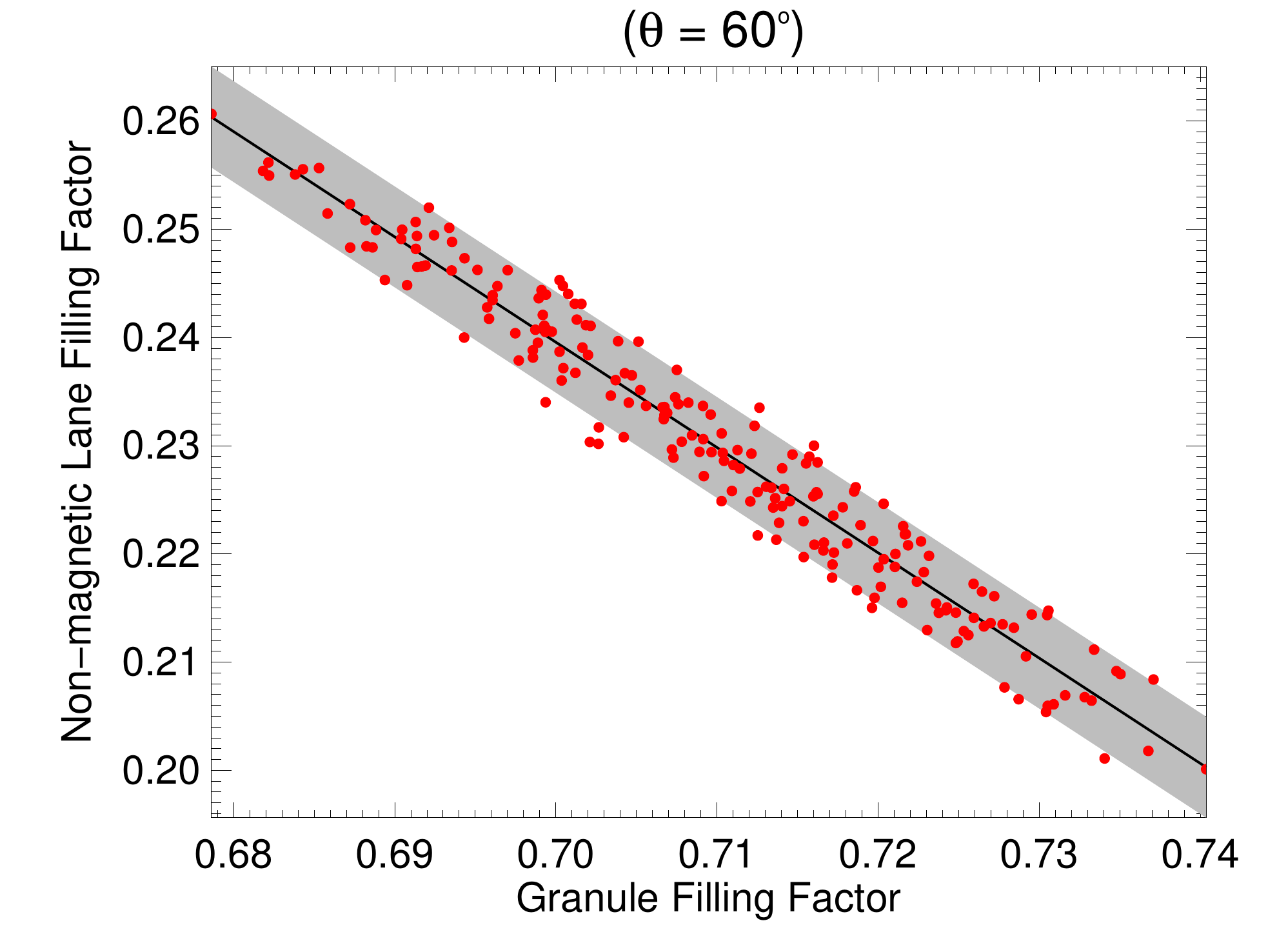}
\includegraphics[width = 8.5 cm]{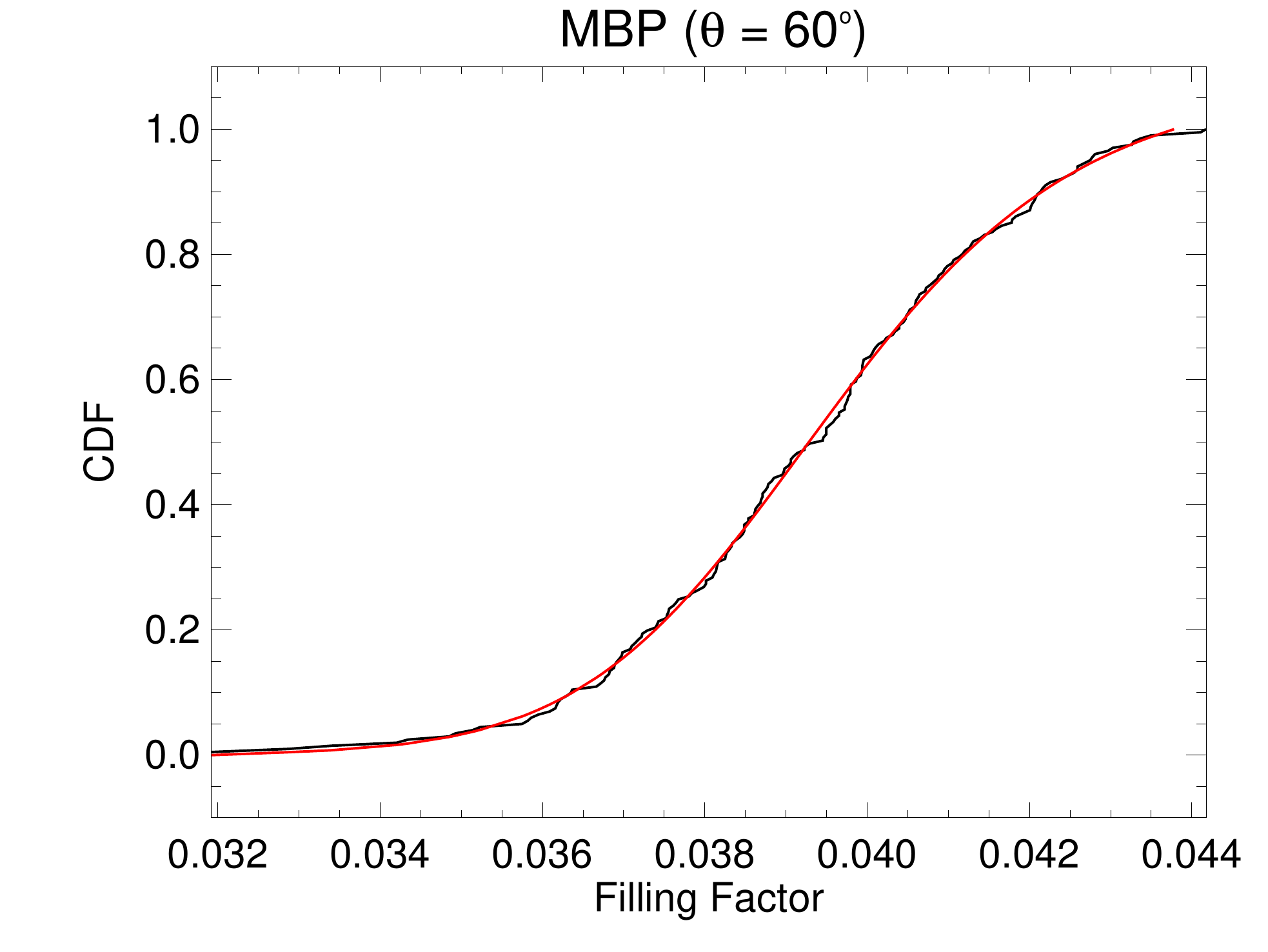} 
\includegraphics[width = 8.5 cm]{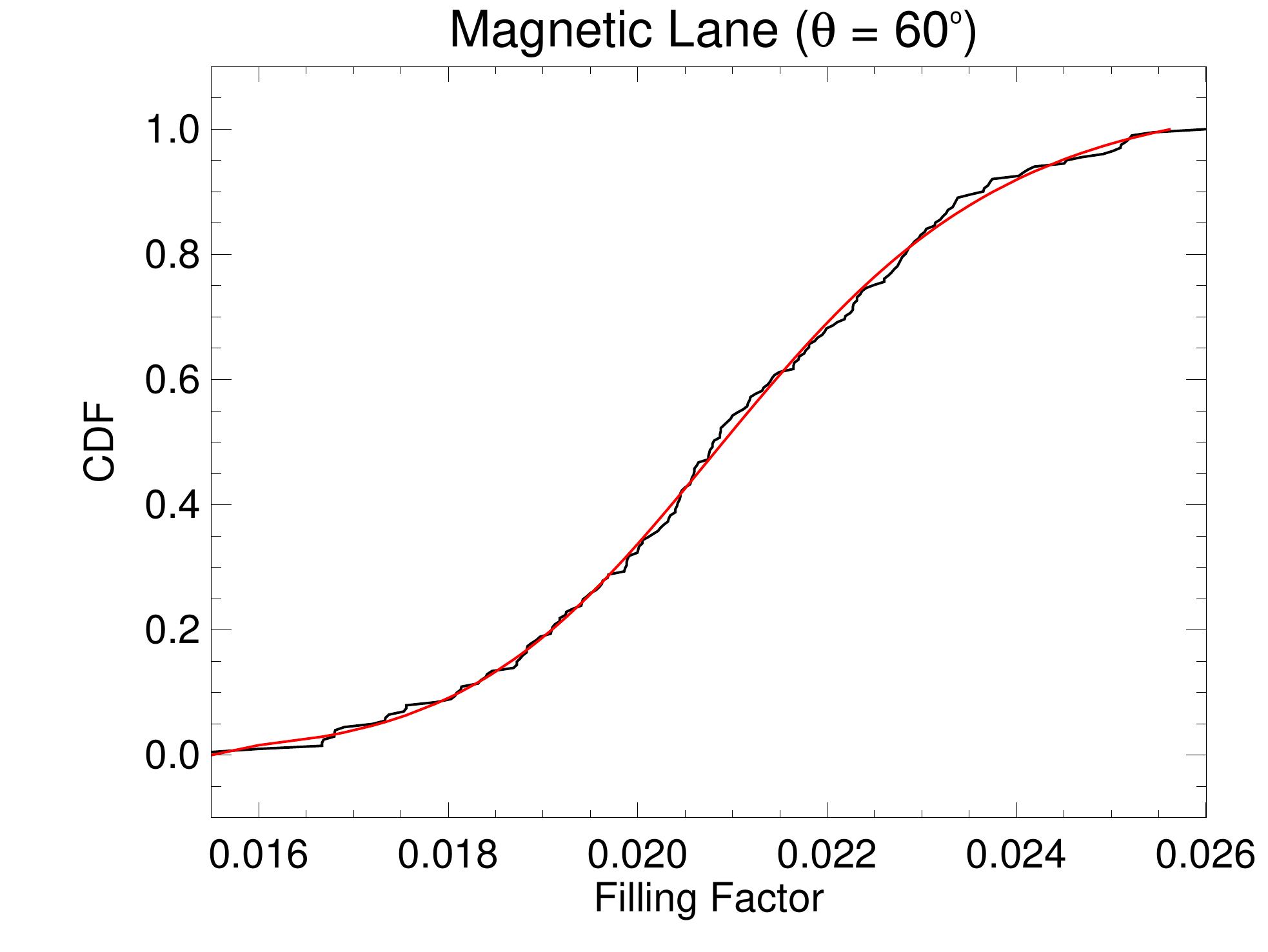} 
\caption{Cumulative distribution functions (CDFs) over the MHD time-series for the granule (top left), MBP (bottom left), and magnetic intergranular lane (bottom right) components are shown in black; fits using the generalised logistic function from Eq.~\ref{eq:logistic} are shown in red. Also displayed is the linear relationship between the filling factors for the granule and non-magnetic intergranular lane components (top right); the shaded area represents a uniform region of width 1.5$\sigma$, where $\sigma$ was determined by a robust bisquares linear regression (fit shown in black). All plots are for a limb angle of 60$^{\rm{o}}$ ($\mu$ = 0.5).}
\label{fig:prob60}
\end{figure*}
\vspace{-10pt}

\subsection{Granulation Component Probability Distributions}
\label{subsec:prob}
To characterise the probability distributions for the four component filling factors we use a non-parametric approach, as several of these distributions failed normality tests (e.g. Kolmogorov-Smirnov, Lilliefors goodness-of-fit, and the Jarque-Bera goodness-of-fit). In line with this, and to avoid a method sensitive to the data's bin size, we use the cumulative distribution function (CDF) to examine the nature of the probability distributions. Naturally, our parameterisation is not perfect, so there is some level of noise in the CDFs, likely due to some small misclassifications. We also note that the MHD time-series corresponds to $\sim$100 minutes of physical time with a cadence of $\sim$30 seconds, which samples well $\sim$15-20 granulation turnovers; as such, there is a small chance some filling factors could be under-sampled.

To overcome any component misclassification, and to  compensate for any potential under-sampling of the filling factors, we fit a generalised logistic function to each CDF; this function is defined as
\begin{equation}
\label{eq:logistic}
F(t) = A + \frac{K - A}{(1 + Qe^{-B(t-M)})^{1/v}}, 
\end{equation}
where $A$ and $K$ are the upper and lower asymptotes, respectively, $B$ is the growth rate, $v$ is a parameter associated with the asymptote near which the maximum growth occurs (when $>$ 0), $Q$ is related to the inflection point in $F(t)$, and $M$ determines the inflection point on the x-axis. The logistic function is well suited to model the S-shaped population of the CDFs, and the generalised version allows us to fine-tune the fit closely to the CDFs. As such, the logistic function allows us to `fill in' any gaps and obtain a smooth probability function from which we can randomly and realistically select the filling factors. We perform the fit to this function with a Levenberg-Marquardt least-squares \citep{markwardt09}, at each limb angle from the MHD simulation (0-80$^{\rm{o}}$, in 2$^{\rm{o}}$ steps -- see Paper~\citetalias{cegla18a} for details on the step choice). 

As the hot, uprising granular plasma eventually cools and sinks down into the intergranular lanes, we expect there to be some anti-correlation between the granule and intergranular lane component filling factors. Indeed, there is a strong anti-correlation between the granule and non-magnetic intergranular lane filling factors (illustrated in Figure~\ref{fig:prob60}). The magnetic components are only weakly correlated with the granule and non-magnetic intergranular lane components, and it is not clear if that weak correlation is physical or a byproduct of creating four components. 
\begin{figure*}[t!]
\centering
\includegraphics[width=17cm]{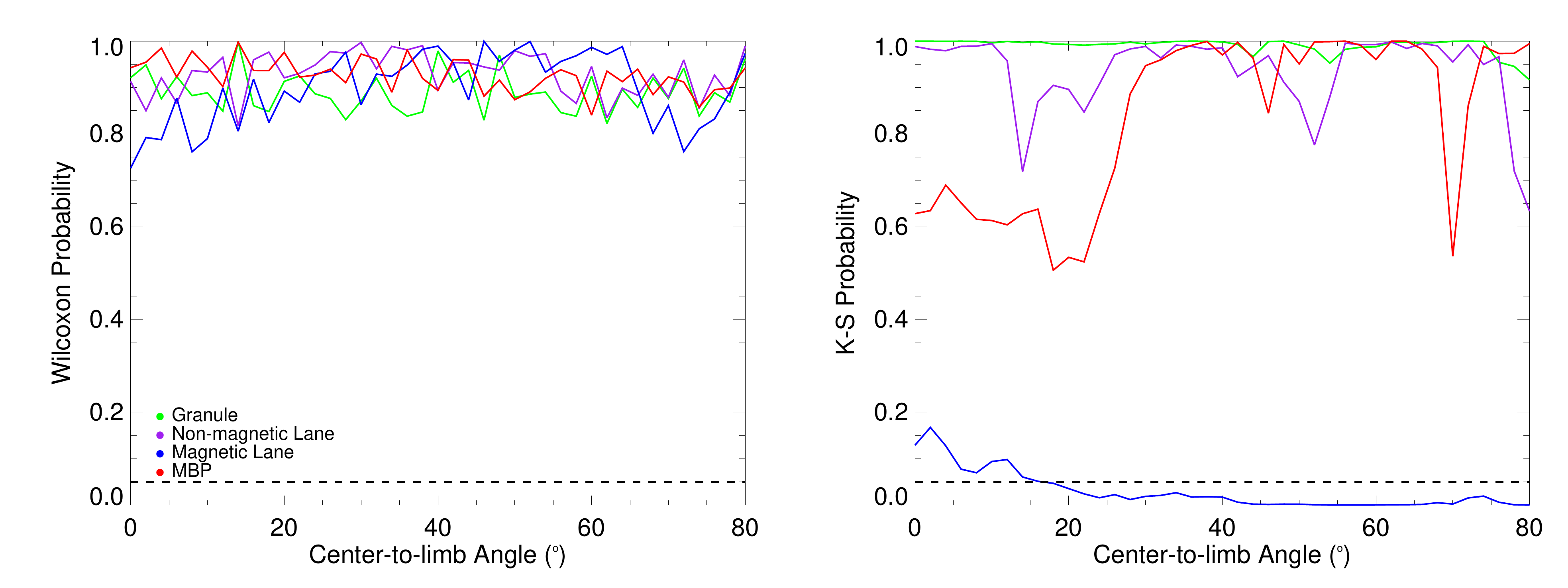}
\caption{The Wilcoxon Rank-Sum (left) and Kolmogorov-Smirnov (right) probabilities for 100,000 artificially generated filling factors as compared to the 201 original filling factors from the MHD simulation for all four granulation components. Probabilities $<$ 0.05 (dashed lines) are typically considered statistically significant results and require the null hypothesis to be rejected.}
\label{fig:wc_ks_60}
\end{figure*}

To create new line profiles, we first generate the granule filling factor. This is because the granule component is the most well characterised and dominates the overall line profile. Then we can use the aforementioned correlation to select the non-magnetic intergranular lane filling factor. To obtain the remaining filling factors for the MBP and magnetic intergranular lane components, we fit a probability distribution to one of these magnetic components; the final component is then determined from the difference between the other three components and unity, since filling factors must naturally total to one. We test both scenarios, i.e. either drawing the third component from the MBP distribution or drawing it from the magnetic intergranular lane distribution -- note this in contrast to Paper~\citetalias{cegla13}, where we did not fully explore the impact of the generating the magnetic components. An example of the granule and magnetic component CDFs and their fits is shown in Figure~\ref{fig:prob60}, alongside the anti-correlation between the granule and non-magnetic intergranular lane filling factors; Figure~\ref{fig:prob60} shows the results for 60$^{\rm{o}}$, i.e. $\mu = 0.5$, see Appendix~\ref{appen:other_prob} for other limb angles. Note the correlation between the granule and non-magnetic lane components has some non-Gaussian noise, which may originate from the stochastic nature of granulation and/or some small misclassifications. To incorporate this aspect, we add uniform noise to the linear fit between the granule and non-magnetic intergranular lane filling factors. 

To ensure we produce a combination of filling factors that represent the true underlying distributions, we compare newly generated distributions with the original sets from the MHD simulation. For this, we use both a Kolmogorov-Smirnov (KS) and Wilcoxon Rank-Sum test, and compare 100,000 new, artificial filling factors against those from the original 201 MHD snapshots. For the K-S test, the null hypothesis states that the two populations are drawn from the same distribution, while the null hypothesis for the Wilcoxon test states that the two sample populations have the same mean of distribution. 

For all four components, at all limb angles, and regardless of drawing from the MBP or magnetic lane distribution, the null hypothesis for the Wilcoxon test could not be rejected -- indicating both populations likely have the same mean. For the KS-test, the null hypothesis was never rejected for any of the granule and non-magnetic lane distributions. However, the KS probability for the magnetic component distributions was sensitive to which magnetic component was drawn from the CDF fit, as well as to the amount of uniform noise added in the granule versus non-magnetic lane linear fit; the results for these various tests are summarised in Table~\ref{tab:comp_gen}. Figure~\ref{fig:wc_ks_60} displays the the KS and Wilcoxon results when the MBP was drawn, and the granule-non-magnetic lane linear fit had 1.5$\sigma$ noise; see Appendix~\ref{appen:other_prob} for the remaining results.  

\begin{table}[t]
\caption{Summary of the null hypothesis rejection for the magnetic component KS tests.}
\begin{center}
Magnetic Intergranular Lane Generation
\begin{tabular}{c|c|c}
    \hline
    Component & Non-Mag. Noise\tablenotemark{a} & Rejected?\tablenotemark{b} \\
    \hline
    Mag. Lane\tablenotemark{c} & 1.0-2.0$\sigma$ & Never \\
    MBP & 1.0$\sigma$ & 0-36$^{\rm{o}}$ \\
    MBP & 1.5-2.0$\sigma$  &  Always \\
\end{tabular}

\vspace{5pt}
Magnetic Bright Point Generation
\begin{tabular}{c|c|c}    
    \hline
    Component & Non-Mag. Noise\tablenotemark{a} & Rejected?\tablenotemark{b} \\
    \hline
    MBP & 1.0-2.0$\sigma$ & Never \\
    Mag. Lane\tablenotemark{c}  & 1.0$\sigma$ & 80$^{\rm{o}}$ \\
    Mag. Lane\tablenotemark{c}  & 1.5$\sigma$  &  20-80$^{\rm{o}}$ \\
    Mag. Lane\tablenotemark{c}  & 2.0$\sigma$ & Always \\
\end{tabular}
\end{center}
 \tablenotetext{a}{The amount of uniform noise injected into the linear fit used to generate the non-magnetic lane component (e.g. shaded region in Figure~\ref{fig:prob60}).} 
  \tablenotetext{b}{Probabilities $<$ 0.05 and/or 0.01 are typically considered statistically significant results requiring the null hypothesis to be rejected.}
  \tablenotetext{c}{Magnetic intergranular lane component.}
  
 \label{tab:comp_gen}
\end{table}
  
Only by drawing from the MBP distribution first can we obtain realistic distributions for both magnetic components. Moreover, since the MBP is brighter and the filling factors are slightly larger, it is more important to correctly attribute this component. In addition, as there appears to be real variability in the granule to non-magnetic intergranular lane relationship we want to include as much of this as possible, whilst maintaining sensible distributions for the magnetic components. Hence, we argue the most realistic way to generate new line profiles is by: 
\begin{itemize}
\item Randomly selecting a granule filling factor from a fit to the CDF derived from the MHD simulation
\item Drawing a corresponding non-magnetic lane component from the linear relationship in the MHD simulation between the granule and non-magnetic lane components, with 1.5$\sigma$ of uniform noise
\item Randomly selecting the MBP filling factor from a fit to the CDF derived from the MHD simulation
\item Determining the magnetic lane filling factor such that all four components add to unity.
\end{itemize}
Note, towards the limb, there were some instances where the randomly generated MBP component meant the total filling factors were greater than one, without the addition of the magnetic intergranular lane component. In these instances, a new MBP was generated; if more than 100 iterations were made without resolving this issue, then the magnetic lane component was set to zero, and the MBP component was set to one minus the non-magnetic components. As the magnetic components make up less than $\sim$10\% of the total line profile, this small effect will be negligible when integrating across the stellar disc. It is also important to note, as only the granule and MBP components are drawn from probability distributions (since the non-magnetic lane largely depends on the granule filling factor and all components must add to unity), there are essentially only two free parameters used in the generation of new line profiles; see Paper~\citetalias{cegla13} for further justifications on the four physical components used herein.

\subsection{Tiling the Stellar Grid}
\label{subsec:grid}
Now that we can generate new line profiles that represent realistic granulation patterns, we can place these onto a stellar surface grid and construct synthetic `Sun-as-a-star' observations. The granulation line profiles are based on 3D MHD simulations with a physical size of 12$\times$12~Mm$^{2}$ at disc centre; as such this sets the tile size for the stellar grid. To appropriately consider the geometrical effects, we construct and populate a 3D stellar grid. The synthetic stellar surface is modelled as a sphere and split into a number of latitudinal slices, which are further sub-divided into smaller surface elements or `tiles'. The number of latitude slices was determined by taking half the circumference of the sphere and dividing by the tile size. The subsequent number of tiles, $t_n$, in each latitude slice is dependent on latitude and defined:
\begin{equation}
\label{eq:tile}
t_n =  \frac{2 \pi R_{\star} \cos \phi}{R_t},
\end{equation}
where $R_{\star}$ is the stellar radius (in our case we set this to approximate the solar value at 695,500 km), $R_t$ is the tile width (12~Mm), and $\phi$ is the latitude; note, this value is rounded to the nearest integer. The very top and bottom annuli of the sphere are approximated by triangles rather than square tiles to help minimise gaps. Although the simulation snapshots are square, the effect of approximating these annuli with triangles is negligible as at these extrema the tiles are heavily limb darkened and have very small projected areas. 

\begin{figure}[t]
\centering
\includegraphics[trim=0cm 0cm 0cm 0.cm, clip, scale=0.2]{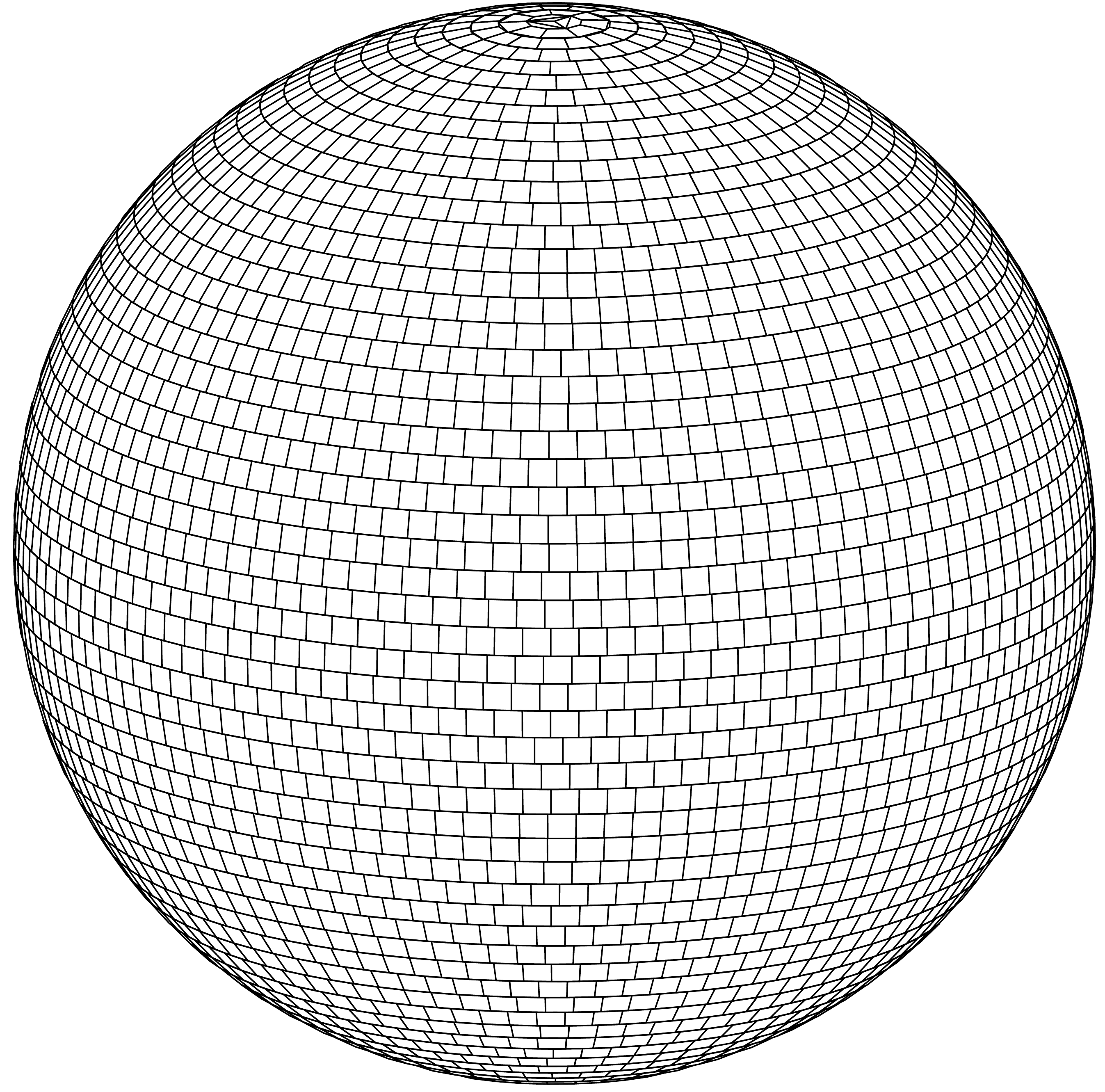}
\caption{The model star grid, shown at a slight inclination, with larger tiles for illustration only.} 
\label{fig:grid}
\end{figure}

Tiles located on the back of the model star are hidden to ensure that they do not contribute to the disc integrated observation. The line-of-sight (LOS) to an individual tile is determined by calculating the angle between the Earth vector (the vector pointing towards the observer) and the normal vector to the tile. For a coordinate system $x,y,z$, with origin at the centre of the star, $z$ pointing along the rotation axis, $y$ pointing towards the observer, and $x$ orthogonal to both $z$ and $y$ (using a left-hand rule), the Earth vector, $\hat{E}$ is defined:
\begin{equation}
\label{eq:Ex}
\hat{E}_x = \cos \theta \ \sin i
\end{equation}

\begin{equation}
\label{eq:Ey}
\hat{E}_y = - \sin \theta \ \sin i
\end{equation}

\begin{equation}
\label{eq:Ez}
\hat{E}_z = \cos i,
\end{equation}
where $i$ is the stellar inclination and $\theta$ it the rotation angle (i.e. 180$^{\rm{o}}$ = 0.5 phase). If the angle between a tile's normal vector and Earth vector is greater than $\pi$/2 then the tile will be hidden from view; this angle is calculated for each corner of the tile. If a tile falls only partially on the stellar disc, then it is sub-divided to calculate the fractional area on the disc; each such tile is sub-divided into 40 latitude slices and equivalent longitude slices such that the sub-tiles are as close to square as possible. The centre of each sub-tile is used to calculate the angle between the normal and Earth vector, and the visible sub-tiles are used to scale the line profile assigned to the full tile accordingly, in terms of both projected area and limb darkening. 

Points ($x,y,z$) on this 3D model star can then be projected onto the plane for the sky as a 2D surface with coordinates $X$ and $Y$ via,
\begin{equation}
\label{eq:X}
X = x \sin \theta + y \cos \theta
\end{equation}

\begin{equation}
\label{eq:Y}
Y = \cos i \ (-x \cos \theta + y \sin \theta) + z \sin i.
\end{equation}
An illustration of this grid, with an exaggerated tile size for viewing ease, is shown in Figure~\ref{fig:grid}. 

Each of the tiles on the model stellar surface can then be assigned an appropriate absorption line profile. Each physical tile corresponds to approximately 1$^{\rm{o}}$ on the stellar disc. The 3D MHD simulation was parameterised in even 2$^{\rm{o}}$ steps from disc centre to 80$^{\rm{o}}$; hence, tiles with a (rounded) even limb-angle in this range are assigned a randomly generated profile corresponding to the same limb-angle. For tiles with odd limb-angles, we randomly assign a line profile corresponding to a limb angle either 1$^{\rm{o}}$ lower or higher (such that half are higher and half are lower). All tiles beyond 80$^{\rm{o}}$ are assigned a line profile corresponding to the 80$^{\rm{o}}$ parameterisation. This region is heavily limb-darkened, with a small projected area, which contributes less than 5\% to the total integrated flux; hence, this small approximation should be negligible. The MHD simulation naturally includes limb-darkening; we interpolate between the 2$^{\rm{o}}$ steps, and those beyond 80$^{\rm{o}}$, to correct the limb darkening in the line profiles assigned to these tiles. This is done by fitting the non-linear Claret limb darkening law to the intensities output by the simulation. The line profiles in each tile must be multiplied by the projected area of the tile to measure the total intensity; this is because the profiles output by the simulations are generated in terms of flux per unit area. To include a solar-like rotation, we shift the line profiles in each tile corresponding to a projected, solid-body\footnote{A solar-like differential rotation, as opposed to solid-body, will introduce slight differences in the line profiles asymmetries \citep{beeck13}, but this effect is minimal, compared to photospheric magnetoconvection effects, for slow rotators like the Sun -- see Appendix~\ref{appen:other_noise} for further details.} rotational velocity of 2~km~s$^{-1}$. Finally, to generate the artificial observations, the contributions from all the visible tiles are summed together to create disc-integrated Sun-as-a-star observations. For this study, this process was repeated to create 1,000 independent, model observations. As such, these represent instantaneous observations (i.e. zero exposure duration), separated by more than one granulation lifetime.

\subsection{Comparison to Solar Observations}
\label{subsec:obs}
In Paper~\citetalias{cegla18a}, we compared the simulated line profiles extensively to empirical solar observations at discrete limb angles across the stellar disc. Now that we can create Sun-as-a-star observations, we can further validate our parameterisation, and new line profile generation, by comparing the model observations to disc-integrated solar observations. For this, we compare against three different solar atlases, constructed at very high resolution, from: the Fourier Transform Spectrometer (FTS) on the McMath-Pierce Solar Telescope at Kitt Peak National Observatory \citep{kurucz84}, the FTS at the Institut f{\"u}r Astrophysik, G{\"o}ttingen (IAG) \citep{reiners16}, and the PEPSI spectrograph on the SDI telescope located at the Large Binocular Telescope (LBT) \citep{strassmeier18}. The FTS instruments provide the highest resolution, at $\sim$500,000 and 670,000 for the McMath and IAG, respectively, while the PEPSI echelle spectrograph still has a very high resolution of 270,000. On the other hand, it is the two most recent instruments, from the IAG and PEPSI, that provide the highest absolute RV precision, of $\sim$10~m~s$^{-1}$; the McMath FTS can have RV deviations as large as $\sim$100~m~s$^{-1}$ \citep[][and references therein]{molaro12,reiners16,strassmeier18}. 

\begin{figure}[t!]
\centering
\includegraphics[trim=1.9cm 0.7cm 0.8cm 1.4cm, clip, scale=0.2]{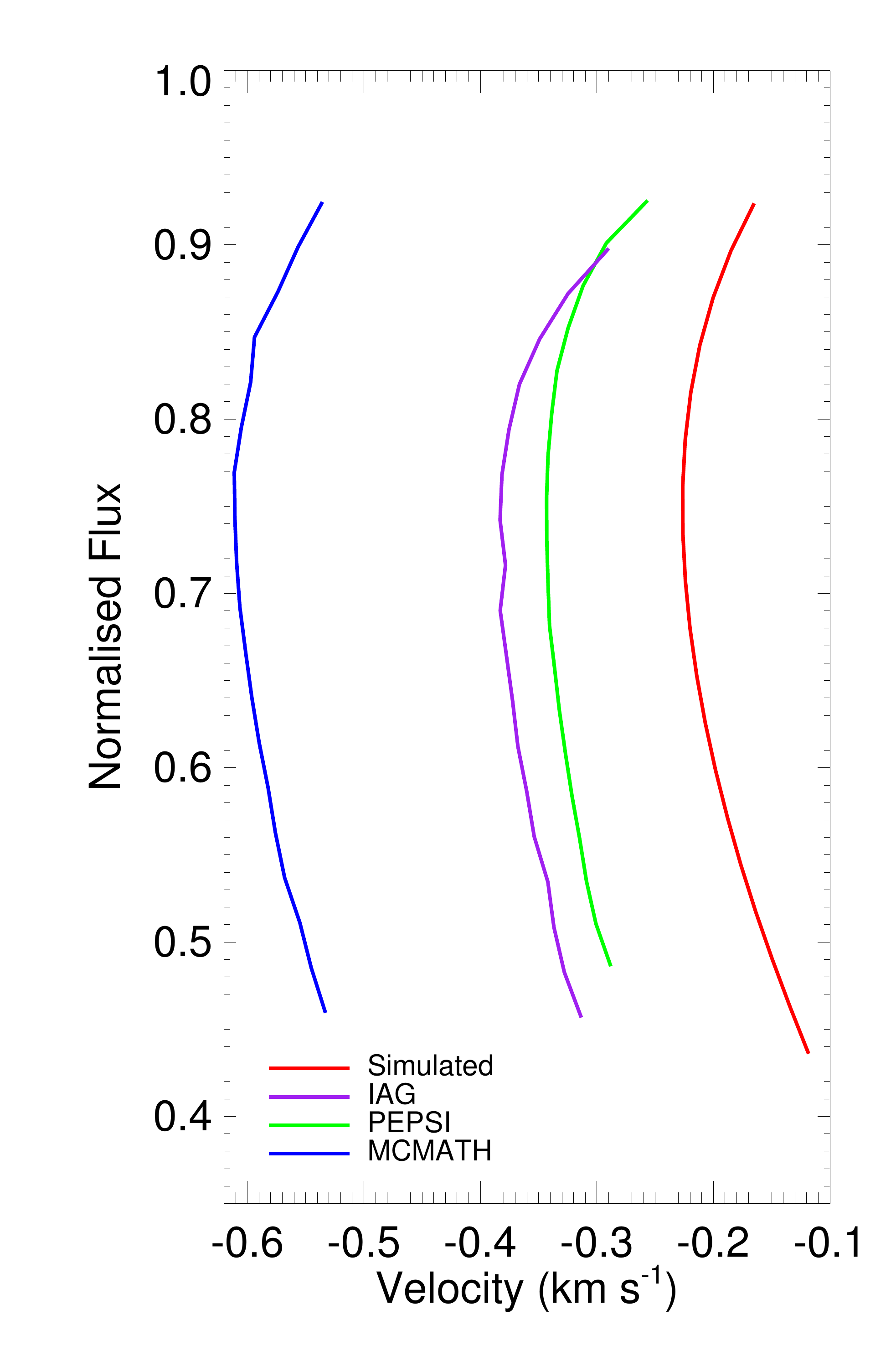} 
\includegraphics[trim=1.9cm 0.7cm 0.8cm 1.4cm, clip, scale=0.2]{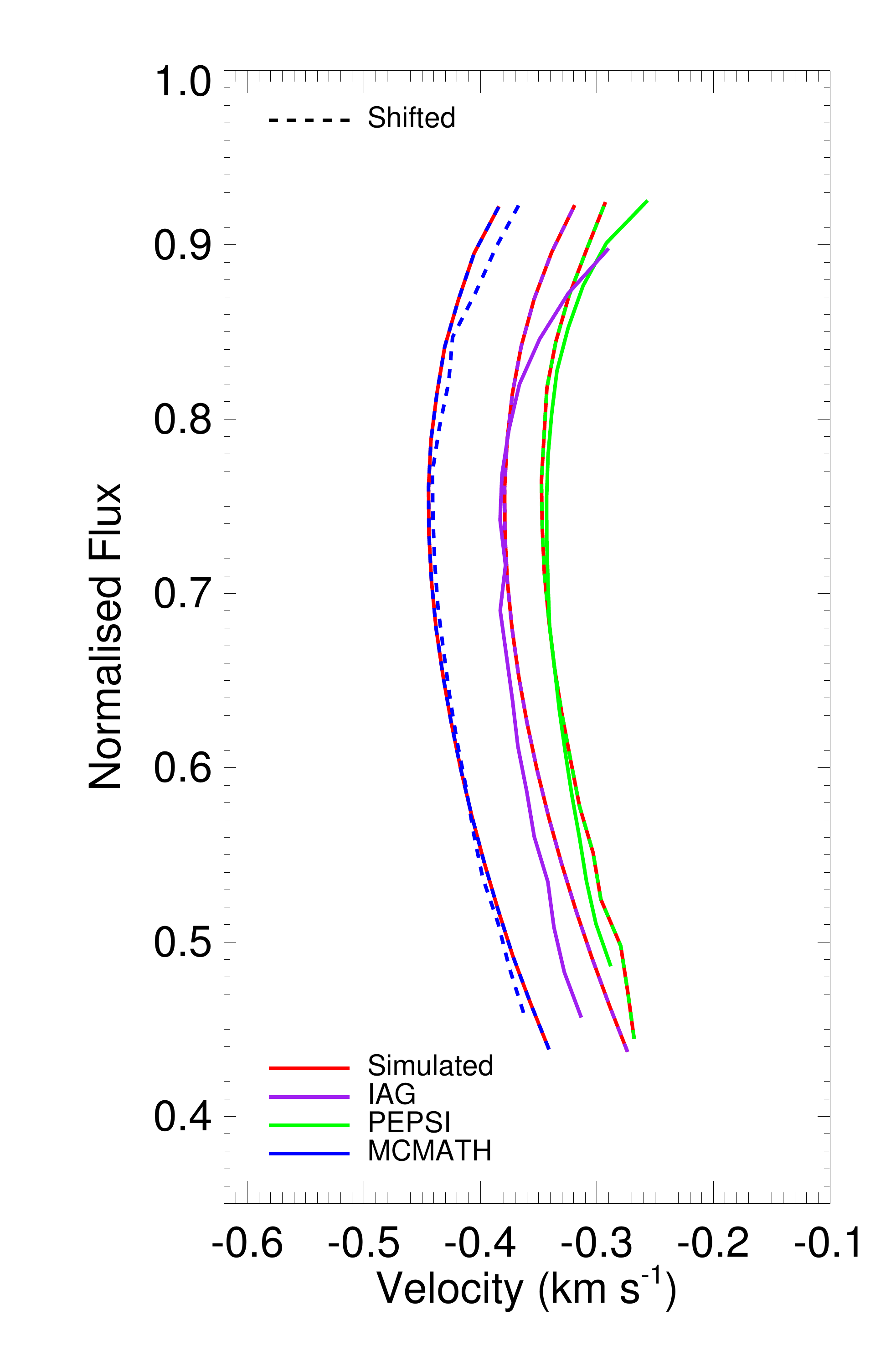} 
\caption{Left: average line bisector from the disc-integrated model observations (shown in red), alongside those from empirical data taken with the McMath (blue), IAG (purple), and PEPSI (green) instruments. Right: same, but model observations are convolved with instrumental profiles, and shifted for shape comparison; red-blue, red-purple and red-green lines correspond to the McMath, IAG, and PEPSI instruments, respectively. The McMath data is also shifted (dashed blue).}
\label{fig:obs_bi}
\end{figure}
Figure~\ref{fig:obs_bi} compares the line bisector of our Sun-as-a-star simulations against those from the solar atlases for the Fe~I~$6302~\mathrm{\AA}$ line. For the model star, we averaged together all 1,000 simulations before constructing the bisector. The models and observations were all normalised to their respective continuum intensities in the region of -14 to -15~km~s$^{-1}$, which corresponds to the continuum of the bluest region in the simulations; however, it is important to note that each solar atlas was already continuum normalised prior to this, so it is possible there may still be small deviations from this normalisation. The lefthand side of Figure~\ref{fig:obs_bi} shows the line bisectors at their absolute velocities (after correction for the gravitational redshift in the empirical data). The righthand side of Figure~\ref{fig:obs_bi} shows the model observation after convolutions with the appropriate instrumental profiles (IPs), with line shifts for viewing ease; dashed red-blue, red-purple, and red-green lines indicate the simulation was convolved with the McMath, IAG, and PEPSI IPs, respectively. For the IP, we assume a Gaussian with a full-width half-maximum (FWHM) based off the instrument resolution, centred on zero, and with a area normalised to one. 

As shown in Figure~\ref{fig:obs_bi}, the line bisector shape from the simulation agrees extremely well with the empirical data, and provides further evidence that the granulation parameterisation and model star generation is realistic. The slight increase in the downward trend of the bisector from the IAG likely originates from better resolving the partial blend with a neighbouring oxygen line in the empirical spectrum. It is important to note that the model star line profile was slightly deeper and wider than the observed lines, regardless of convolution with the IPs. These mismatches could come from small differences in the: continuum normalisation, stellar activity level, differences in the average magnetic field (as these will impact average flow velocities, contrasts between granules and intergranular lanes, and also the level of Zeeman splitting), and/or if the nominal resolution for this particular region was lower than the average for the atlas. Regardless, the good match between the line bisectors shows we capture the same shape/curvature characteristics, and can thus use these aspects with confidence when searching for granulation noise diagnostics. 

Similar to what we found in Paper~\citetalias{cegla18a} at discrete limb angles, the absolute velocity of the simulated bisector is more redshifted than the IAG and PEPSI data (as shown in Figure~\ref{fig:obs_bi}), which we again attribute to the increased magnetic field strength in the simulated data (200~G) compared to the quiet Sun ($\sim$100~G; \citealt{Trujillo04}) -- see Paper~\citetalias{cegla18a} for more details. The simulated bisector is considerably more redshifted than the McMath data; however, we attribute this to the lower absolute RV precision of McMath dataset as it deviates from the other empirical solar data by $\sim$300~m~s$^{-1}$. We note that placing the bisectors on an absolute scale is very difficult due to uncertainties in the tellurics and line blends, as well as uncertainties in the laboratory wavelengths \citep{dravins08}; even a small difference of 0.005~$\mathrm{\AA}$ leads to an offset $>$200~m~s$^{-1}$. Moreover, as we are interested in the application to exoplanet data, our main concern is the relative RV precision.
\begin{figure*}[t!]
\centering
\includegraphics[width = 8.5 cm]{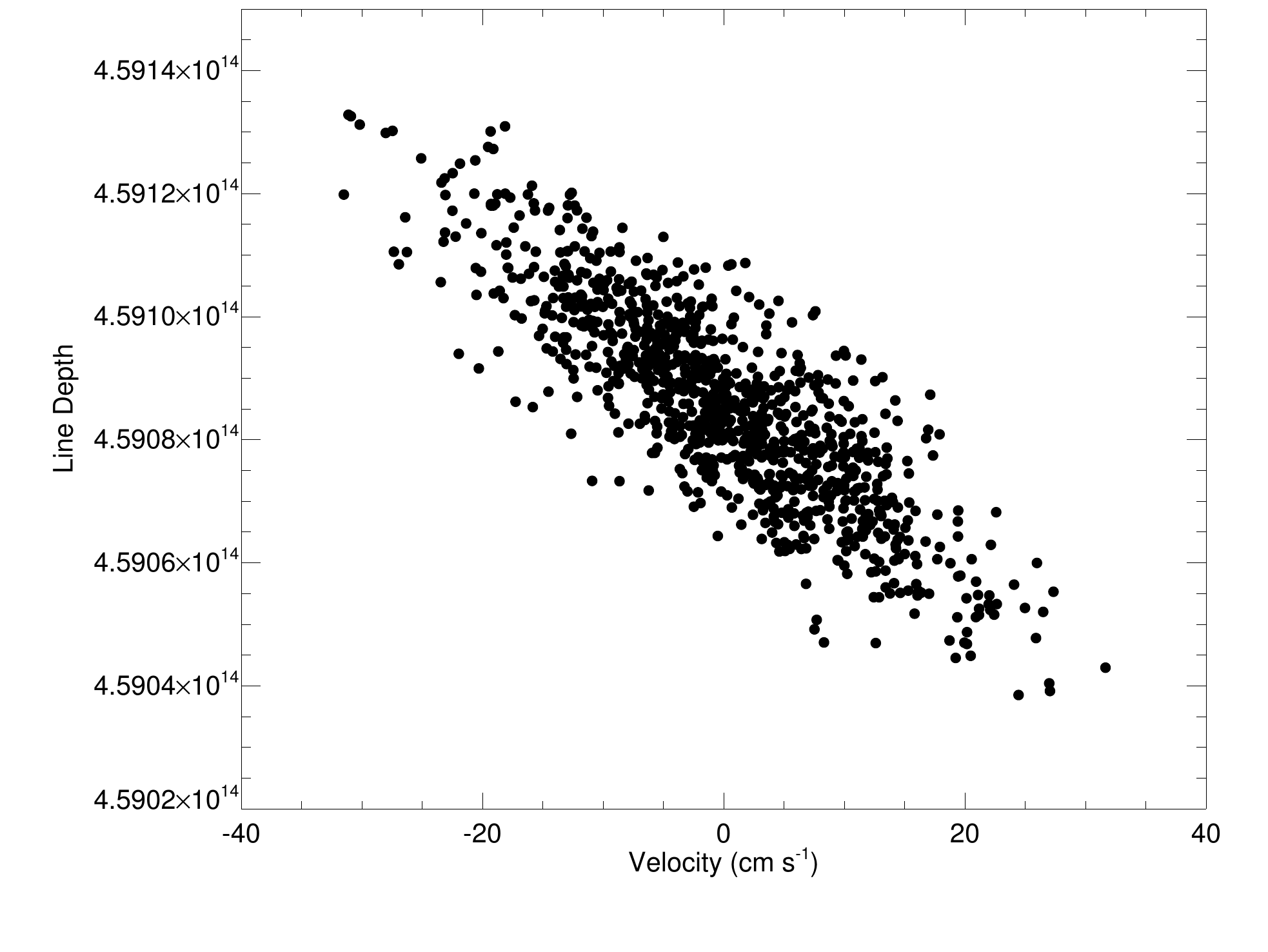} 
\includegraphics[width = 8.5 cm]{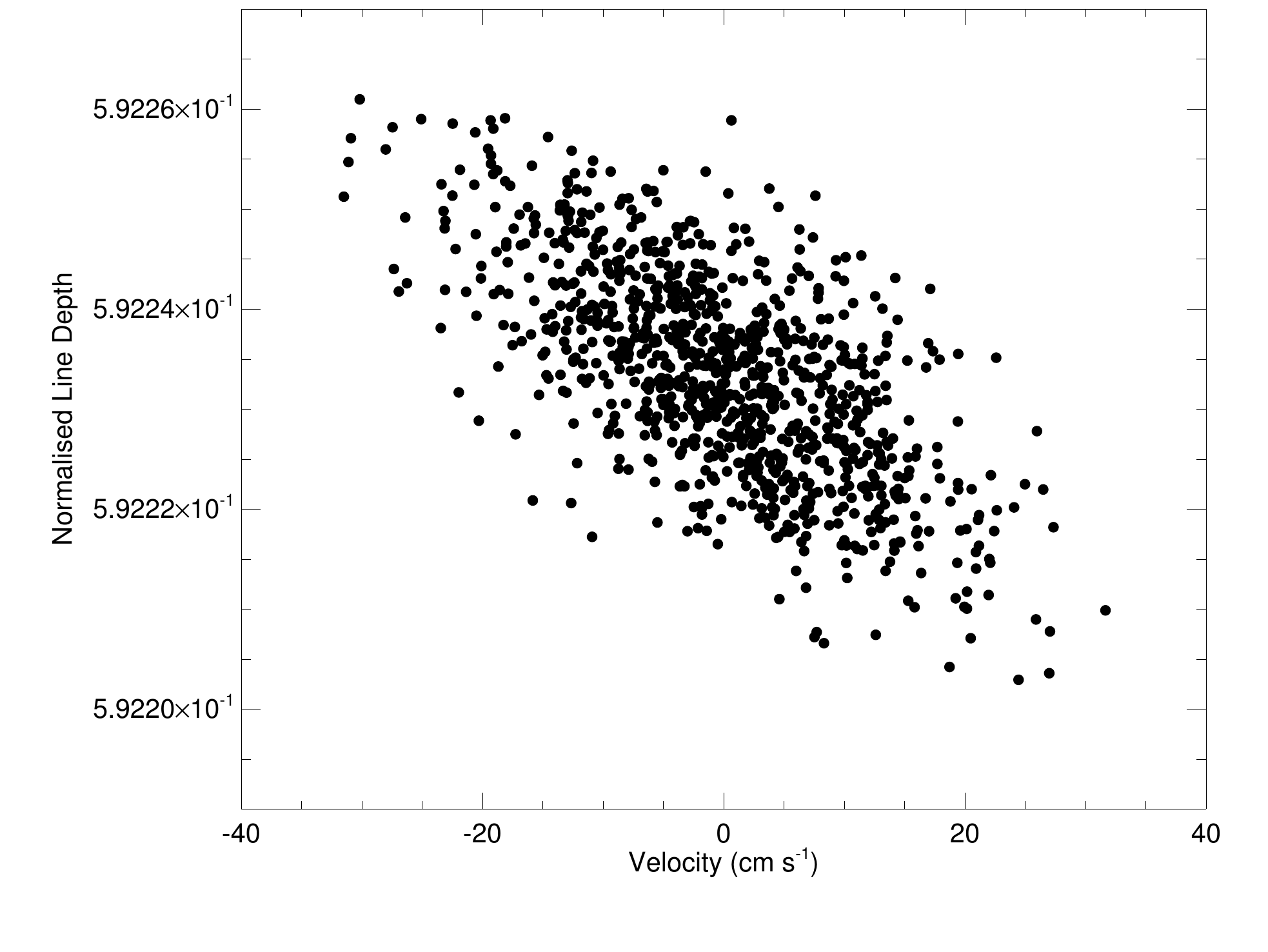} 
\vspace{-15pt}
\caption{Absolute (left) line depth and continuum normalised line depth (right) versus the convection-induced RVs; absolute flux is measured in units of erg s$^{-1}$ cm$^{-2}$ sr$^{-1}$ \AA$^{-1}$.}
\label{fig:depth}
\end{figure*}

To measure the RV variability, we cross-correlate each disc-integrated model observation with one template profile; for the template we use one randomly, selected disc-integrated profile. The RVs are then determined by fitting a polynomial to the peak of the cross-correlation function. The peak to peak RV variation of the 1,000 Sun-as-a-star model observations is $\sim$63~cm~s$^{-1}$ and the root-mean-square (rms) is $\sim$10~cm~s$^{-1}$. To compare with empirical observations, we turn to measurements of the solar background in helioseismology. \cite{harvey84,harvey85} proposed that this `background noise' can be modelled as a superposition of the temporal evolution from magnetic active regions, supergranulation, mesogranulation, and granulation. The assumption is that the time evolution of each phenomena can be approximated by an exponential decay with various characteristic timescales. Note today, that this often described by a linear combination of Lorentzian and super-Lorentzian functions (as done in Paper~ \citetalias{cegla18a}); at least two components are necessary to adequately fit the background signal from the granulation phenomena in empirical power spectra, though we highlight that the physical origin for a component on the mesoscale (not discussed here) remains an open question \citep[e.g. see][and references therein]{matloch10,kallinger14, corsaro17,rincon18,kessar19}. Such an approach provides a characteristic timescale, $\tau$, and a corresponding disc-integrated rms for each phenomenon, $\sigma$.

\cite{harvey84} found that the contribution from granulation was  $\sigma \approx$~70~cm~s$^{-1}$. However, \cite{elsworth94} later examined the same potassium line, at 770~nm, and report that, with their increased instrumental precision, they find the solar background power is almost five times lower and the $\sigma$ from granulation was only 31.9~$\pm$~9~cm~s$^{-1}$. \cite{palle99} also measured the solar background, but using the sodium doublet at 589~nm, and found the granulation rms to be 46.1~$\pm$~10~cm~s$^{-1}$; this measurement was determined over a period of 804 days, where it varied by $\sim$10\% according to the periodic variation in the depth of the operating point along the year \citep{palle99}. In addition to the helioseismology analyses, recent observations from the HARPS-N solar telescope indicate a daily correlated `noise' term that is also near 40~cm~s$^{-1}$, and is believed to be due to granulation \citep{cameron19}.

The exact granulation rms will depend on both the observed line choice, and the average magnetic field. Nonetheless, from the results above it appears our Sun-as-a-star observations underestimate the total RV rms by 3-4 times compared to the quiet Sun. This is not altogether unexpected given the increased magnetic field strength in our simulations, relative to the quiet Sun. For instance, in Paper~ \citetalias{cegla18a} we found that the increase in magnetic field strength relative to the quiet Sun reduced the net convective blueshift at disc centre by approximately a factor of 3. The convective blueshift is reduced in our simulations because the magnetic flux inhibits the convective flows; hence, it is natural to expect a similar decrease in the overall RV rms of our model observations. Going forward, we operate under the hypothesis that our granulation parameterisation is capturing the fundamental convection physics, and as such, that our model observations represent a scaled version of the quiet Sun. In future work, we will test this hypothesis by fully exploring the temporal variability of lower magnetic field strength simulations.

\section{Correlations}
\label{sec:corr}
In this section we analyse the influence of magnetoconvection on the disc-integrated Sun-as-a-star observations of the Fe~I~$6302~\mathrm{\AA}$ line profile (with an average magnetic field of 200~G). In particular, we examine a number of different line profile characteristics and their relation to the convective-induced RV shifts, where the RV is determined in the same manner as Section~\ref{subsec:obs}. See Sections~\ref{sec:add_fact} and \ref{sec:reduct} for the impact of instrumental resolution and stellar rotation.
 
\subsection{Line Depth and Width}
\label{subsec:depth_fwhm}
To begin, we examine the behaviour the line depth and width. The absolute line depth would only be available from space, as ground-based data will need to be continuum normalised to remove fluctuations due to the Earth's atmosphere; nonetheless, it is interesting to examine the normalised case in order to inform ourselves of potential degeneracies. Ultimately, the shape of the disc-integrated line profile depends on the underlying granulation pattern across the star. Since line profiles originating from the granules are deeper than those from the components in the intergranular lanes, we expect that a higher granule filling factor will induce both a greater blueshift and a deeper disc-integrated profile. Nonetheless, if there are instances when different combinations of filling factors can produce the same ratio from continuum to line core, then the normalised line depth may not correlate with the induced RV as strongly as the absolute depth.

If the line depth changes, this may be accompanied by a change in line width. To explore this, we examine the full width half maximum (FWHM), as it has been shown to be a strong indicator of large amplitude stellar activity signals, such as that from starspots \citep[e.g. see][]{queloz09, hatzes10,boisse11}. However, it is also important to note that the granule and non-magnetic intergranular lane (which dominate the profiles with a combined filling factor near $\sim$0.9), actually have quite similar half-maximum brightness measurements and similar FWHM measurements. Hence, variations in the ratio between these two components may change the RVs with very little change in FWHM. On the other hand, the intergranular lane components are redshifted relative to the granular component and thus could potentially contribute to width variations when the components are summed to create the average line profiles.

As shown in Figure~\ref{fig:depth} find there is a strong linear correlation, with a Pearson's R coefficient of -0.84 between the absolute line depth and the RVs, which decreases to -0.66 for the normalised line depth -- indicating that there may be some degeneracies introduced by the continuum normalisation. However, in both cases the change in line depth is incredibly small ($<< $1\%), and unlikely to be discernible in the presence of photon noise. That said, we expect that the RV rms may be underestimated in our models, compared to the quiet Sun, and as such we may expect a larger, more discernible, variation in reality -- future work with lower magnetic field simulations will be used to test this hypothesis. On the other hand, we find the FWHM does not correlate with the RV, wtih a Pearson's R coefficient of only 0.07 (figure shown in Appendix~\ref{appen:gran_plots}). Hence, the FWHM is unlikely to provide a good diagnostic for the short-term convection behaviour. However, if the magnetic field changes then the flow velocities and contrasts between the granular and intergranular lane components are likely to change and this may lead to FWHM changes. As such, in future work we intend to re-examine the impact of FWHM as a plage indicator. 

\subsection{Bisector Analysis}
\label{subsec:bisect}
\begin{figure}[t!]
\centering
\includegraphics[width = 8.5 cm]{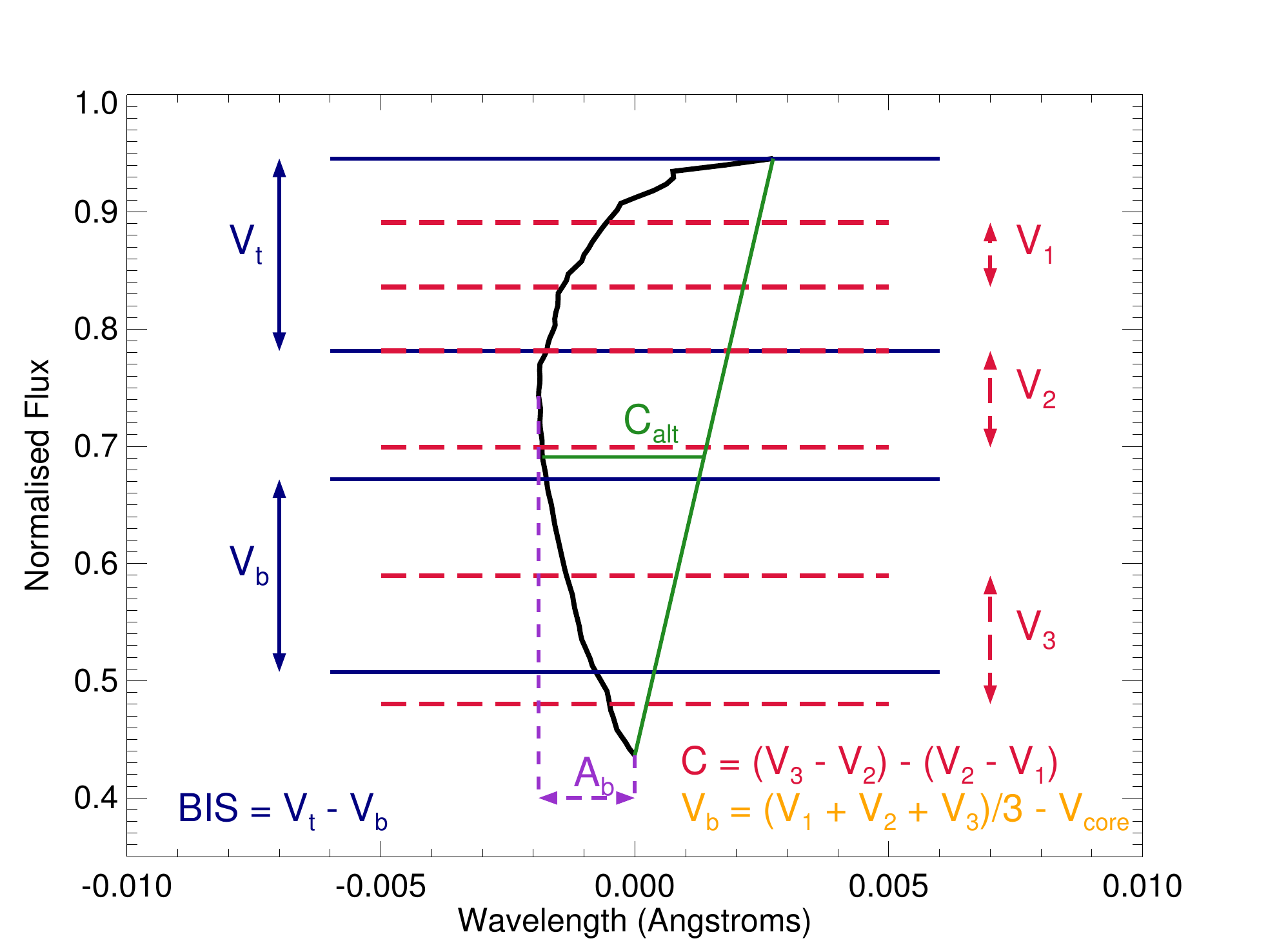} 
\vspace{-15pt}
\caption{Schematic to illustrate the various bisector diagnostics, with an example bisector \cite[from][]{kurucz84} shown in black. $BIS$ is defined as the difference between an average region near the top and bottom of the bisector (shown in blue).  $C$ is the difference between three average regions (shown in red), and the $V_b$ is average of those relative to the line core (defined in orange). $A_b$ is the difference between the line core and the most blue-shifted region of the bisector (shown in purple), and $C_{alt}$ is the perpendicular distance between the bisector and the midpoint of a line connecting the top and bottom of the bisector (shown in green).}
\label{fig:b_def}
\end{figure}

The line depth and FWHM only explore two particular regions of the line profiles. To go a step further, we can analyse the shape of the line profile through its bisector (defined as the midpoints of horizontal slices in the profile). There are a number of ways to characterise the bisector shape; herein, we utilise the bisector inverse slope\footnote{Also refereed to as bisector velocity span in the literature.} ($BIS$), curvature ($C$ and $C_{alt}$), bisector velocity displacement ($V_b$), and bisector amplitude ($A_b$) -- see Figure~\ref{fig:b_def} and \cite{queloz01, povich01, dall06} for definitions. All correlation plots for the bisector analysis can be found in Appendix~\ref{appen:gran_plots}.

Both the $BIS$ and $C$ look at the difference between average regions in the bisector and were designed to probe the impact of starspots, that effectively take a bite out of the line profile as the spot rotates in and out of view. If using these diagnostics as a granulation probe, then it may be better to fine-tune the regions to the C-shape of the observed line bisector. To test this, we used both the standard regions ($V_t$, $V_b$, $V_1$, $V_2$, $V_3$) and also iteratively varied them to maximise the correlation with the net convective RV shift; regions from 5 -- 95\% of the line depth, in steps of 5\% were explored (with regions defined with a width at least 5\% of the depth).

When using the standard $BIS$ regions (10-40\% and 55-90\% the line depth), we find a Pearson's R correlation coefficient of 0.57. Maximising this correlation, we find the strength increases dramatically to a Pearson's R of 0.93 if we probe the difference between 40-45\% and 80-95\% of the line depth. Hence, we find the strongest correlation when $V_b$ is bounded by a small region just above the line core and $V_t$ covers a similar region at the C-bend in the bisector (i.e. when the bisector has changed from increasing blueshift to decreasing in blueshift). Since the $V_t$ region is tied tightly to the `C'-bend in the bisector, it is entirely possible that the strength of this correlation will change based on which stellar lines are observed. The correlation strength will also likely be impacted by the stellar rotation and instrumental resolution, as these will change the bisector shape.  

For the bisector curvature, $C$, we find the standard regions (20 -- 30\%, 40 -- 55\%, and 75 -- 95\% of the line depth) result in a moderate-to-weak anti-correlation with the RVs, with a Pearson's R of -0.6. After fine-tuning, we find an optimal correlation when the upper, middle, and lower regions are defined as 15 -- 25\%, 25 -- 30\%, and 90 -- 95\%, respectively; with these we probe the bisector just above the C-bend and also near the line core, and the correlation increases substantially to a Pearson's R of -0.93. Similar to the BIS, the strength of this correlation will depend on the actual shape of the bisector, and will be subject to stellar rotation, instrumental resolution, and the innate line properties. 

\cite{povich01} argue that because the bisector curvature is essentially a second derivative of the line profile, it can require a high signal-to-noise ratio to measure accurately. While the signal-to-noise ratio is not a problem for our model observations, it could prove a limiting factor for the bisector analysis in real observations. This is our motivation for exploring the bisector velocity displacement, $V_b$, which averages the three regions from $C$ and measures their position relative to the line core. Here, the line core was determined by fitting a parabola to the bottom three points of the profile. Note, we differ from \cite{povich01}, in that our bisector curvature values are created by averaging over select regions rather than individual points.

Since the velocity displacement averages the bisector curvature values together, we investigate both standard and optimised curvature regions previously mentioned. We find a much stronger correlation between $V_b$ and RV than the original $C$-RV correlation (i.e. with the standard values), with a Pearson's R of 0.89. As such, the additional dimension of measuring the bisector shape relative to the line core seems to strengthen the correlation with convection-induced RV. On the other hand, when using the regions designed to optimise the curvature-RV relationship, we find the $V_b$-RV correlation is still very strong, but slightly weaker than the corresponding $C$-RV relation, with a Pearson's R of 0.90. 

We also examine an alternate measurement of the bisector curvature, $C_{alt}$, defined as the perpendicular distance between the bisector and the midpoint of a line that connects the bluest and reddest points of the bisector \cite{povich01}. As such, this provides a measure of how much the C-shape bends towards the blue-end of the spectrum. In this search, we limit the bisector to 10 -- 95\% of the line depth to avoid the very bottom and top of the line; we also average over individual points corresponding $\pm$~2\% of the flux value. We find $C_{alt}$ curvature measure has a much weaker correlation with the induced RV than that from $C$ above, with a Pearson's R of only -0.39. Moreover, because this definition of the curvature is based on very small regions, the signal-to-noise ratio necessary to construct such measurements is higher.

The last bisector measurement analysed was the bisector amplitude, $A_b$, exploring the difference between the line core and the bluest part of the bisector (determined from an average over a small region covering $\pm$~2\% of the flux). As such, this is a measure of the C-bend in the bisector, and since the C-shape is a direct result of granulation, $A_b$ should correlate with the convection-induced RV shifts. Indeed, we find a strong correlation, with a Pearson's R coefficient of -0.89. Comparing this strength to the $C_{alt}$-RV relation indicates that the bend in the C-shape, where the profile is most blueshifted, happens at a slightly different region than that probed by $C_{alt}$. 

While the centre of the line can be measured to very high precision observationally with the cross-correlation technique, this is not necessarily the case for the most blueshifted point of the bisector. As a result, even if spectrographs reach cm~s$^{-1}$~precision, we may need to average over a slightly larger region of the most blueshifted point to beat down the photon noise. Regardless, the strength of this correlation clearly indicates that the bisector amplitude is a very simple, fast and potentially powerful granulation noise diagnostic.

\subsection{Full Profile Analysis}
\label{subsec:full_prof}
With a view toward combating photon noise in future empirical observations, and maximising the information content from the available spectrum, we also want to investigate diagnostics that utilise the entire line profile. We start this investigation by looking at the two stellar activity indicators introduced by \cite{figueira13}. The first is termed bi-Gaussian fitting; it was originally developed by \citeauthor{nardetto06} (2006 -- and references therein) to analyse the line centres and asymmetries in the line profiles of pulsating stars. Bi-Gaussian fitting involves simultaneously fitting the left and right side of the line profile (or CCF) with a Gaussian that contains an additional asymmetry parameter ($A$); for this we use a Levenberg-Marquardt least-squares fitting. The difference between RV centroids from the bi-Gaussian and the pure Gaussian fits, $\Delta V$, then acts as a proxy for the RV induced from the line/CCF asymmetry\citep[see Appendix~\ref{appen:gran_plots} and][for more details]{figueira13}. We find a moderately strong anti-correlation between the convection-induced RV shifts and $\Delta V$, with a Pearson's R coefficient  of -0.71. 

Since $\Delta V$ originates from the line asymmetry, we also examined the relationship between RV and the asymmetry fraction of the FWHM, $A$. As one might expect, the strength of the correlation is the same as that for $\Delta V$ (with a Pearson's R of 0.70). The interesting thing to note about the $A$-RV relation is the variation in $A$ seen for the disc-integrated model profiles. In \cite{figueira13}, the authors used bi-Gaussian fitting to explore the impact of starspots, where their simulations indicated changes in $A$ of $\sim$2\%. Since convection is a smaller amplitude effect, compared to spots, we expect a lower variation in $A$; however, we find $A$ only changes by $\sim$0.01\%, which means this correlation may be very difficult to see in the presence of photon noise. 
 
The second activity indicator introduced in \cite{figueira13} is the velocity asymmetry, $V_{asy}$. The goal of the $V_{asy}$ indicator is to compare the RV spectral information on the blue wing to that on the red wing for the entirety of the line profile and/or CCF. It allows us to evaluate the gradient in the spectral line for both the blue and red wings separately at equal flux values, and condenses these gradients into one (weighted) average measurement per observation. We follow the updated definition from \cite{lanza18} -- see Appendix~\ref{appen:gran_plots} for more details. For our models, there is a strong correlation with RV, with a Pearson's R of -0.92. Given the strength of this correlation, the $V_{asy(mod)}$ diagnostic has potential to help disentangle convection and should be further pursued in attempts to remove activity-related noise in stellar observations. However, further investigation from solar observations and simulations of varying magnetic field will help quantify how much we expect $V_{asy(mod)}$ to vary and whether this can be discernible with future instrumentation. 

Another way to probe the behaviour of an entire line profile is to measure its equivalent width (EW). Since the EW essentially measures the area of a spectral line profile, it could be affected by convective-induced variability. In fact, it is well known that the net convective blueshift of spectral lines correlates with their EWs \citep{AllendePrieto98, ramirez08, reiners16}, and that both quantities are altered by magnetic fields \citep{meunier10a,fabbian10}. The EW has also long-been used as a diagnostic of pressure-mode oscillations, excited by surface convection, and evidence for granulation can be seen in the power spectra derived from EW variations \citep{kjeldsen95,kjeldsen99}. Power spectra from RVs indicate a similar behaviour, which suggests convection may cause the two quantities to manifest in a similar frequency structure; yet, it remains to be seen if the EWs and RVs correlate with one another contemporaneously. 

 In particular, we know a granule line profile has a larger EW than those originating from the intergranular lanes, partially because these profiles are formed higher in the atmosphere and have larger line depths. Hence, instances when more granules are present, could lead to profiles with larger EWs and greater blueshifts. This is indeed what we find, where there is a strong correlation between the EW and RV and a Pearson's R coefficient of -0.91. The EW is a particularly powerful diagnostic for ground-based data since it is independent of both the continuum intensity and the stellar rotation. However, RV precision is not independent of rotation as it is more difficult to determine precise RVs at higher stellar rotation rates; hence, the strength of this correlation could still be slightly impacted by stellar rotation. 

\subsection{Photometric Brightness}
\label{subsec:area}
Although we lose the absolute continuum information in ground-based spectra, we can retrieve it from photometry. This could potentially be useful as a granulation noise diagnostic, as one would naively expect a star with a greater granule filling factor to be both brighter and more blueshifted. However, in reality this may not be the case if the stochastic nature of granulation washes out such a correlation. In line with this latter point, \cite{meunier15} failed to see such a correlation in their granulation models, which were based off a combination of empirical solar relations and velocities derived from a hydrodynamical simulation. \cite{meunier15} argued that the relationship between observed granule size, velocity, and brightness was noisy and lost its coherence when integrating across the entire stellar disc. Our approach differs from \cite{meunier15} in two key ways: our model backbone is purely MHD-based (and therefore not limited by instrumental precision) and the velocities are derived from the line profile shapes (rather than the raw x, y, z velocities from the HD/MHD simulation). 

\begin{figure}[t]
\includegraphics[width = 8.5 cm]{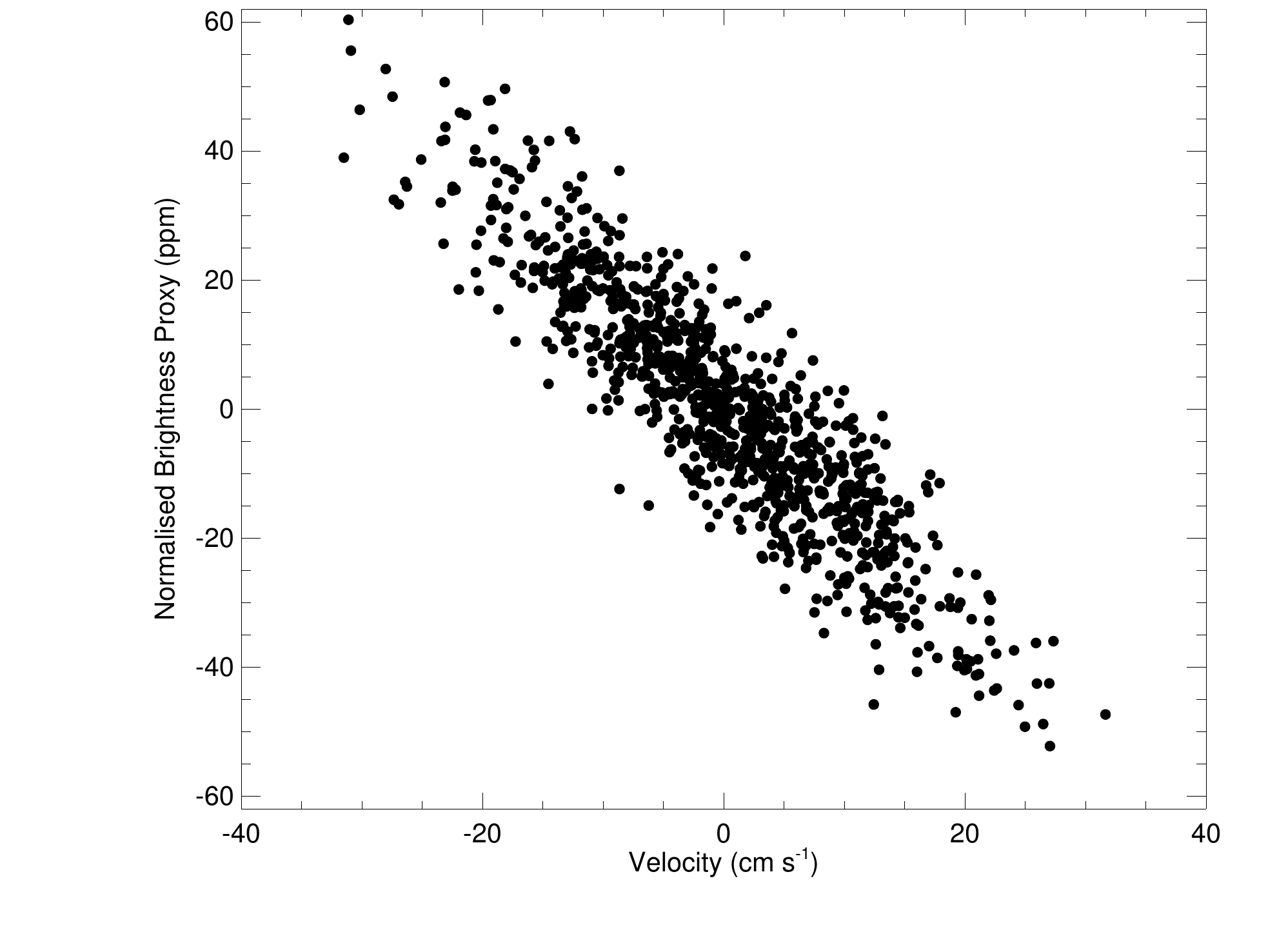} 
\vspace{-15pt}
\caption{A brightness proxy versus RV, where the proxy is determined by the integrated area underneath the disc-integrated model line profiles. This area has been normalised by its maximum value, mean-subtracted, and converted to parts per million (ppm).}
\label{fig:area}
\end{figure}

To emulate photometric measurements, we integrate the area under the disc-integrated model line profiles and use this as a proxy for the photometric brightness. It is important to note that we are `simulating' photometric variability over a very small wavelength range, and operating under the assumption the changes seen here will also be present across the observed wavelength region. To have a feel for the magnitude of the changes in this brightness proxy, we normalise by the maximum area, subtract off the mean, and convert to parts per million (ppm). This normalised brightness proxy is shown in Figure~\ref{fig:area}, where we see the predicted correlation, with the largest brightness corresponding to the largest blueshifts. In fact, this is one of the strongest correlations with convective-induced RV, with a Pearson's R of -0.90. The peak-to-peak variation is $\sim$110~ppm, with an rms of  20~ppm, which would be measurable with current space-based photometric technology. For instance, the Kepler satellite was designed to detect Earth-analog planets, with a dip in brightness corresponding to 84 ppm for an Earth-size planet around a Sun-sized star, and has demonstrated 10~ppm precision. The limiting factor to discern this correlation comes down to the RV precision, which should be attainable with ESPRESSO, and potentially EXPRES, for bright targets -- such as those that will be targeted with TESS and PLATO.  

It is not entirely clear why we see such a strong brightness-RV correlation with our models, while \cite{meunier15} do not see it in their model stars. It could potentially stem from the instrumental noise inherent in the empirical relationships governing the granular evolution in  \cite{meunier15}, which are circumvented here. However, it could also be due to a flaw in our granulation parameterisation, that could be missing some physics. For example, the nature of the four-component parameterisation assumes fixed flow velocities and continuum contrasts for each component; this was based off little variation/impact of these parameters in our MHD time-series, but may not be reflective of a global pattern across the stellar disc or for other line profiles. This parameterisation is also tied to the physical size of the MHD snapshots, which is incorporated in the stellar grid, while \cite{meunier15} populate their star with individual granules; hence, there could be some over-averaging in our approach, but as each tile is independent and their the physical area is quite small ($\sim$0.005\% of visible stellar disc) this seems unlikely. Regardless, these aspects will be explored in future work, and we will seek to empirically validate this relationship.  

It is also interesting to note that the rms of the brightness fluctuations seen here are of similar magnitude to those observed on the Sun. For example, \cite{kallinger14} report that the rms intensity fluctuations from solar granulation fall just shy of 30 ppm, as measured by the VIRGO instrument aboard the SOHO satellite. On top of this, the Sun's photometric variability extends from $\sim$100 -- 1800 ppm over the course of a solar cycle with an average variability near $\sim$400 ppm, as seen from the SOHO data presented in \citeauthor{bastien13} (2013 -- and references therein). Our model star observations have an average magnetic field of $200~\mathrm{G}$, which is more active than the quiet Sun, but does not produce the sunspots found in the more active parts of the solar cycle. In addition, we have evidence that the magnetic field in our simulations is large enough to restrict flow velocities (e.g. redshifts relative to the quiet Sun), hence it may actually act to inhibit the overall photometric variability. Consequently, the brightness fluctuations of $\sim$110~ppm peak-to-peak and  20~ppm rms seen in our model stars is credible, given the magnetic field, and acts as further evidence that the granulation parameterisation and noise diagnostics are representative of the physics taking place in photospheric magnetoconvection.

If this brightness-RV relationship is confirmed, its strength here implies that photometry may be among the best ways to identify granulation-induced RVs shifts. Accordingly, high precision simultaneous photometry from space-based transit missions could potentially be a key to disentangling the granulation noise within the spectroscopic observations.

\section{Additional Factors}
\label{sec:add_fact}
In addition to the inherent stellar line properties (e.g. formation height, excitation potential, Land\'{e} factor etc.), there are a number of additional factors that can impact the line profile shape and our ability to discern that shape. Here we will explore the impact of instrumental resolution and the inherent stellar rotation. 

\subsection{Impact of Instrumental Resolution}
\label{subsec:Ip}
\begin{figure}[t]
\includegraphics[width = 8.5 cm]{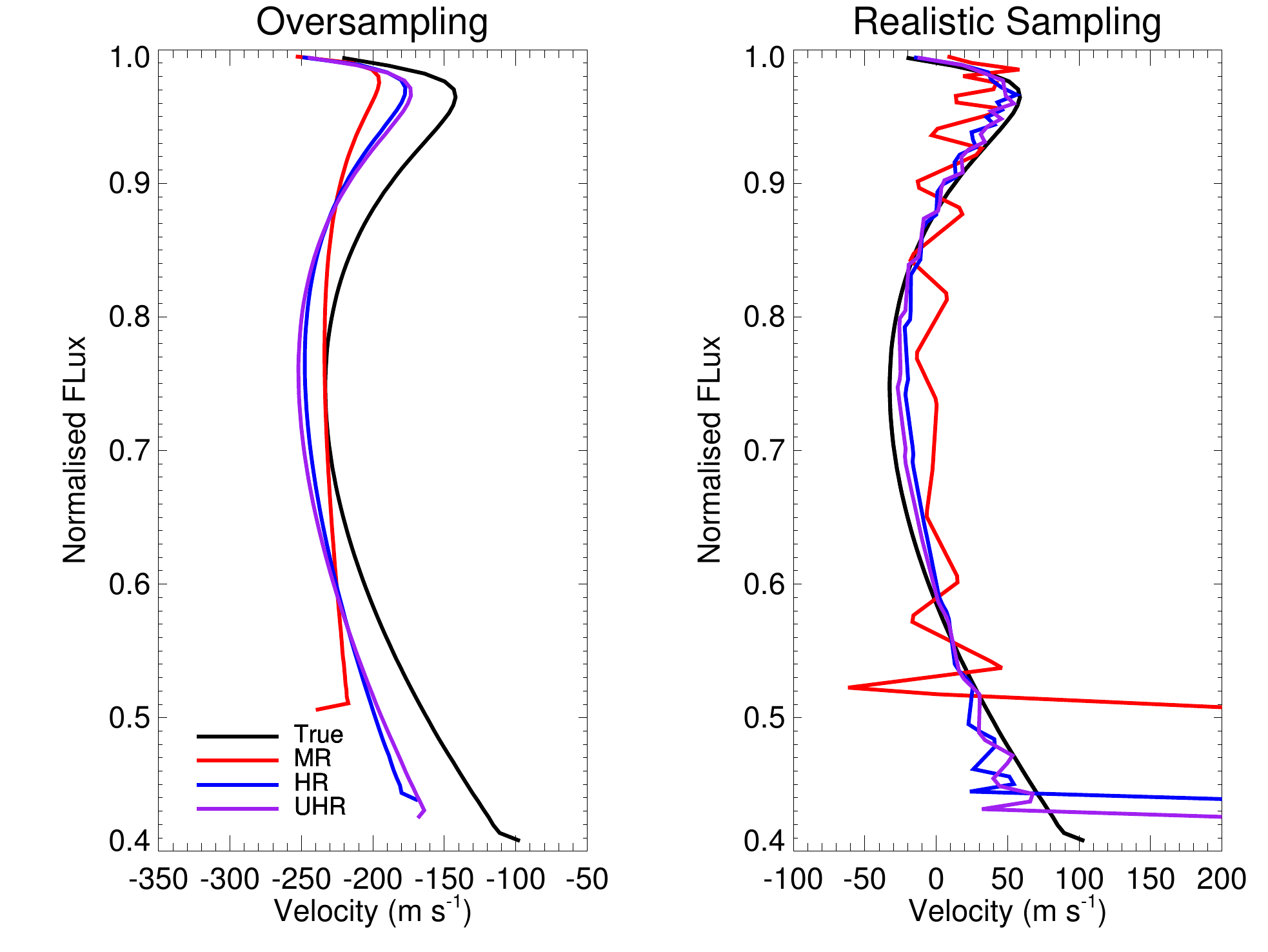} 
\vspace{-15pt}
\caption{Line bisector of the average model line profile before (in black) and after it was convolved with an ESPRESSO-like IP corresponding to its three modes: Medium Resolution (70,000), High Resolution (140,000), and Ultra-High Resolution (190,000). Left: bisectors when the profile sampling has not changed; right: when the sampling changes corresponding to the resolution mode.}
\label{fig:Res}
\end{figure}
\begin{figure*}[t]
\centering
\includegraphics[scale = 0.45]{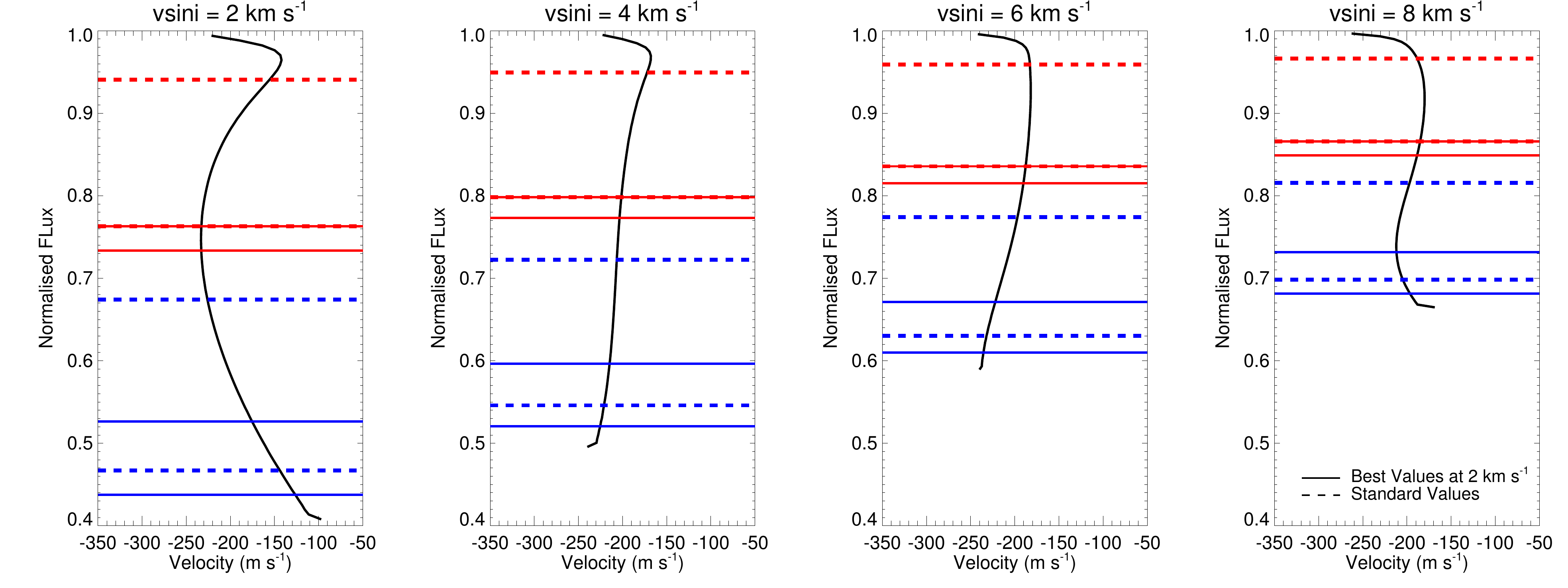} 
\vspace{-15pt}
\caption{Average line bisectors from model observations with stellar rotation rates corresponding to $v \sin i$ = 2, 4, 6, and 8~km~s$^{-1}$ (left to right, respectively). The $\overline{V_t}$ and $\overline{V_b}$ regions of the BIS are over-plotted in red and blue, respectively;  dashed lines show the standard regions from the literature and solid lines show the regions fine-tuned to the solar rotation (2~km~s$^{-1}$).}
\label{fig:stel_rot}
\end{figure*}
The instrumental resolution of a spectrograph determines how well we are able to measure the shape of any given stellar line. A low instrumental resolution will smear out line asymmetries due to a broader point-spread function; this can make it difficult to measure small asymmetries, including those induced by granulation. To explore this we assume ESPRESSO-like hypothetical instrumental profiles (IPs). The ESPRESSO spectrograph is highly stabilised and its IP can be approximated by a Gaussian. The width of each Gaussian corresponds to ESPRESSO's three modes: Medium Resolution (MR), High Resolution (HR), and Ultra-High Resolution (UHR), with a resolving power of 70,000, 140,000, and 190,000, respectively. We convolve each of these IPs with an average line profile from the 1,000 Sun-as-a-star model observations. If we sample the line profile at the same rate as the true, simulated line profile, then we can isolate the impact of the line profile shapes. The corresponding line bisectors for this case are shown in the lefthand side of Figure~\ref{fig:Res}. We can clearly see that the lower resolution smears out the asymmetries, as the bisector from the MR case is nearly straight. Fortunately, we see both the HR and UHR modes can capture most of the line asymmetries, with only small deviations from the original C-shape. However, we note that this may not necessarily be the case for line profiles with different shapes. For example, hotter stars will have the C-bend occur much lower down in the line profile, which will be more heavily impacted by the resolving power. 

Naturally, the measured line profile shapes are also impacted by how finely we can sample line profiles (which depends on the Echelle grating and the number of pixels on the CCD). For the HR and UHR modes of ESPRESSO, the true sampling rate is 0.5~km~s$^{-1}$. The MR mode is binned by a factor of two, so the true sampling is 1~km~s$^{-1}$. Taking both of these sampling rates into account by interpolating onto grids with the respective samplings, we re-do the convolutions and plot their resultant bisectors in the righthand side of Figure~\ref{fig:Res}. From this, it is clear that the sampling rate of the lower resolution mode adds significant noise to the bisector calculation. Hence, it will be very important to have high to ultra-high resolution if we want to measure the minute asymmetry variations induced by convection. This aspect will be further discussed in Section~\ref{sec:reduct}.

\subsection{Impact of Stellar Rotation}
\label{subsec:rot}
If the local photospheric line profile is symmetric, then an increase in stellar rotation will result in a broader, shallower disc-integrated profile, with symmetric changes on both line wings and zero change to the line core. An asymmetric local line profile complicates this picture significantly. Ultimately, the disc-integrated profile will still be shallower and broader, but the changes in the line wings will not be symmetric and the line core will also be impacted. The exact nature of the stellar rotation impact will be dependent on the underlying line profile shape, which is dependent on the particular stellar line and the magnetic field of the photosphere. Even if the line profile is not magnetically sensitive, the magnetic field can alter the thermodynamic structure of the convection, which will impact the line shape \citep{beeck15b}. To explore how this impacts our simulated Fe~I~6302$\AA$ line profile (with 200~G average field strength), we re-ran all 1,000 model star observations for various solid body stellar rotation rates corresponding to a $v \sin i$ of : 4, 6, 8, and 10~km~s$^{-1}$. The average bisector from these model observations are shown in Figure~\ref{fig:stel_rot} for $v \sin i$ of 4 -- 8~km~s$^{-1}$, alongside the original 2~km~s$^{-1}$. For slow-rotators like the Sun, differential rotation has a minimal influence on the line profiles and does not significantly modify our final results -- see Appendix~\ref{appen:other_noise} for more details.

With increasing stellar rotation, the line core first increases in blueshift and then increases in redshift, while the line wings continue to blueshift; this is in agreement with the behaviour seen by \cite{beeck15b} for their 100~G MHD simulations of the Fe~I~6165 and 6173~$\AA$ lines. Even a small change in $v \sin i$ from 2 to 4~km~s$^{-1}$ makes a remarkable change in the line profile, with far less asymmetry seen in the bisector when increasing the $v \sin i$ to 4~km~s$^{-1}$. At higher rotations we begin to see more asymmetry in the line bisector, but its behaviour is quite different from the solar rotation case. To further highlight these differences, we over-plotted the bisector regions probed by the two different BIS definitions in Section~\ref{subsec:bisect} (from the standard literature values and when fine-tuned to the solar rotation case). From Figure~\ref{fig:stel_rot}, it is clear that granulation-noise diagnostics probing the line shape and corresponding bisector will perform very differently depending on the stellar rotation (further discussed in Section~\ref{sec:reduct}).

\section{Noise Reduction}
\label{sec:reduct}
In this section, we analyse the effectiveness of each line profile characteristic as a granulation-noise mitigation tool. From Section~\ref{sec:corr}, it is clear that many of the line characteristics are linearly correlated with the granulation-induced RV shifts. Hence, we can heavily mitigate this noise by subtracting off a linear fit between each diagnostic and the RVs. To determine the noise mitigation success, we compare the correlation-corrected RV rms to the original, uncorrected rms. Even though the total RV variability may be underestimated in our model observations as compared to the quiet Sun (see Section~\ref{subsec:obs}), the fractional reduction in the rms should still scale with the true value if the fundamental physics is correct. Hence, in Table~\ref{tab:noiseremove_tab1}, we present the fractional reduction in the RV rms; negative fractions represent cases where an attempt to remove the correlation actually increased the RV scatter, as is the case when the correlation was weak or non-existent. 
\begin{table*}[t]
\caption{Granulation noise reduction for various diagnostics, alongside their Pearson's R correlation coefficient.}
\centering

\begin{tabular}{c|cc|cc|cc|cc}
    \hline
    $v \sin i$ = 2~km~s$^{-1}$ & & & \multicolumn{2}{c|}{ Resolution = 70,000} &  \multicolumn{2}{c|}{ Resolution = 140,000}  &  \multicolumn{2}{c}{ Resolution = 140,000}   \\
    \hline
     Diagnostic & Reduction (\%) & R & Reduction (\%) & R  & Reduction (\%) & R & Reduct. (\%) & R\\
    \hline
 Abs. Depth & 36.37 & -0.84 & 35.65 & -0.84 & 32.42 & -0.83 & 32.66 & -0.83 \\
Norm. Depth & -13.38 & -0.66 & -20.02 & -0.64 & -28.97 & -0.61 & -28.04 & -0.62 \\
FWHM & -1387.91 & 0.07 & -586.67 & 0.14 & -831.93 & 0.11 & -4468.55 & -0.02 \\
BIS$_{\rm{std}}$ & -42.31 & 0.57 & -81.40 & 0.48 & -48.05 & 0.56 & -44.98 & 0.57 \\
BIS$_{\rm{best}}$ & 61.20 & 0.93 & 11.16 & 0.75 & 52.37 & 0.90 & 57.41 & 0.92 \\
C$_{\rm{std}}$ & -34.29 & -0.60 & 39.12 & -0.85 & -51.39 & -0.55 & -6.30 & -0.69 \\
C$_{\rm{best}}$ & 61.63 & -0.93 & 36.46 & -0.84 & 45.50 & -0.88 & 50.80 & -0.90 \\
V$_{\rm{b,std}}$ & 49.58 & 0.89 & -0.77 & 0.70 & 56.72 & 0.92 & 59.24 & 0.93 \\
V$_{\rm{b,best}}$ & 52.32 & 0.90 & -58.94 & 0.53 & 36.78 & 0.85 & 43.54 & 0.87 \\
C$_{\rm{alt}}$ & -135.77 & -0.39 & -694.33 & 0.12 & -234.10 & -0.29 & -170.33 & -0.35 \\
A$_{\rm{b}}$ & 48.10 & -0.89 & 3.93 & -0.72 & 54.01 & -0.91 & 60.68 & -0.93 \\
$\Delta$V & -0.53 & -0.71 & -66.31 & -0.52 & -25.83 & -0.62 & -15.42 & -0.65 \\
A & -0.63 & 0.70 & -66.35 & 0.52 & -25.81 & 0.62 & -15.44 & 0.65 \\
V$_{\rm{asy}}$ & 57.85 & -0.92 & 49.40 & -0.89 & 36.71 & -0.85 & 36.54 & -0.84 \\
EW & 53.91 & -0.91 & 52.83 & -0.90 & 52.74 & -0.90 & 52.66 & -0.90 \\
Brightness & 52.70 & -0.90 & 54.30 & -0.91 & 53.02 & -0.91 & 52.76 & -0.90 \\   
    \hline
    $v \sin i$ = 4~km~s$^{-1}$\\
    \hline
Abs. Depth & -51.21 & -0.55 & -42.63 & -0.57 & -66.16 & -0.52 & -70.06 & -0.51 \\
Norm. Depth & -258.48 & -0.27 & -305.51 & -0.24 & -466.26 & -0.17 & -493.11 & -0.17 \\
FWHM & -932.73 & -0.10 & -270.23 & -0.26 & -377.25 & -0.21 & -521.71 & -0.16 \\
BIS$_{\rm{std}}$ & -212.15 & -0.31 & -235.44 & -0.29 & -201.93 & -0.31 & -201.09 & -0.32 \\
BIS$_{\rm{best}}$ & -300.44 & -0.24 & -260.68 & -0.27 & -235.40 & -0.29 & -257.31 & -0.27 \\
C$_{\rm{std}}$ & -235.87 & -0.29 & -2655.09 & -0.04 & -213.01 & -0.30 & -213.20 & -0.30 \\
C$_{\rm{best}}$ & -316.02 & 0.23 & -414.35 & 0.19 & -454.37 & 0.18 & -241.24 & 0.28 \\
V$_{\rm{b,std}}$ & -428.24 & -0.19 & -307.87 & -0.24 & -192.87 & -0.32 & -212.02 & -0.31 \\
V$_{\rm{b,best}}$ & -251.70 & -0.27 & -272.93 & -0.26 & -178.59 & -0.34 & -182.96 & -0.33 \\
C$_{\rm{alt}}$ & -167.97 & -0.35 & -404.79 & -0.19 & -189.73 & -0.33 & -182.47 & -0.33 \\
A$_{\rm{b}}$ & -1962.35 & -0.05 & -1641.84 & -0.06 & -741.17 & 0.12 & -648.14 & 0.13 \\
$\Delta$V & -196.16 & 0.32 & -223.28 & 0.30 & -199.24 & 0.32 & -196.04 & 0.32 \\
A & -196.81 & -0.32 & -223.97 & -0.29 & -199.92 & -0.32 & -196.72 & -0.32 \\
V$_{\rm{asy}}$ & -275.98 & 0.26 & -27.84 & 0.62 & -307.18 & 0.24 & -203.44 & 0.31 \\
EW & 25.66 & -0.80 & 28.86 & -0.81 & 26.40 & -0.81 & 25.94 & -0.80 \\
Brightness & 19.88 & -0.78 & 23.50 & -0.79 & 20.67 & -0.78 & 20.17 & -0.78 \\
    \hline
   $v \sin i$ = 6~km~s$^{-1}$\\
   \hline
Abs. Depth & -210.79 & -0.31 & -180.38 & -0.34 & -262.05 & -0.27 & -280.50 & -0.25 \\
Norm. Depth & -3529.02 & -0.03 & -202717.06 & 0.00 & -1835.29 & 0.05 & -1589.69 & 0.06 \\
FWHM & -375.79 & -0.21 & -193.26 & -0.32 & -244.97 & -0.28 & -60.50 & -0.53 \\
BIS$_{\rm{std}}$ & -40.16 & -0.58 & -42.70 & -0.57 & -38.33 & -0.59 & -39.24 & -0.58 \\
BIS$_{\rm{best}}$ & -38.26 & -0.59 & -41.13 & -0.58 & -36.81 & -0.59 & -37.04 & -0.59 \\
C$_{\rm{std}}$ & -82.91 & 0.48 & -38.43 & 0.59 & -65.99 & 0.52 & -70.29 & 0.51 \\
C$_{\rm{best}}$ & -37.59 & 0.59 & -41.10 & 0.58 & -39.79 & 0.58 & -39.92 & 0.58 \\
V$_{\rm{b,std}}$ & -47.48 & -0.56 & -43.14 & -0.57 & -30.48 & -0.61 & -32.53 & -0.60 \\
V$_{\rm{b,best}}$ & -46.71 & -0.56 & -44.04 & -0.57 & -31.45 & -0.61 & -33.04 & -0.60 \\
C$_{\rm{alt}}$ & -521.42 & -0.16 & -4466.65 & 0.02 & -932.33 & -0.10 & -729.72 & -0.12 \\
A$_{\rm{b}}$ & -170.79 & 0.35 & -42.86 & 0.57 & -43.48 & 0.57 & -75.93 & 0.49 \\
$\Delta$V & -33.02 & 0.60 & -40.75 & 0.58 & -33.90 & 0.60 & -32.86 & 0.60 \\
A & -33.47 & -0.60 & -41.10 & -0.58 & -34.32 & -0.60 & -33.29 & -0.60 \\
V$_{\rm{asy}}$ & -40.52 & 0.58 & -44.78 & 0.57 & -40.70 & 0.58 & -33.70 & 0.60 \\
EW & -6.59 & -0.68 & -4.23 & -0.69 & -6.26 & -0.69 & -6.71 & -0.68 \\
Brightness & -14.76 & -0.66 & -11.20 & -0.67 & -14.37 & -0.66 & -14.90 & -0.66 \\   
 \end{tabular}

\label{tab:noiseremove_tab1}
\end{table*}  

We find that very strong correlations are needed to reduce the RV variability. For example, diagnostics that produced correlations with a Pearson's R of $\sim |0.7|$ resulted in effectively no change in the RV rms (e.g. those from the bi-Gaussian fitting, $\Delta V$ and $A$). This is likely because the RVs were not uniformly distributed, but instead heavily clustered around 0 shift, and subtracting off a linear fit between the diagnostic and RV will most heavily impact the extremes. Nonetheless, there are many diagnostics that show a very strong correlation and significant noise mitigation. For example, the fine-tuned BIS and bisector curvature, (both) velocity displacement(s), bisector amplitude, $V_{asy}$, equivalent width, and brightness all remove around 50\% of the granulation noise; this could mean up to four times less observing time is required to reach desired RV precisions. 

It is also interesting to note that the increase in scatter in the RV correlation when changing from absolute to continuum normalised line depth is sufficient to kill any noise reduction in the model observations; this may not be the case if the RV variability were greater, but should be carefully considered for ground-based observations. We also want to draw attention to the significant increase in noise reduction ability when fine-tuning the BIS and bisector curvature regions, as compared to using the standard ranges in the literature; in both instances, the diagnostic goes from zero noise reduction to a reduction $>$60\%. This highlights the importance of examining the particular stellar line characteristics rather than applying a blanket approach to all lines. However, we do see that the velocity displacement, which utilises the same regions as the curvature, performs well when using both the standard and fine-tuned regions. As such, in the future we will  explore this behaviour in additional stellar lines to confirm if the additional information from the line core is sufficient to make these diagnostic more independent of the particular line shape than the bisector curvature. Additionally, observations of starspots have taught us that we often need to use many diagnostics to determine the stellar activity behaviour, and we recommend the same for granulation; this may be especially important as the total variation in each diagnostic is small and may be difficult to discern empirically. 

In Section~\ref{sec:add_fact}, we also explored the impact of various ESPRESSO-like instrumental resolutions, as well as the effect of varying the stellar rotation from solar-like up to a $v \sin i$ of 10~km~s$^{-1}$. The noise mitigation results at each instrumental resolution are also shown in Table~\ref{tab:noiseremove_tab1}, alongside the results for models with $v \sin i$ of 4 and 6~km~s$^{-1}$ (see Appendix~\ref{appen:other_noise} for the $v \sin i$ = 8 and 10~km~s$^{-1}$ results). We note that when exploring the impact of the resolving power, we use only the cases where we also altered the sampling of the line profile accordingly. 

As expected, we see that a decrease in resolution results in a decrease in noise reduction. This is because the convolution with the instrumental profile smooths out the asymmetries in the observed line profile. On top of this, these asymmetries are even more difficult to measure when the profile is sampled more coarsely. The medium resolution (R = 70,000) reduced the noise reduction for many diagnostics by more than half, while the high and ultra-high resolutions (R = 140,000 or 190,000) only saw a marginal decrease. There are a few exceptions to this behaviour.  For example, moving to the medium resolution led to an increase in correlation strength for the standard bisector curvature definition, but the correlation strength in the high and ultra-high resolution modes behaves as expected. A similar effect is seen for the $V_{asy(mod)}$, where in this instance the correlation is weaker in the medium resolution mode, but the medium mode still has a stronger correlation than the high and ultra-high resolutions. As such, for both it is unlikely the increased correlation strengths at medium resolution will hold up in the presence of photon noise and/or slightly different line sampling. We also see a very slight increase in the correlation/noise reduction for the standard velocity displacement and bisector amplitude when moving to the high and ultra-high resolutions (as compared to the original case with no convolution). However, these improvements are very small and both measurements consider small regions of the line profile (e.g. the line core), which are hard to discern when the line sampling is reduced. We also find the line depth, equivalent width, and brightness proxy to be almost independent of the resolution. This is not surprising as a convolution with a Gaussian instrumental profile should conserve the equivalent width and therefore also the area/brightness. Additionally, a convolution will change the absolute value of the line depth, but should preserve the behaviour relative to the RV because it is less impacted by the convolution-altered asymmetries than the line wings. Consequently, the equivalent width and simultaneous photometry may provide some of the best diagnostics to disentangle granulation noise. Given the future wealth of photometric data from missions like TESS, CHEOPS, and PLATO, and the ease of calculating the EW from the necessary ground-based spectroscopic follow-up, there is strong potential to disentangle the granulation effects from the Doppler wobble of low-mass, long-period planets. 

From Figure~\ref{fig:stel_rot}, we see that even a small change in the stellar rotation can have a significant impact on the line profile shape. Moving from a $v \sin i$ of 2 to 4 km~s$^{-1}$ significantly reduced the asymmetry seen in the line bisector, and as a result we expect drastic reductions in our noise mitigation attempts. However, as the equivalent width/area should be conserved when the rotation is increased, we expect the this diagnostic, and the brightness, to be relatively unaffected by the $v \sin i$. Regardless of the diagnostic, we also expect a decrease in the correlation with RV as rotation increases because the RV becomes more difficult to precisely determine. This is indeed what we find, as shown in Table~\ref{tab:noiseremove_tab1}. At a $v \sin i$ of 4~km~s$^{-1}$, only the equivalent width and brightness maintain a strong enough correlation to provide any granulation noise reduction, and at $v \sin i$ = 6~km~s$^{-1}$ even these diagnostics fail. We compare the equivalent width and area measurements between each stellar rotation case and confirm that they are conserved; it is the increase in scatter amongst the RV measurements that decreases their correlation strength. Ultimately, these results will depend on the inherent shape of the particular stellar lines observed, and the overall variability. Nonetheless, the Fe~I~6302~$\AA$ line simulated here is fairly representative of most lines observed in a given solar-like star, so we may expect the corresponding CCFs to behave similarly. Consequently, the slowest rotating stars should provide the most ideal cases for granulation noise mitigation --  both because of the increased RV precision available, and because we can measure the asymmetries to diagnose the underlying granulation behaviour. 

\section{Summary and Concluding Remarks}
\label{sec:conc}
Stellar surface magnetoconvection has the potential to induce spurious RV shifts that can completely mask the Doppler-reflex signal induced by low-mass, long-period planets. Throughout this work we have shown that it may be possible to disentangle and therefore correct the convective-induced RV shifts. Herein, we use a realistic granulation parameterisation, with a 3D MHD backbone, from \cite{cegla13} and \cite{cegla18a}, to generate new stellar absorption line profiles that contain the same fundamental convection characteristics as those from the computationally heavy radiative 3D MHD simulations. Independent granulation profiles are then tiled across a model star and disc-integrated to create 1000 realistic Sun-as-a-star model observations that contain magnetoconvection with an average magnetic field strength of 200~G for the Fe~I~6302~$\AA$ line. The shapes of the disc-integrated line profiles match very well those from empirical solar observations from the McMath, IAG, and PEPSI spectrographs. The net convective blueshift in our simulation is approximately a third of that from the quiet Sun, but this is in agreement with what we expect due to the inhibition of convection from the increased magnetic field strength in our MHD simulation \citep{cegla18a}. Along with this, the RV rms from the granulation in our model observations is only $\sim$10~cm~s$^{-1}$, which is a factor of 3-4 lower than what we expect in the quiet Sun \citep{elsworth94, palle99}. We attribute the reduced rms to the increased inhibition of the convective flows from the higher magnetic field strength, and operate under the hypothesis that our models represent a scaled version of the quiet Sun behaviour; this hypothesis will be further tested in future work where we explore the temporal variability of lower magnetic field strength MHD simulations. 

Using these Sun-as-a-star model observations, we search for correlations between line shape characteristics and the convection-induced RV shifts. We find many line profile characteristics show a strong linear correlation with the induced RVs, and that by subtracting off this correlation we can significantly reduce the RV variability. We find that the line depth is well correlated with the RV, but that this correlation is only strong enough to be useful as a noise mitigation tool if the line profiles are not continuum normalised, with a $\sim$35\% of reduction in the RV rms. Unfortunately, ground-based data must be normalised so this may not be the most ideal diagnostic; on top of this, the total variation in the line depth is very small and likely undetectable. However, other stellar lines may have a larger variation in depth due to the granulation evolution, so it could potentially be a useful diagnostic for other lines. Nonetheless, we find the BIS, bisector curvature ($C$), bisector velocity displacement ($V_b$), bisector amplitude ($A_b$), $V_{asy(mod)}$, and equivalent width may all be capable of removing up to $\sim$50-60\% of the RV noise, which could mean approximately four times less observing time is required. Each of these diagnostics had peak-to-peak variations of $\sim$10-30~cm~s$^{-1}$, which may be detectable with instruments like ESPRESSO and EXPRES, especially if the variability is a factor of 3-4 times larger in the quiet photosphere. We also integrated the area underneath the disc-integrated line profiles to act as a proxy for photometric brightness, which we found to correlate very strongly with RV and may also offer a $\sim$50\% reduction in the granulation RV noise. It is important to keep in mind that we were approximating the brightness over a very small wavelength region, but if confirmed, this could mean simultaneous photometry from current and future missions like TESS, CHEOPS, and PLATO may play a key role in disentangling granulation-induced RV noise. It is also important to note that we are not simply advocating for a `one-off, flicker-style' photometric rms measurement to predict the level of stellar RV noise; we are advocating for simultaneous photometric and spectroscopic measurements, as this could allow us to actually mitigate the granulation-induced RV variability and push our detection limits to lower-mass, longer-period planets. While these space-missions will provide thousands of potential targets, it is clear that the future long-term monitoring by PLATO will make such simultaneous ground-based RV follow-up far more amenable.

We also explore the impact of instrumental resolution and stellar rotation. Even at a resolution of 70,000, the instrumental profile will act to smooth out the asymmetries in the observed line profiles; this decreases the amplitude of the variability in each line diagnostic, making it more difficult to discern and increasing the scatter in the correlation with the RV, thereby decreasing the noise mitigation success. Fortunately, a high resolution of 140,000 or 190,000 is sufficient to capture most line asymmetries with only minimal impact on the noise mitigation. However, this result is contingent on the shape of the line bisector; a bisector with a `C-bend' closer to the line core (as expected for hotter stars) will be more severely impacted. The equivalent width and brightness are mostly independent of the instrumental resolution. The stellar rotation also strongly impacts the line profile shape, and even a small increase from a $v \sin i$ of 2 to 4~km~s$^{-1}$ causes the line bisector to be significantly less asymmetric. This means that our bisector diagnostics fine-tuned to the solar rotation are no longer useful noise mitigation tools. The exception to this is the equivalent width and the brightness; however, even these diagnostics perform more poorly due to the decreased RV precision from the increase in the stellar rotation. 

Similar to procedures that correct for starspot/plage contamination, we advocate that all diagnostics found here to produce any noise reduction should be explored in conjunction with one another when analysing empirical data. Additionally, we anticipate that diagnostics that utilise the entire line profile should be most robust against photon noise (e.g. $V_{asy(mod)}$ and equivalent width). In future work we will explore the impact of photon noise, finite exposures, and the possibility to use a linear combination of diagnostics to further increase the noise reduction. This will also be extended across a variety of magnetic field strengths and stellar lines, and eventually various spectral types and astrophysical noise sources (e.g. supergranulation and/or plage regions etc.). Nonetheless, the strong noise mitigation we find here promises great potential for disentangling granulation, which will be critical to reach the RV precision necessary for the future confirmation of true Earth-analogs. Moreover, it is clear that high precision, high resolution observations of slowly rotating stars hold the most promise for such granulation noise mitigation. 

\acknowledgments
H.M.C. acknowledges the financial support of the National Centre for Competence in Research “PlanetS” supported by the Swiss National Science Foundation (SNSF); S.S., H.M.C and C.A.W acknowledge support from the Leverhulme Trust. C.A.W. acknowledges support from the UK Science and Technology Facilities Council (STFC), including grants ST/I001123/1 and ST/P000312/1. This research has made use of NASA's Astrophysics Data System Bibliographic Services.

\software{MURaM \citep{vogler05,shelyag14, shelyag15};  OPAL \citep{rogersOPAL, rogersnayfonovOPAL}; NICOLE \citep{NICOLE1, NICOLE2}}

\bibliographystyle{aasjournal}
\bibliography{abbrev,mybib_crt}

\begin{appendix}
  
\section{Probability Distributions for Additional Limb Angles}
\label{appen:other_prob}  
\begin{figure*}[h!]
\centering
\includegraphics[width = 8.5 cm]{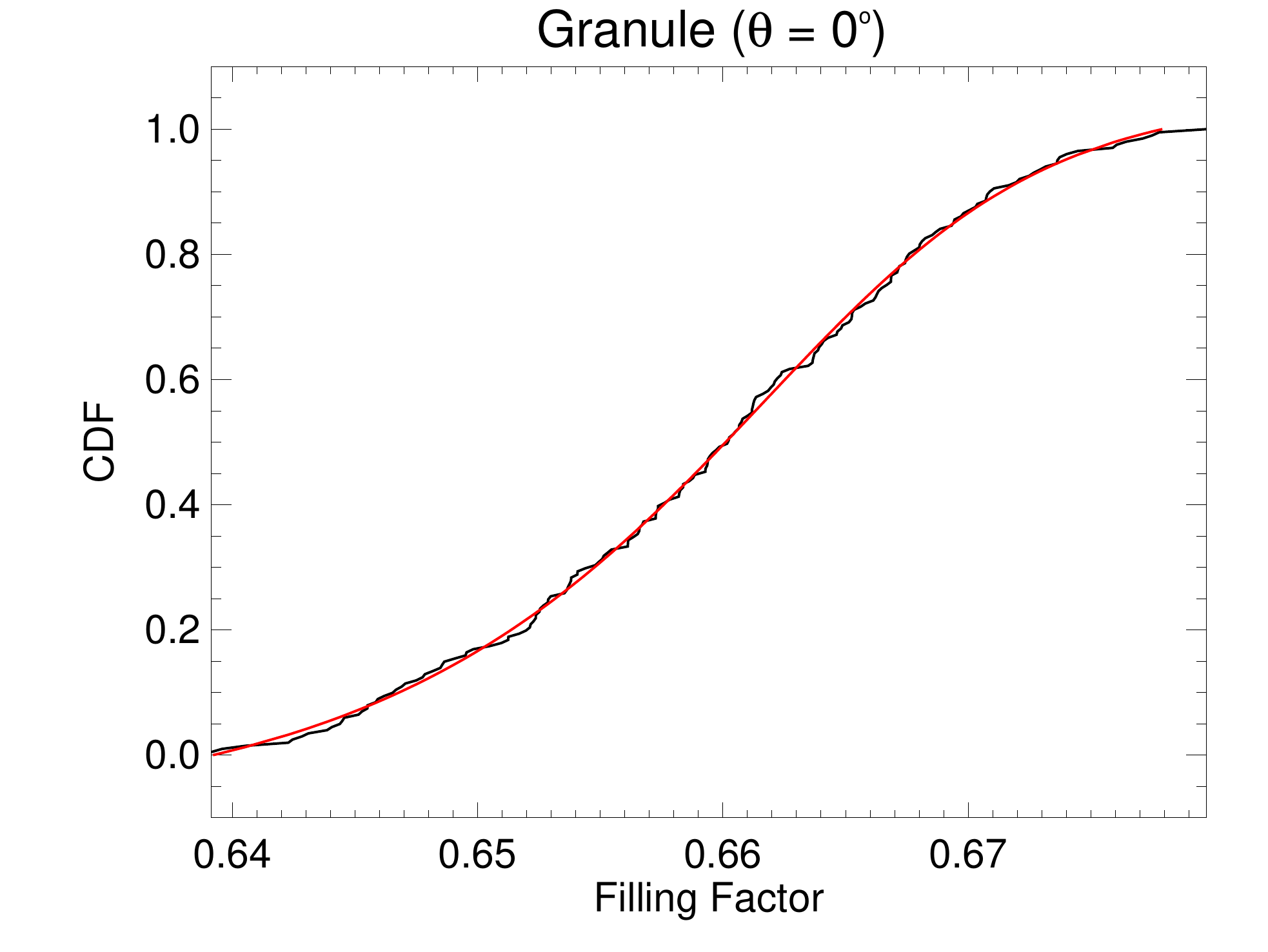} 
\includegraphics[width = 8.5 cm]{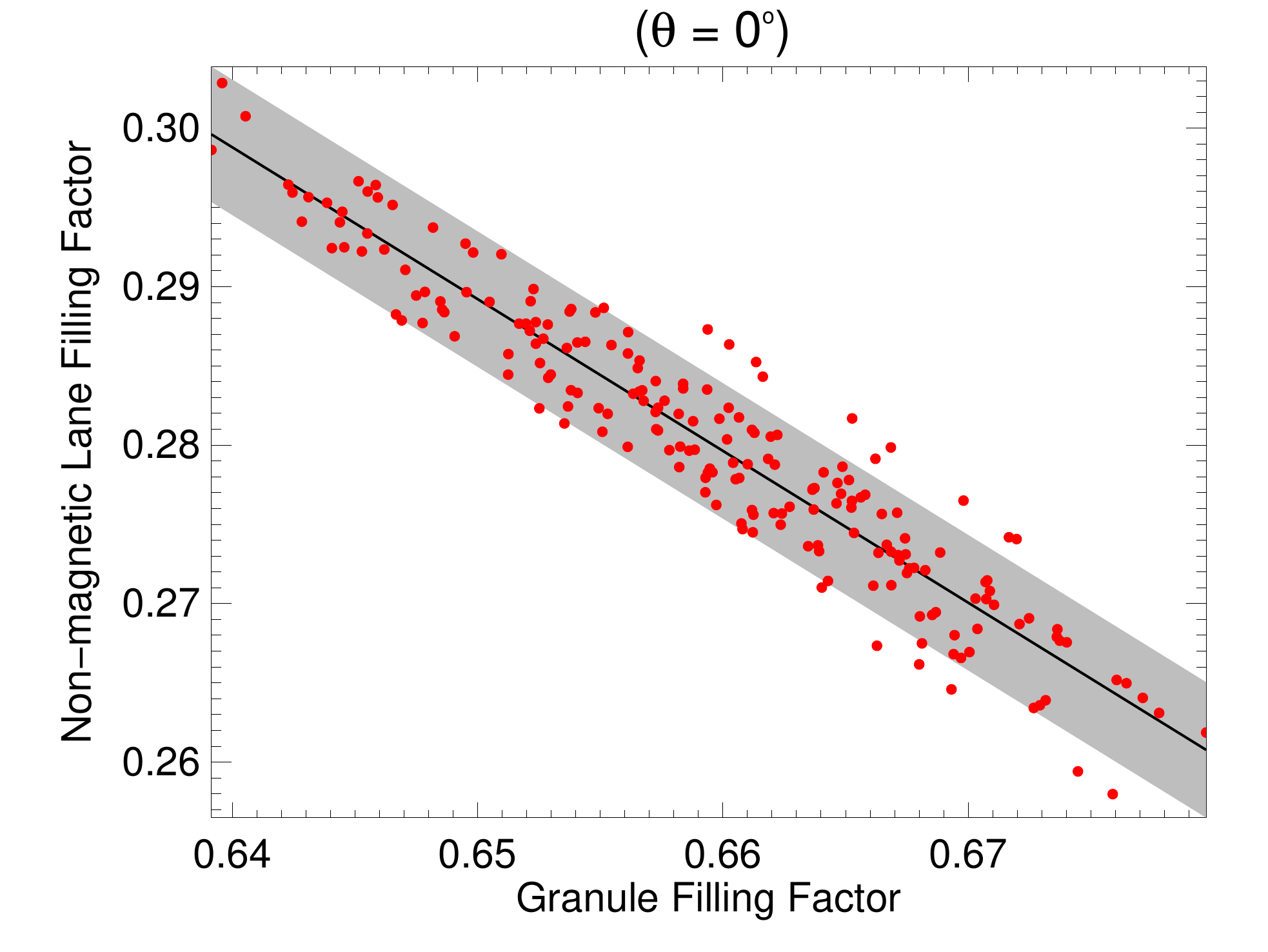} 
\includegraphics[width = 8.5 cm]{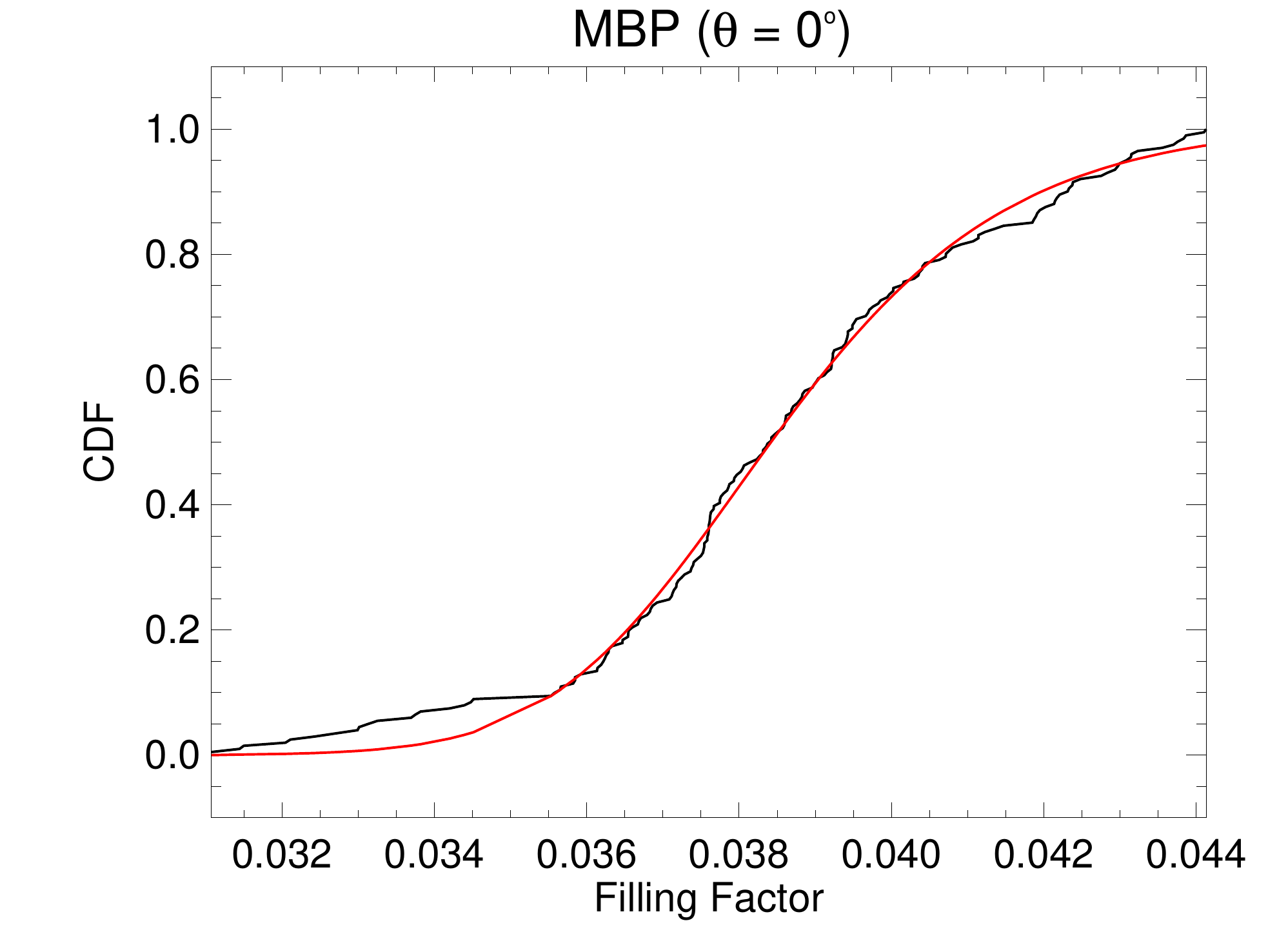} 
\includegraphics[width = 8.5 cm]{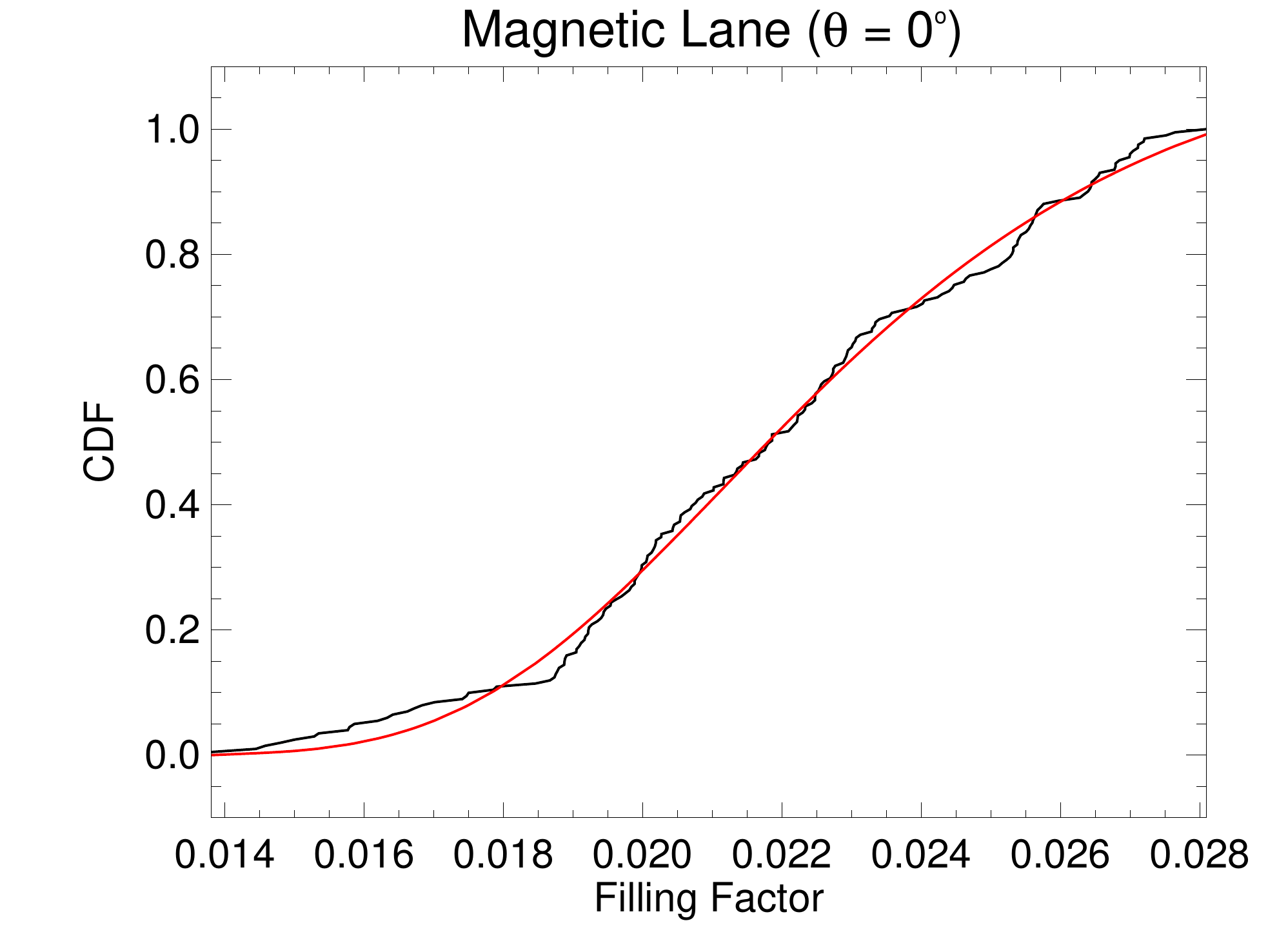} 
\caption{Cumulative distribution functions (CDFs) over the MHD time-series for the granule (top left), MBP (bottom left), and magnetic intergranular lane (bottom right) components are shown in black; fits using the generalised logistic function from Eq.~\ref{eq:logistic} are shown in red. Also displayed is the linear relationship between the filling factors for the granule and non-magnetic intergranular lane components (top right); the shaded area represents a uniform region of width 1.5$\sigma$, where $\sigma$ was determined by a robust bisquares linear regression (fit shown in black). All plots are for a limb angle of 0$^{\rm{o}}$ ($\mu$ = 1.0).}
\label{fig:prob0}
\end{figure*}

\begin{figure*}[h!]
\centering
\includegraphics[width = 8.5 cm]{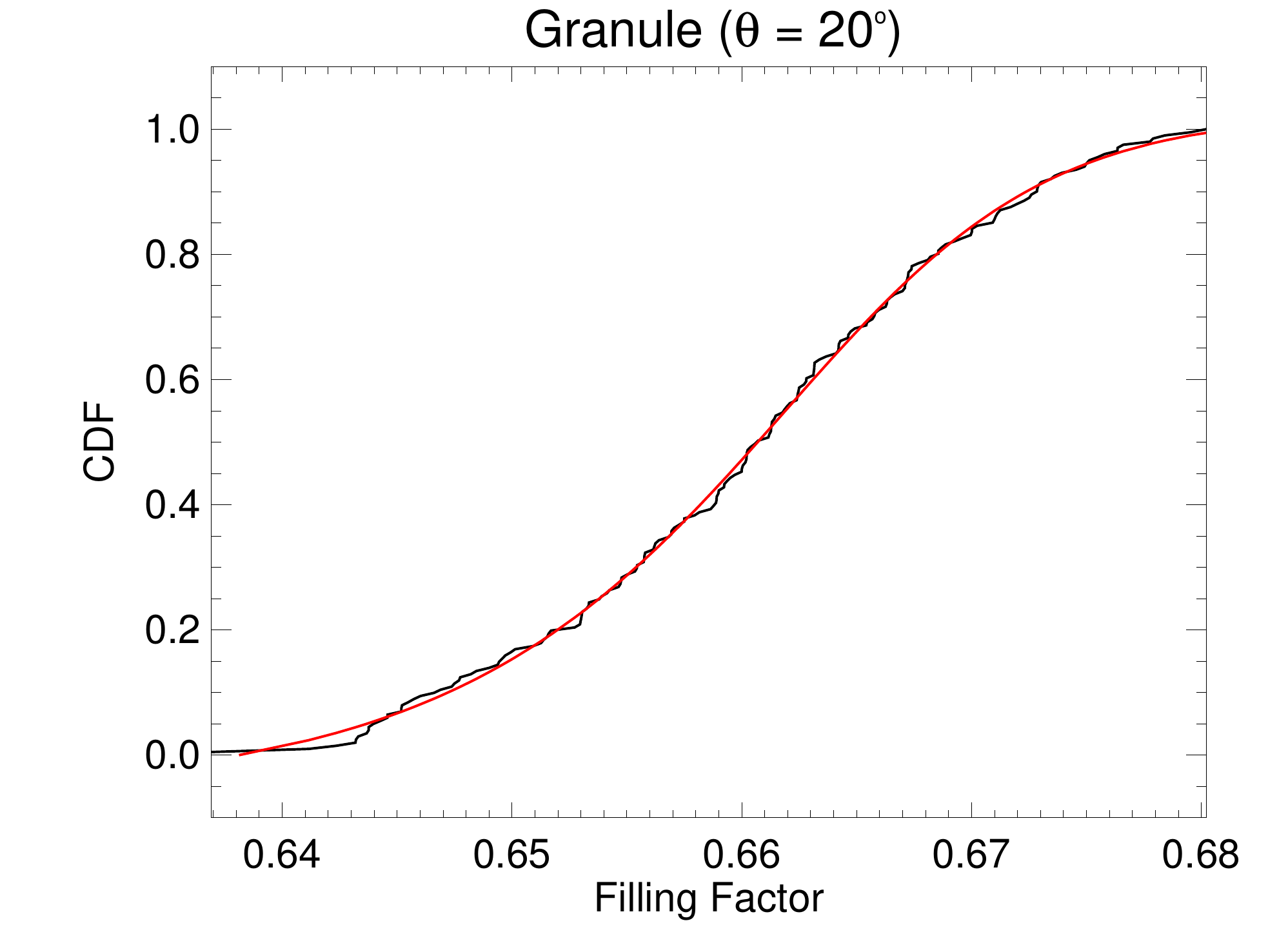} 
\includegraphics[width = 8.5 cm]{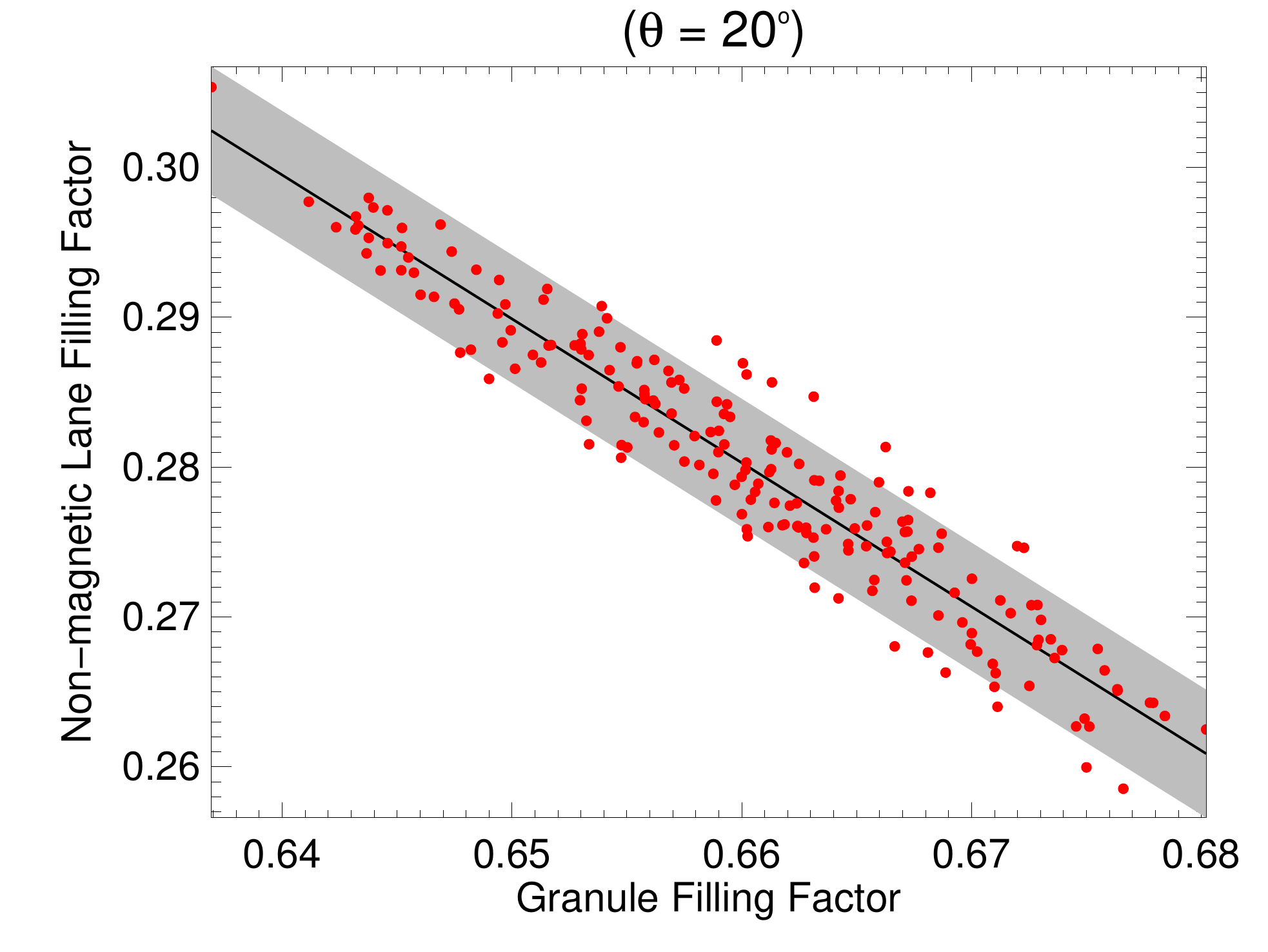} 
\includegraphics[width = 8.5 cm]{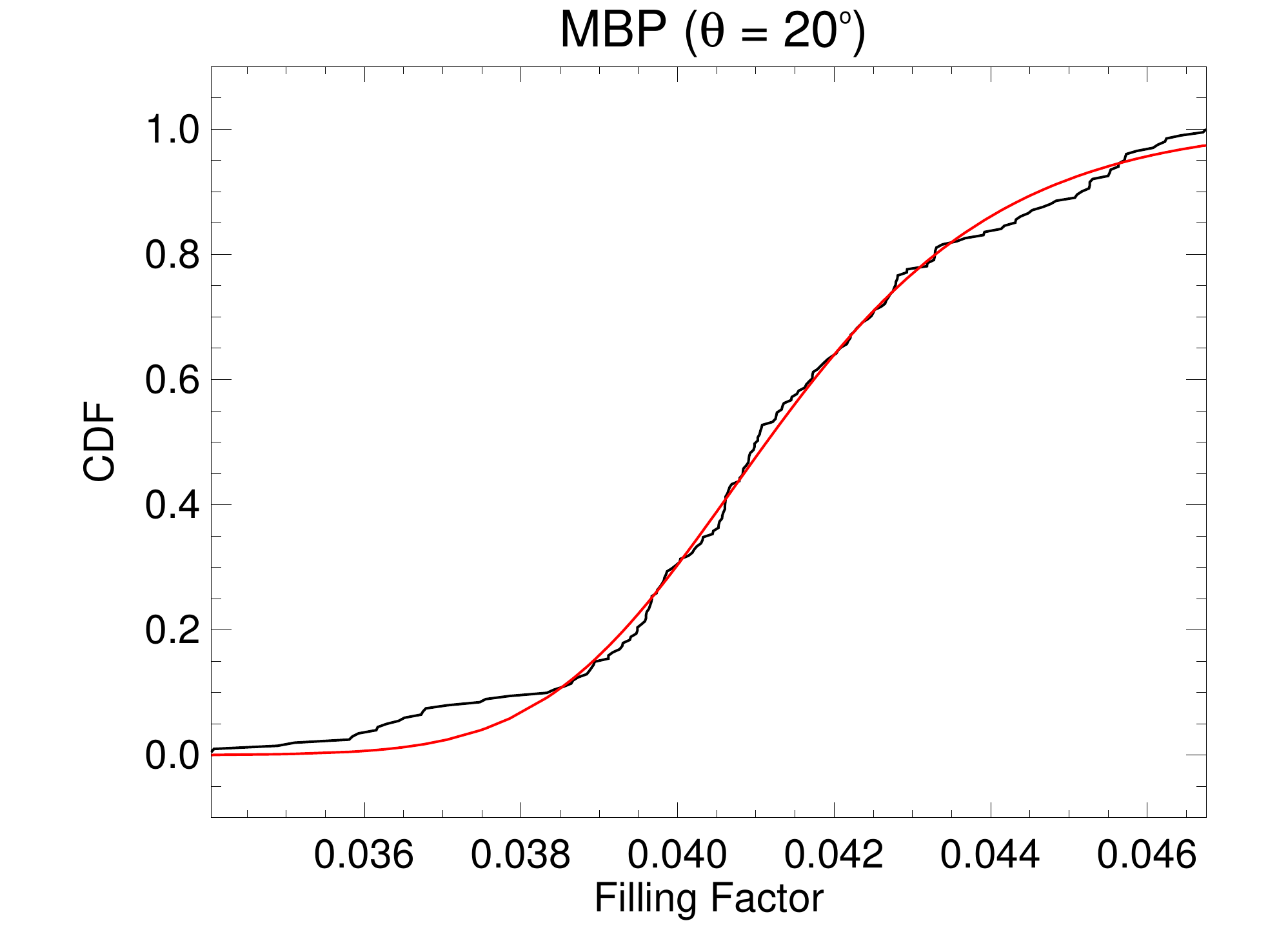} 
\includegraphics[width = 8.5 cm]{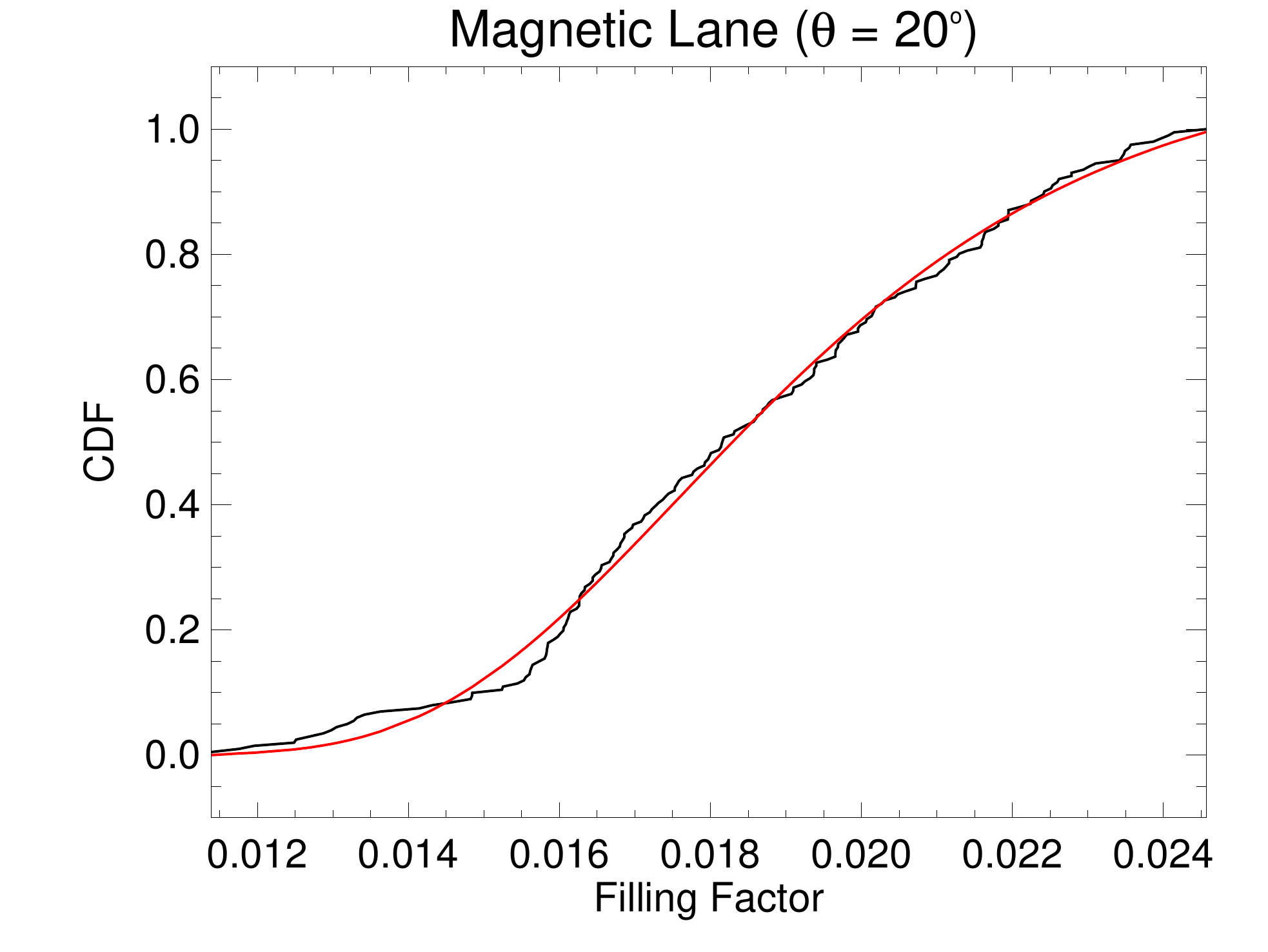} 
\caption{Cumulative distribution functions (CDFs) over the MHD time-series for the granule (top left), MBP (bottom left), and magnetic intergranular lane (bottom right) components are shown in black; fits using the generalised logistic function from Eq.~\ref{eq:logistic} are shown in red. Also displayed is the linear relationship between the filling factors for the granule and non-magnetic intergranular lane components (top right); the shaded area represents a uniform region of width 1.5$\sigma$, where $\sigma$ was determined by a robust bisquares linear regression (fit shown in black). All plots are for a limb angle of 20$^{\rm{o}}$ ($\mu \ \approx$ 0.94).}
\label{fig:prob20}
\end{figure*}

\begin{figure*}[h!]
\centering
\includegraphics[width = 8.5 cm]{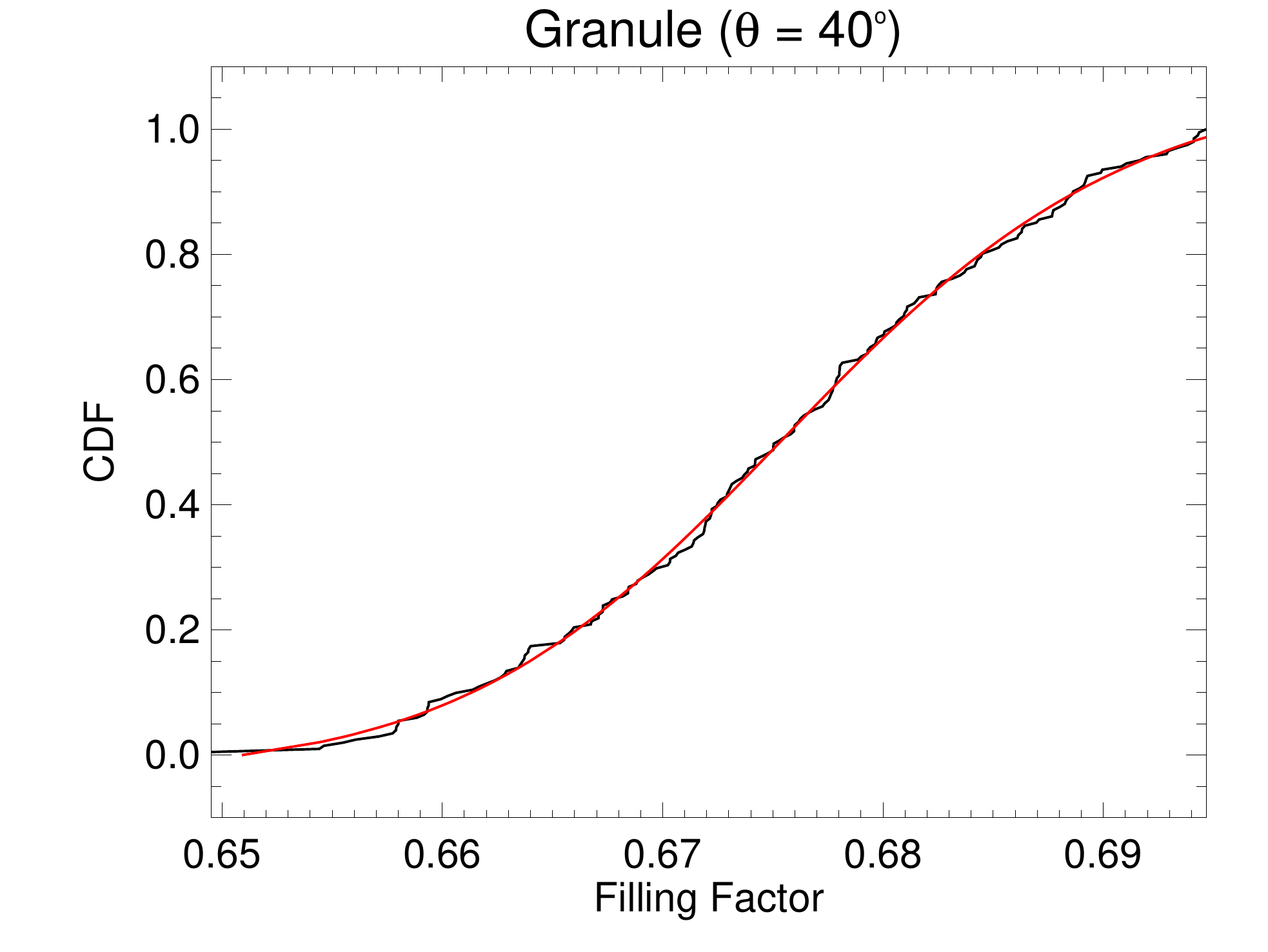} 
\includegraphics[width = 8.5 cm]{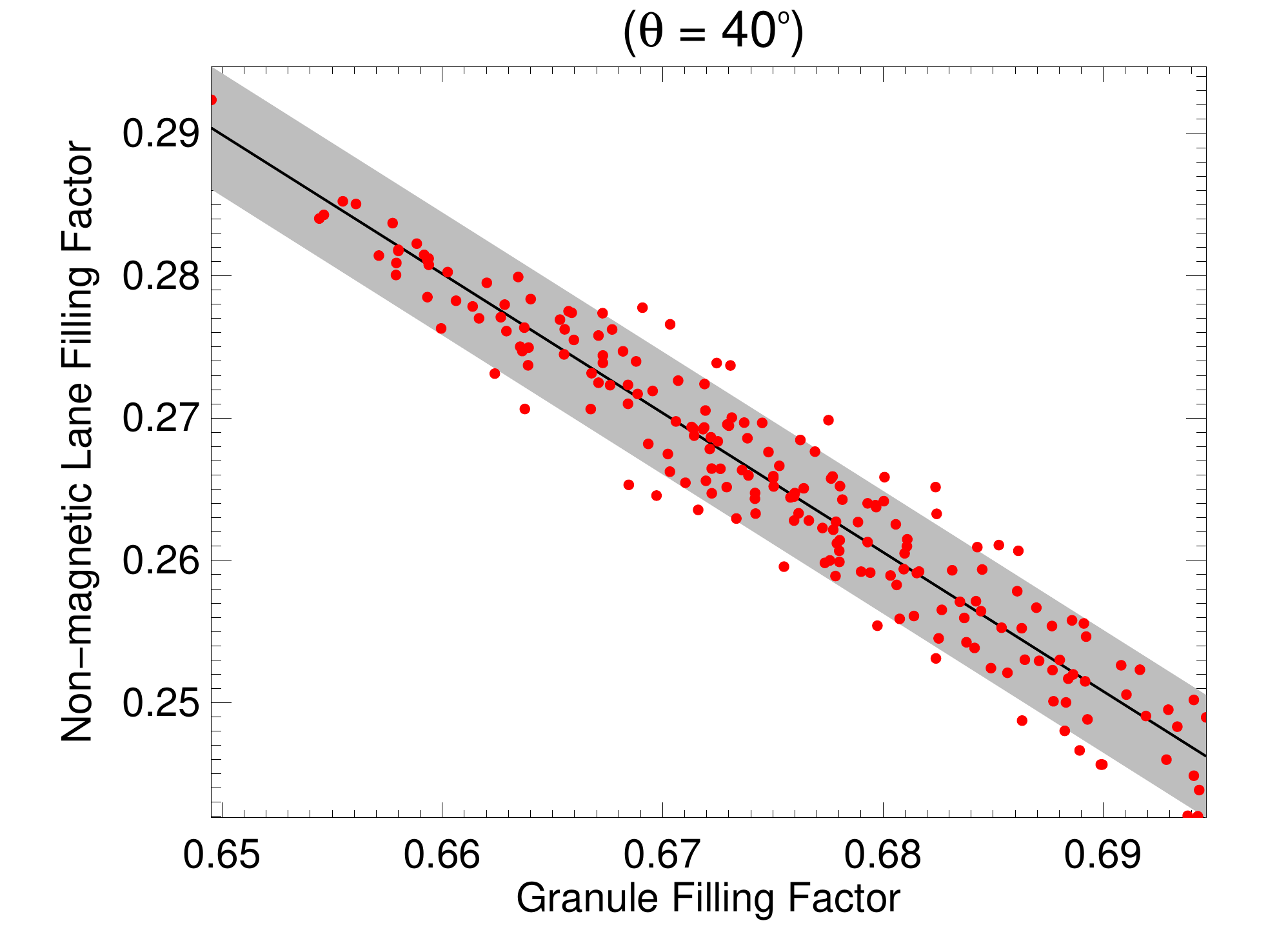} 
\includegraphics[width = 8.5 cm]{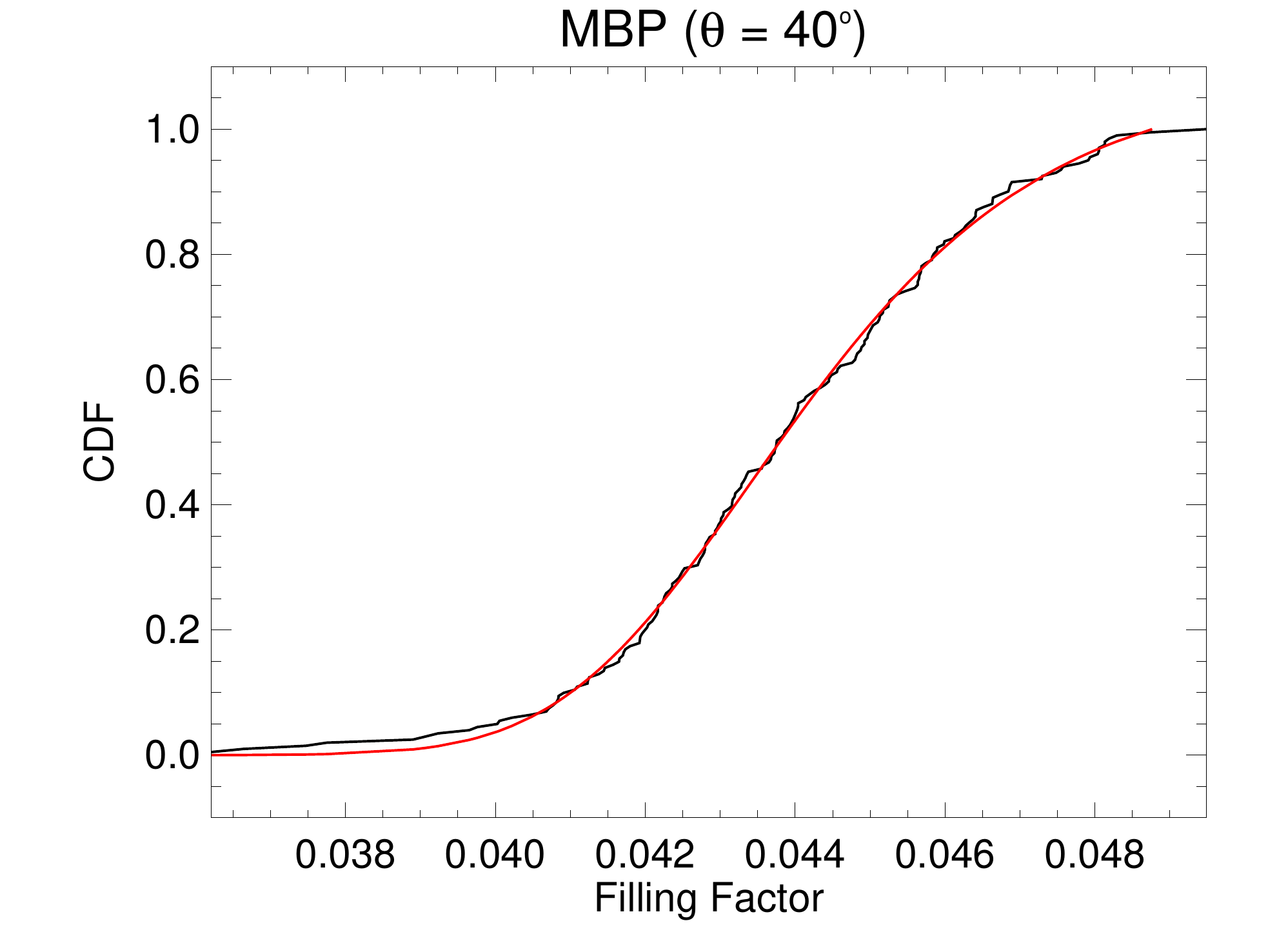} 
\includegraphics[width = 8.5 cm]{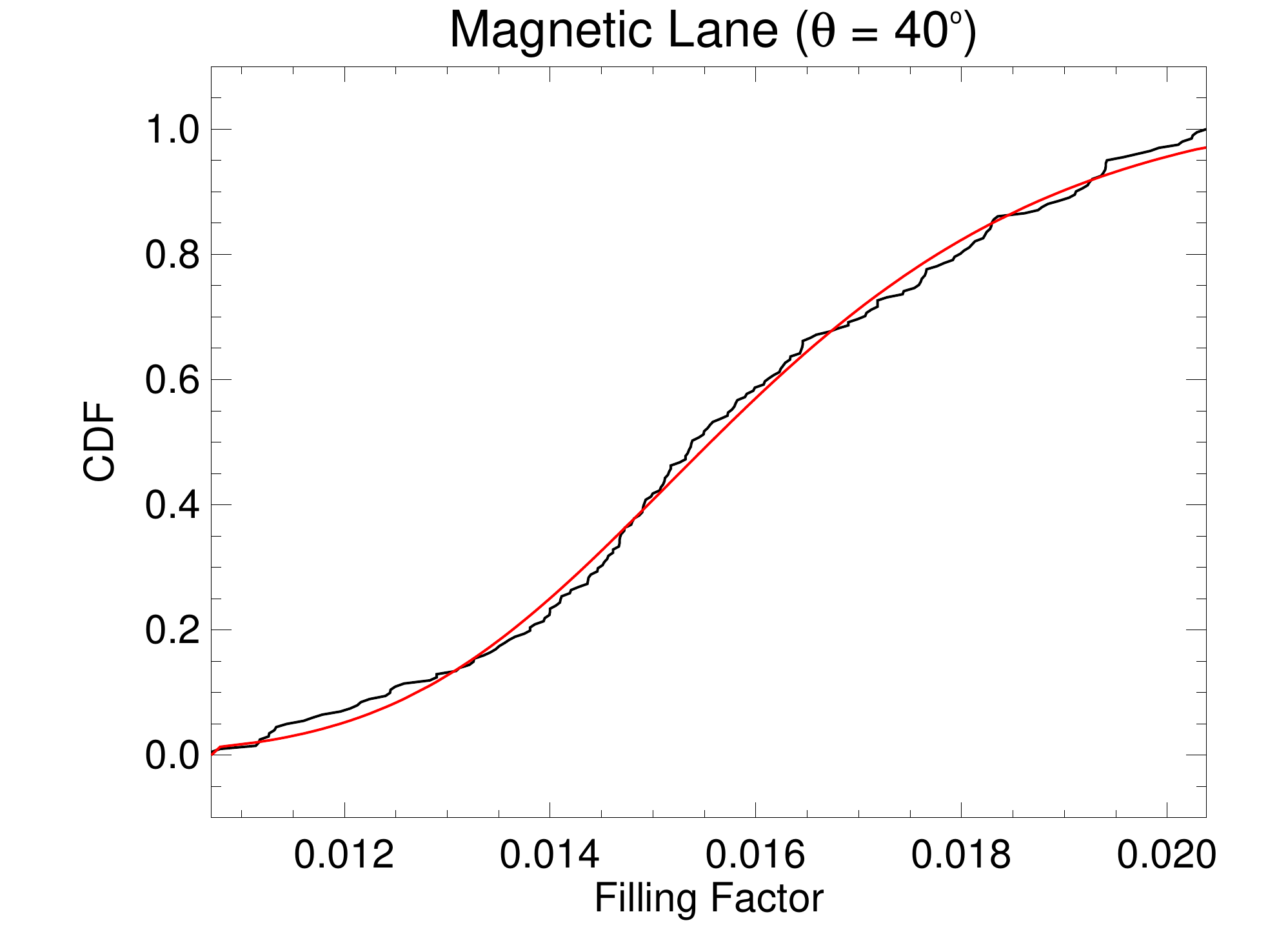} 
\caption{Cumulative distribution functions (CDFs) over the MHD time-series for the granule (top left), MBP (bottom left), and magnetic intergranular lane (bottom right) components are shown in black; fits using the generalised logistic function from Eq.~\ref{eq:logistic} are shown in red. Also displayed is the linear relationship between the filling factors for the granule and non-magnetic intergranular lane components (top right); the shaded area represents a uniform region of width 1.5$\sigma$, where $\sigma$ was determined by a robust bisquares linear regression (fit shown in black). All plots are for a limb angle of 40$^{\rm{o}}$ ($\mu \ \approx$ 0.77).}
\label{fig:prob40}
\end{figure*}

\begin{figure*}[h!]
\centering
\includegraphics[width = 8.5 cm]{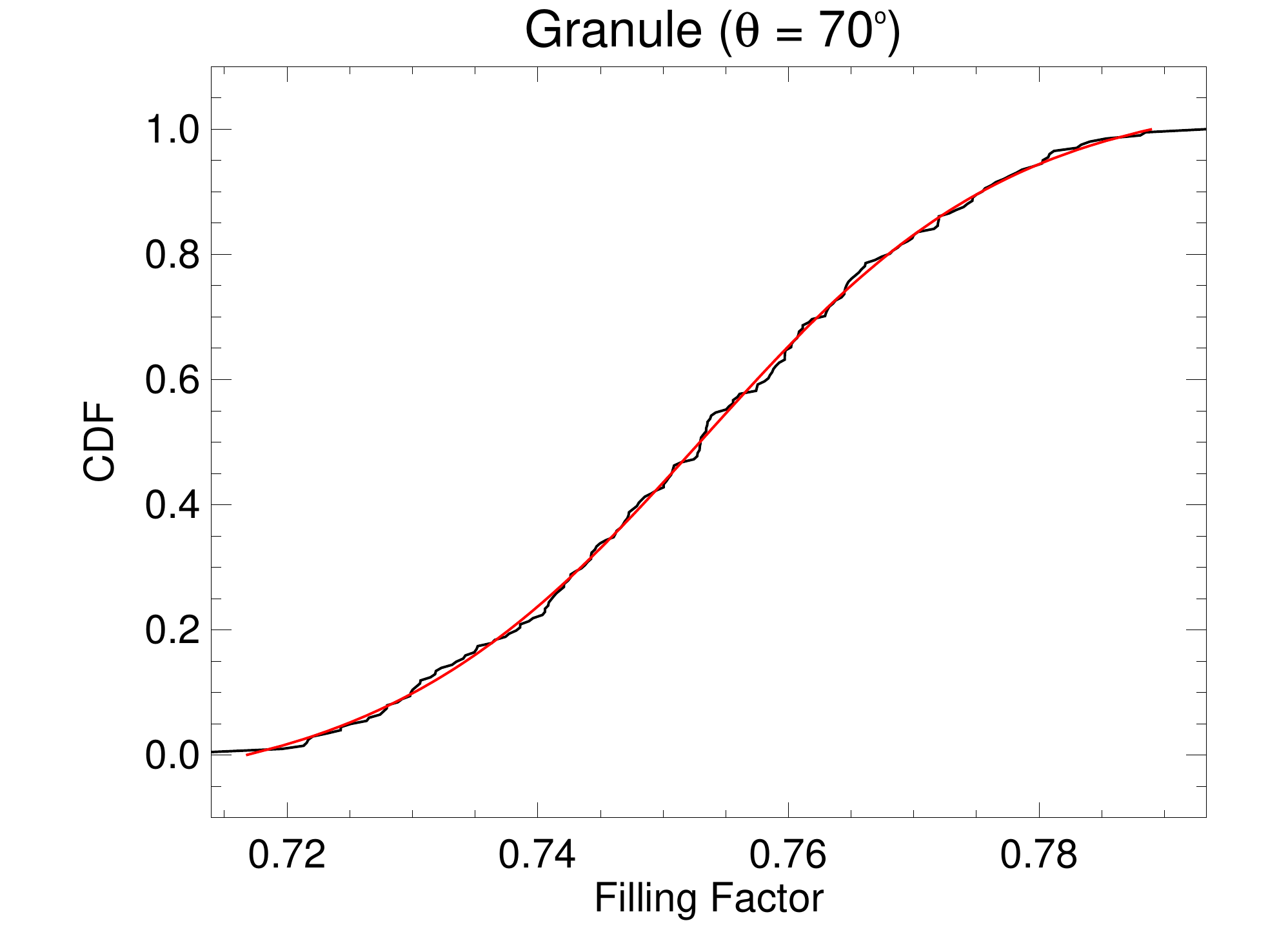} 
\includegraphics[width = 8.5 cm]{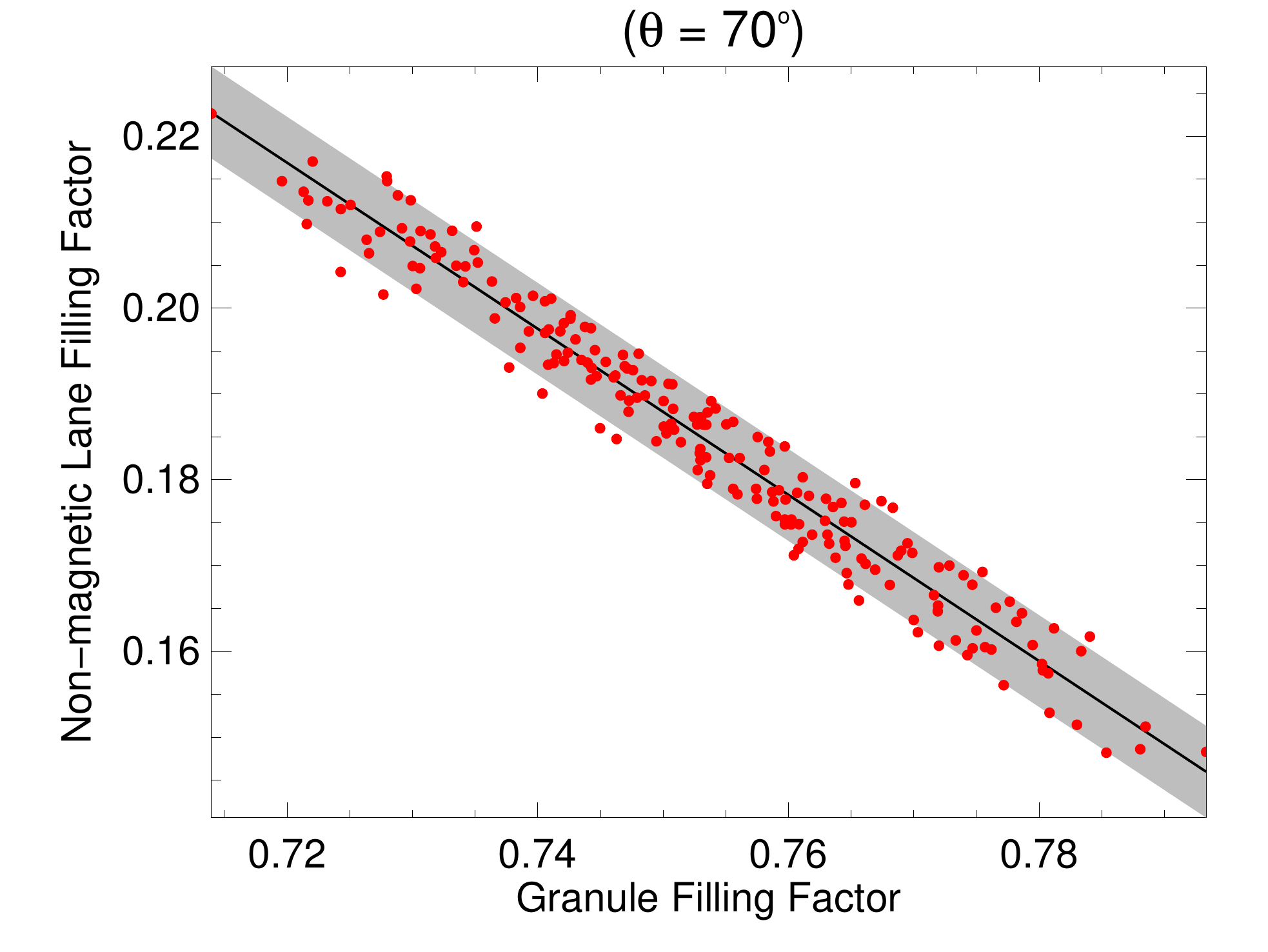} 
\includegraphics[width = 8.5 cm]{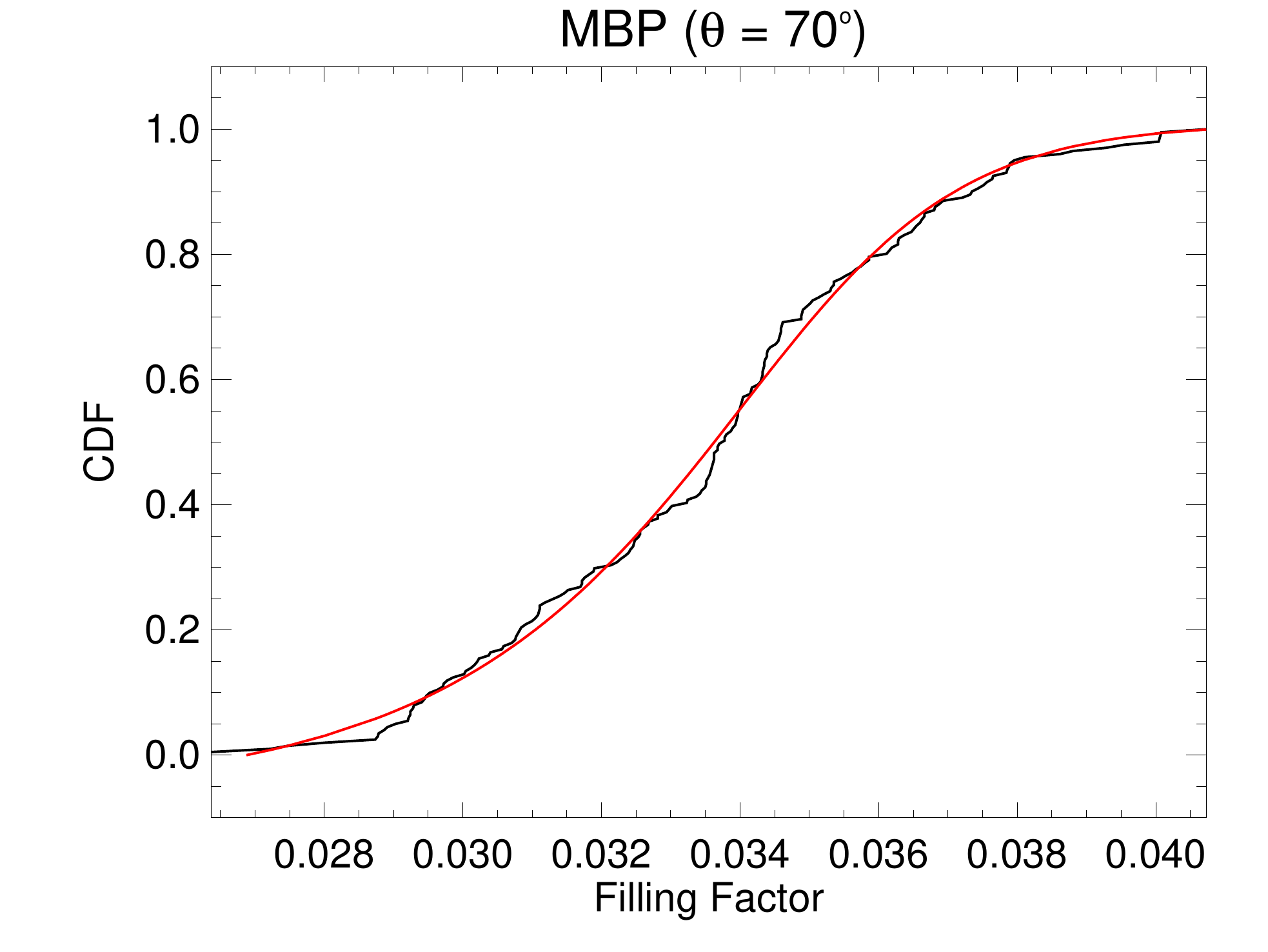} 
\includegraphics[width = 8.5 cm]{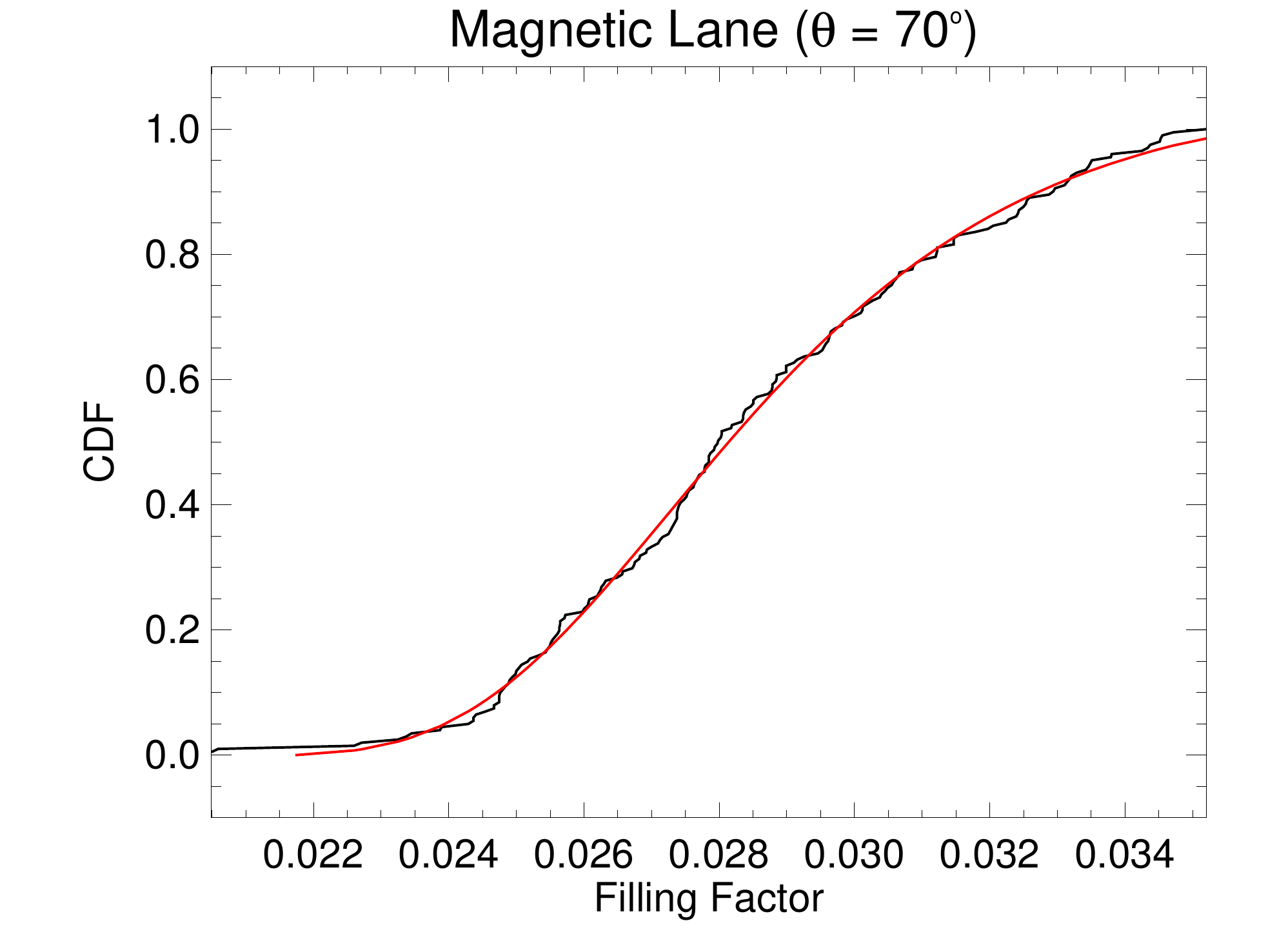} 
\caption{Cumulative distribution functions (CDFs) over the MHD time-series for the granule (top left), MBP (bottom left), and magnetic intergranular lane (bottom right) components are shown in black; fits using the generalised logistic function from Eq.~\ref{eq:logistic} are shown in red. Also displayed is the linear relationship between the filling factors for the granule and non-magnetic intergranular lane components (top right); the shaded area represents a uniform region of width 1.5$\sigma$, where $\sigma$ was determined by a robust bisquares linear regression (fit shown in black). All plots are for a limb angle of 70$^{\rm{o}}$ ($\mu \ \approx$ 0.34).}
\label{fig:prob70}
\end{figure*}

In this section we display additional granulation component filling factor probability distributions and relationships that were used in Section~\ref{sec:creat_mod}, to create new line profiles that contain convective-induced asymmetries and radial velocity shifts. Figures~\ref{fig:prob0}, \ref{fig:prob20}, \ref{fig:prob40}, and \ref{fig:prob70} show these results for the limb angles: 0, 20, 40, and 70$^{\rm{o}}$ ($\mu$~= 1, 0.94, 0.77, 0.34), respectively. In each of these figures we plot the filling factor cumulative distribution functions for the granule, MBP, and magnetic intergranular lane components. In addition, these figures also include the linear relationship between the granule and non-magnetic components. We remind the reader these filling factors are from 201 snapshots of a solar MHD simulation, corresponding to approximately 100 minutes of physical time. See Section~\ref{sec:creat_mod} for more details, including Figure~\ref{fig:prob60} for the equivalent results at 60$^{\rm{o}}$ ($\mu$ = 0.5).

\begin{figure*}[h!]
\centering
\includegraphics[width=17cm]{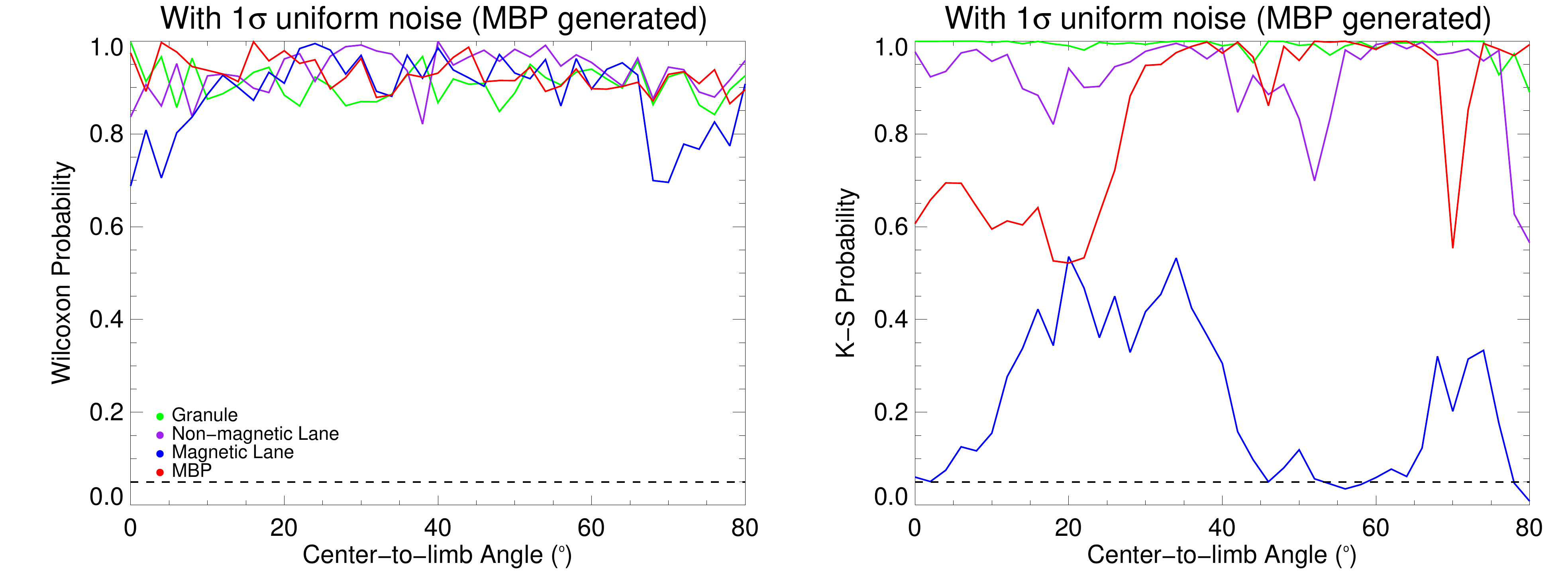}
\includegraphics[width=17cm]{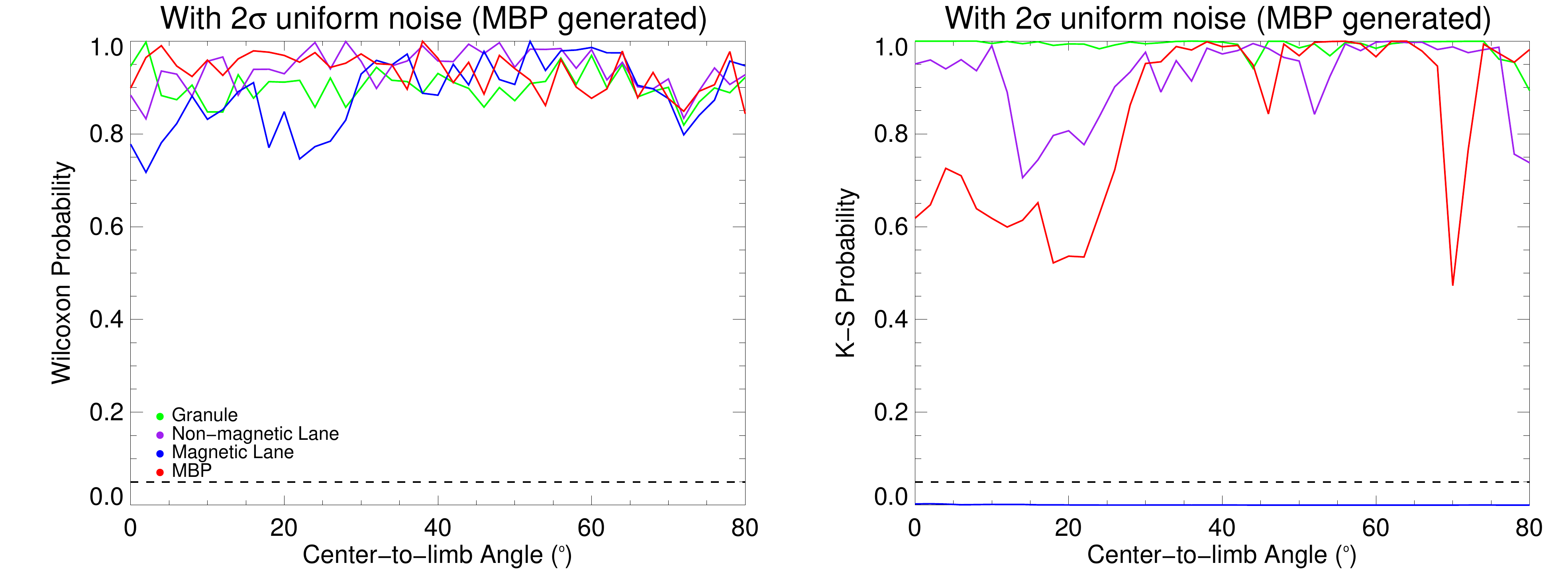}
\caption{The Wilcoxon Rank-Sum (left) and Kolmogorov-Smirnov (right) probabilities for 100,000 artificially generated filling factors as compared to the 201 original filling factors from the MHD simulation for all four granulation components. Probabilities $<$ 0.05 (dashed lines) are typically considered statistically significant results and require the null hypothesis to be rejected.}
\label{fig:wc_ks_12_mbp}
\end{figure*}

\begin{figure*}[h!]
\centering
\includegraphics[width=17cm]{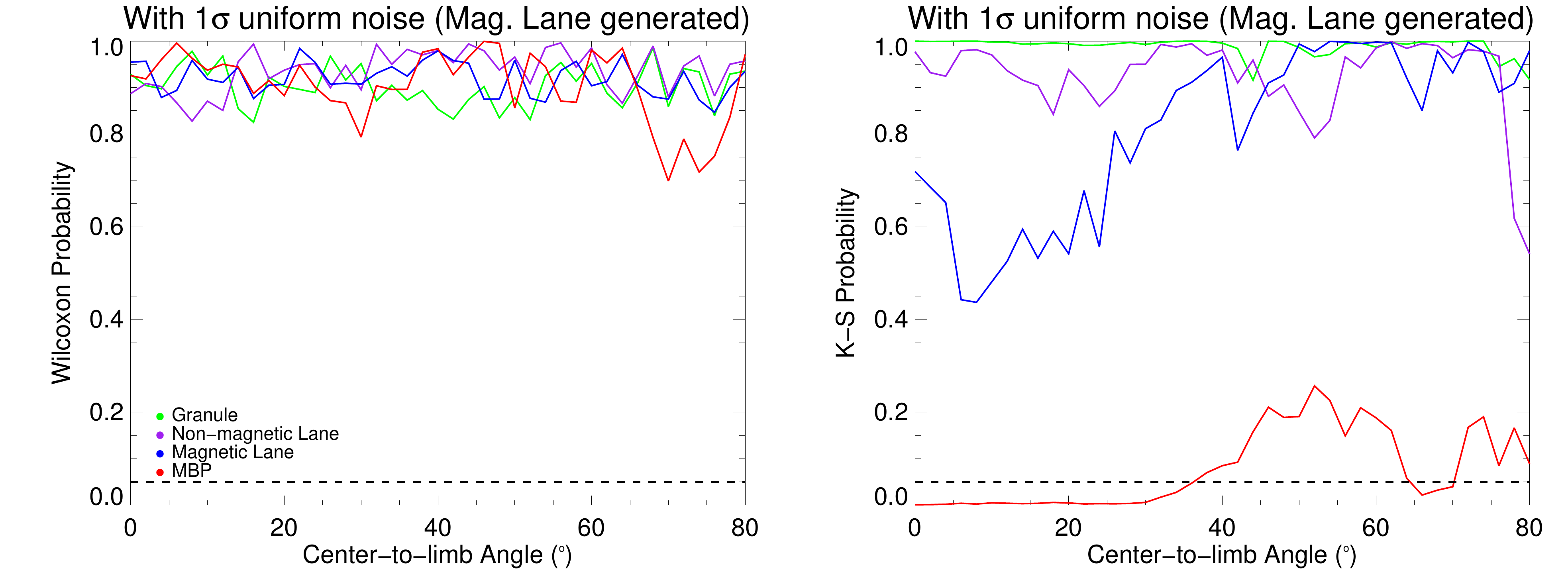}
\includegraphics[width=17cm]{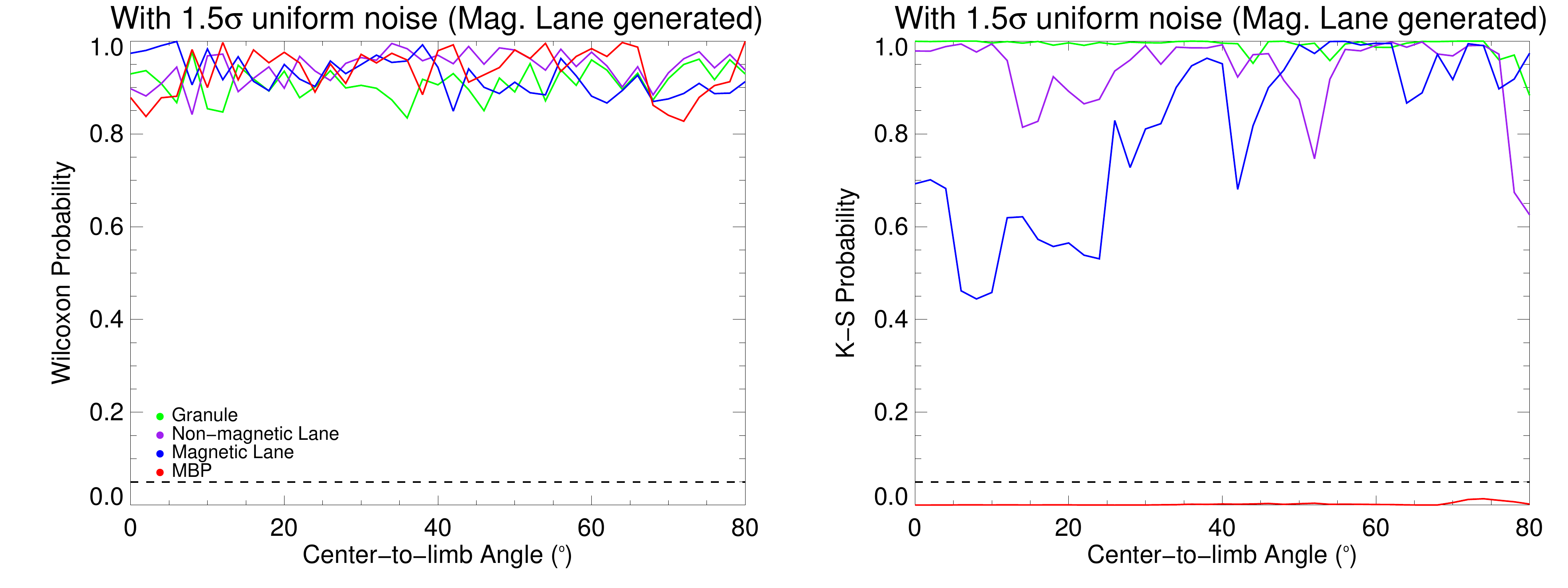}
\includegraphics[width=17cm]{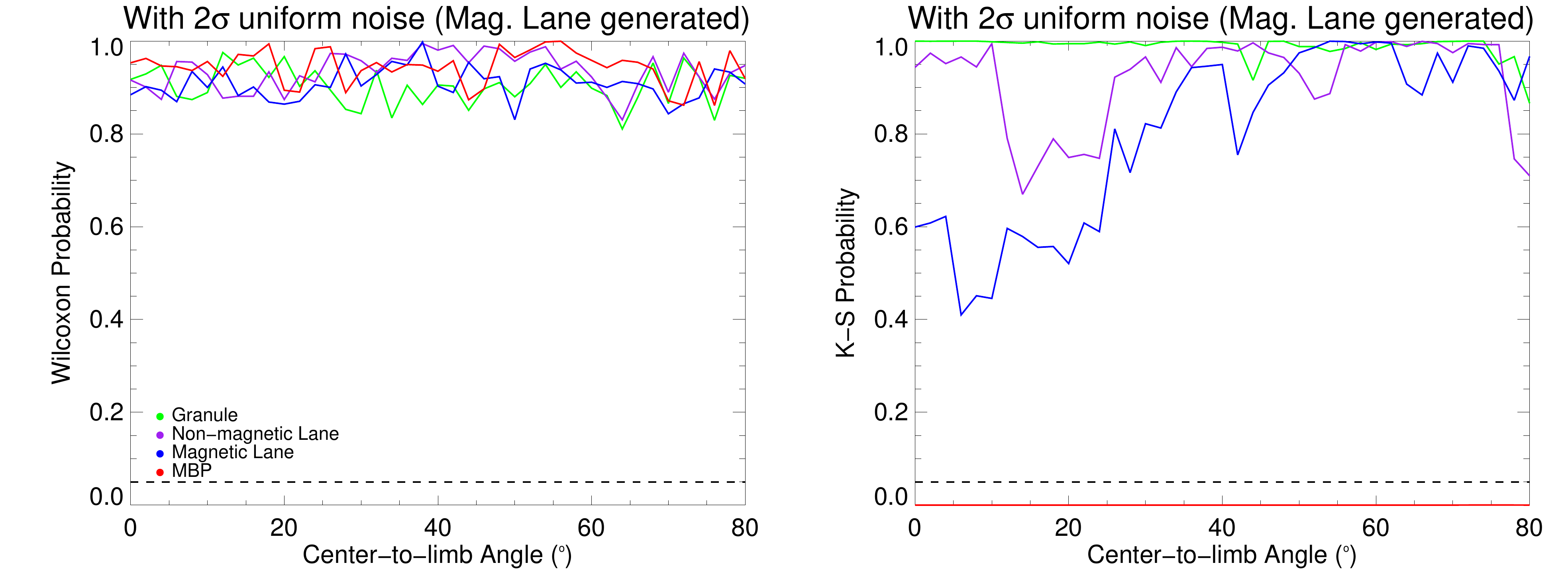}
\caption{The Wilcoxon Rank-Sum (left) and Kolmogorov-Smirnov (right) probabilities for 100,000 artificially generated filling factors as compared to the 201 original filling factors from the MHD simulation for all four granulation components. Probabilities $<$ 0.05 (dashed lines) are typically considered statistically significant results and require the null hypothesis to be rejected.}
\label{fig:wc_ks_1_1.5_2_blane}
\end{figure*}

In Section~\ref{sec:creat_mod}, we also tested how well we could reproduce the original granulation component filling factor distributions based on the KS and Wilcoxon Rank-Sum tests. In each instance, a granule filling factor is selected based on a probability distribution determined from its CDF, and then a non-magnetic intergranular lane filling factor is selected based on the anti-correlation between these components in the MHD simulation. In addition, some uniform noise is added to non-magnetic intergranular lane filling factor; the noise level is based on the observed scatter between these components in the MHD simulation, as determined from a robust bisquares linear regression. Then one of the magnetic components is selected from a probability distribution determined from its CDF, and the final magnetic component is selected such that all components add to unity (see Section~\ref{sec:creat_mod} for more details). Figure~\ref{fig:wc_ks_12_mbp} shows these results when the MBP is generated from its probability distribution, both when 1 and 2$\sigma$ of uniform noise is added to the granule -- non-magnetic lane relationship (see Figure~\ref{fig:wc_ks_60}, for the results if selecting 1.5$\sigma$ uniform noise). On the other hand, Figure~\ref{fig:wc_ks_1_1.5_2_blane} displays these results, alongside the case for 1.5$\sigma$ uniform noise, when the magnetic intergranular is generated instead. See Section~\ref{sec:creat_mod} for more discussion of these results. 

\section{Additional Granulation Correlations Plots}
\label{appen:gran_plots}
\begin{figure}[h]
 \centering
\includegraphics[width = 8.5 cm]{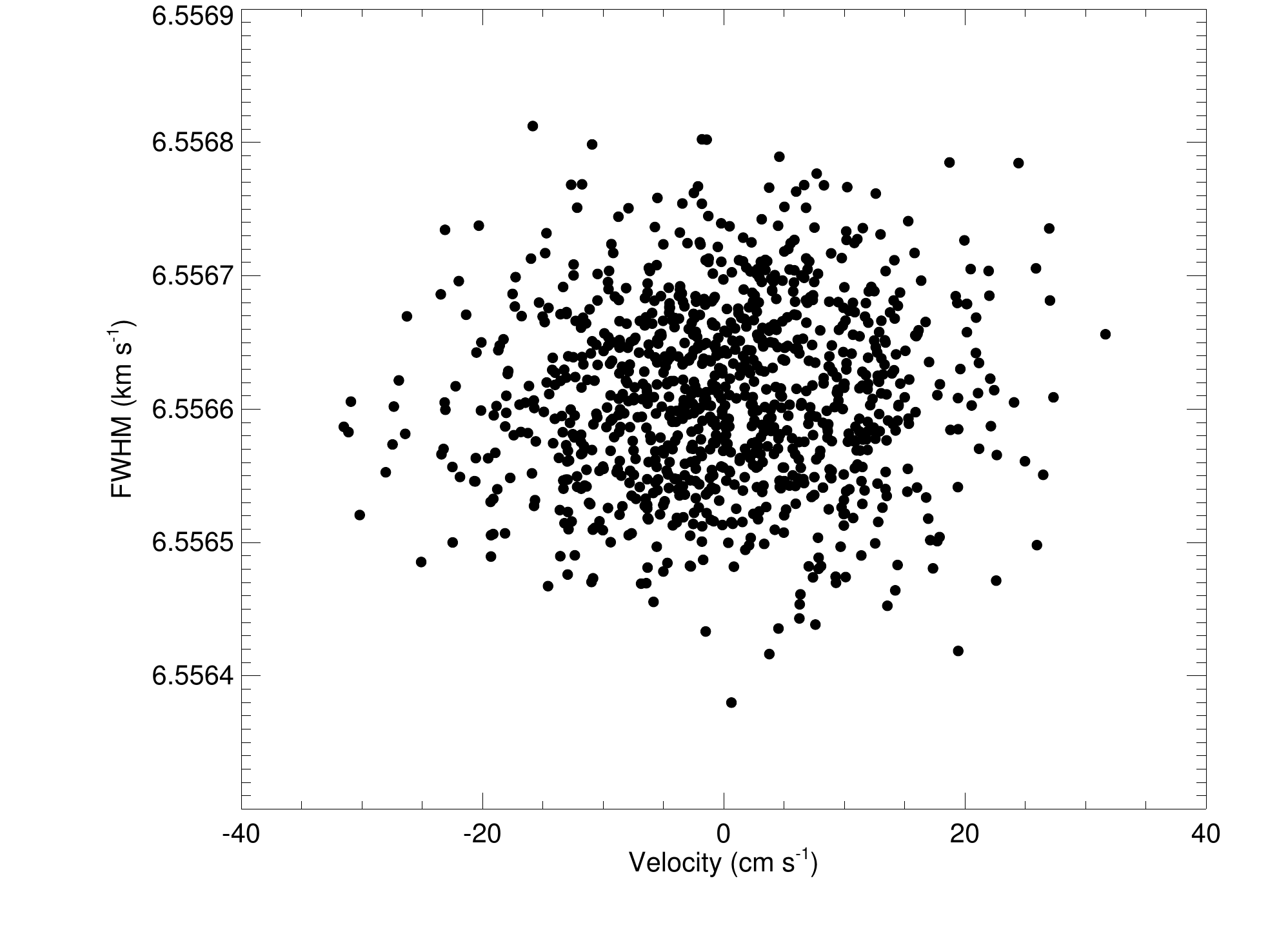} 
\vspace{-20pt}
\caption{Full-width half-maximum (FWHM) versus RV.}
\label{fig:fwhm}
\end{figure}

\begin{figure*}[t!]
\centering
\includegraphics[width = 8.5 cm]{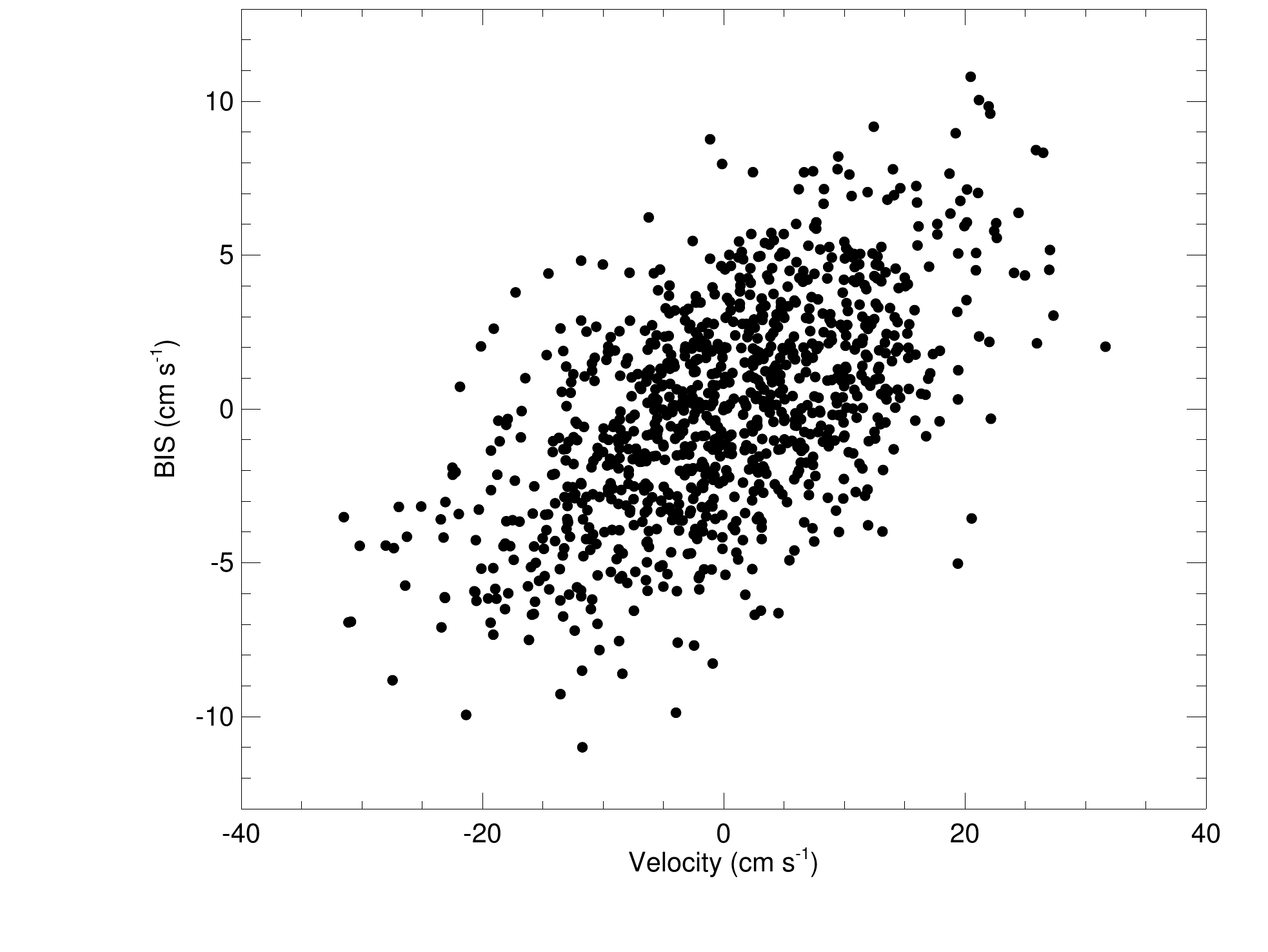} 
\includegraphics[width = 8.5 cm]{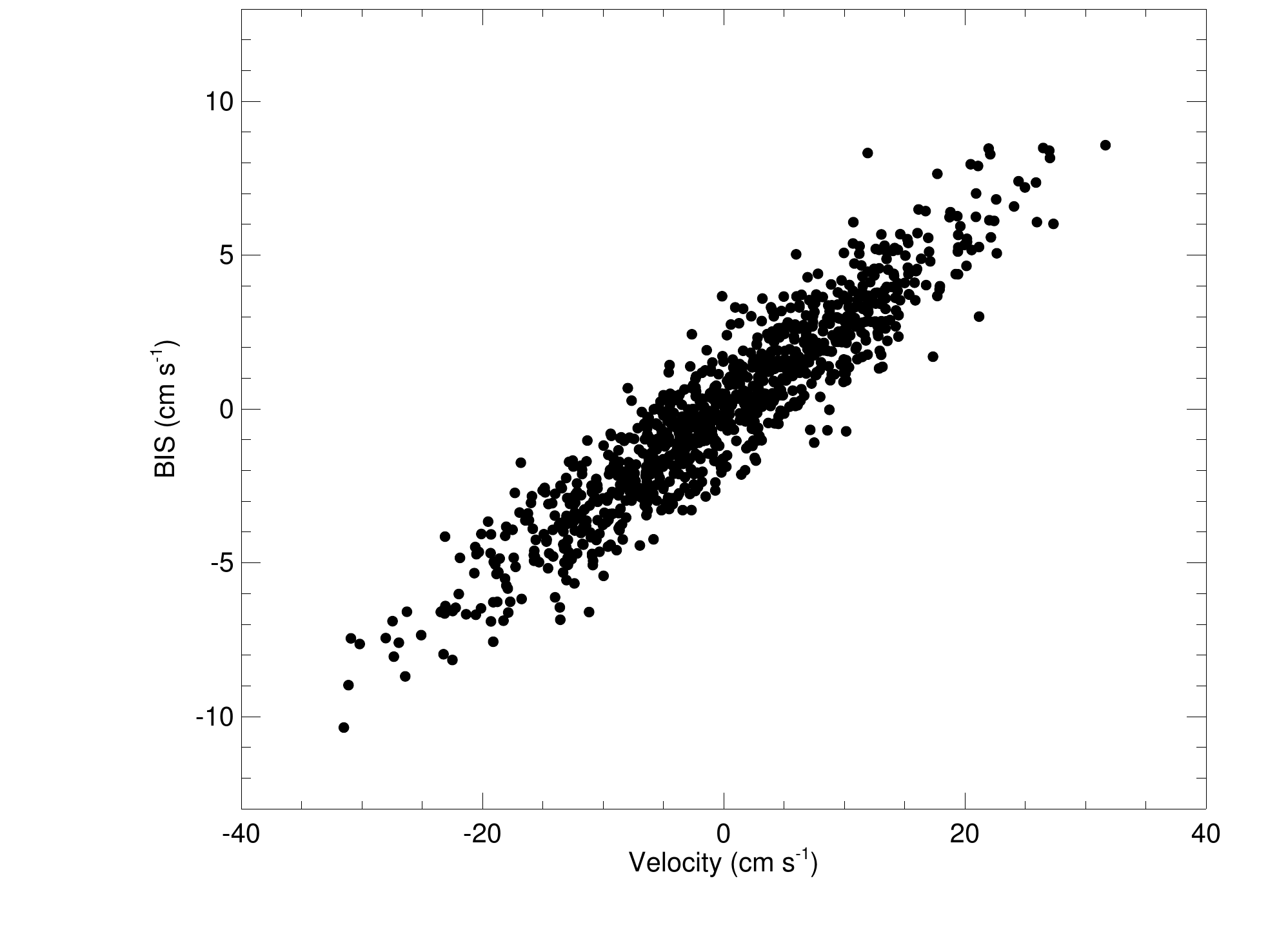} 
\vspace{-15pt}
\caption{The bisector inverse span (BIS) versus RV for the case when the top and bottom regions are defined by the standard values in the literature (left) and also when they are fine-tuned to the disc-integrated model line bisector shape (right). The mean values of the BIS was subtracted for each, to more easily see their net variation (-15.26 and -81.25~m~s$^{-1}$, respectively).}
\label{fig:bis}
\end{figure*}

\begin{figure*}[t]
\includegraphics[width = 8.5 cm]{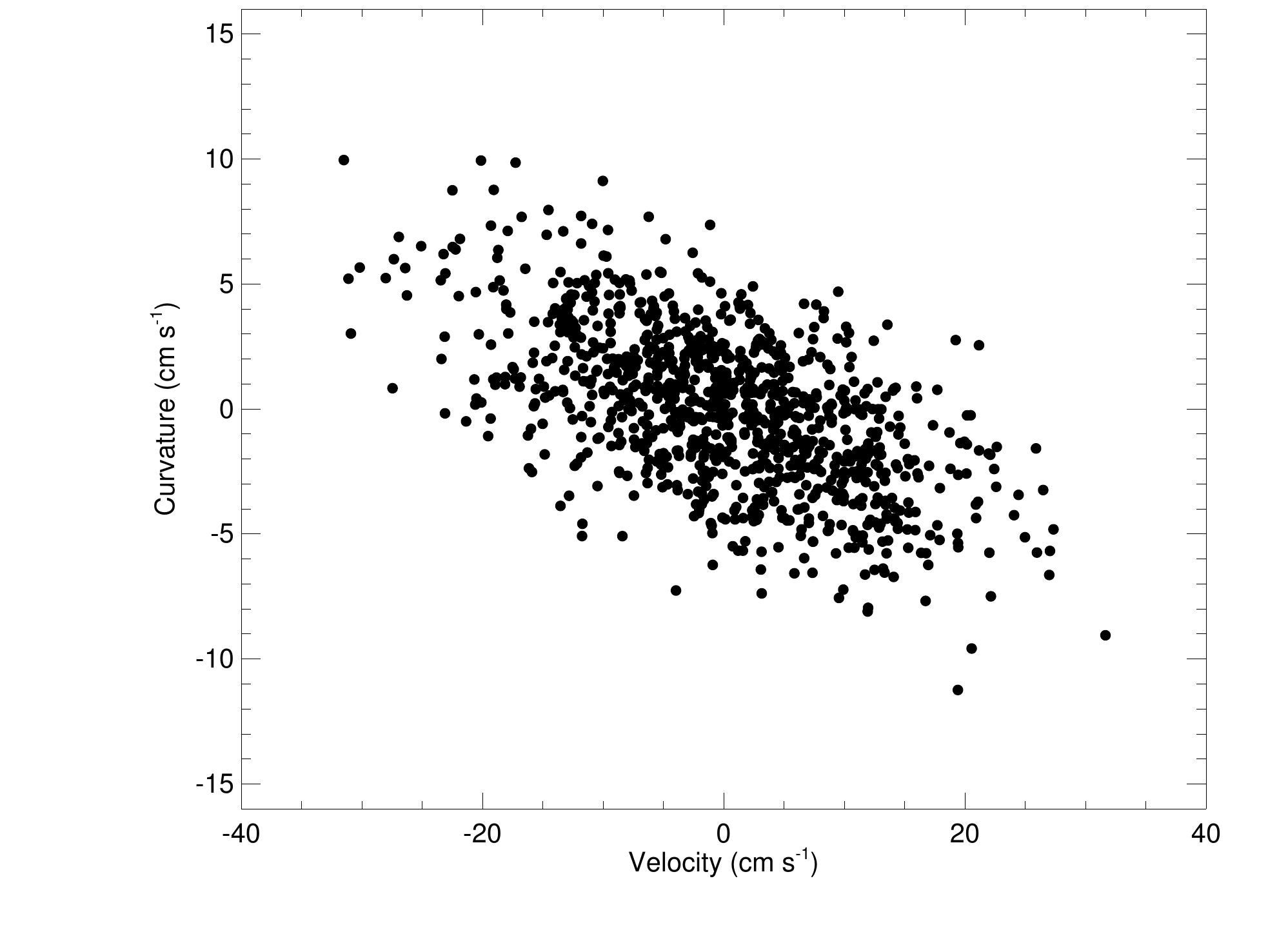} 
\includegraphics[width = 8.5 cm]{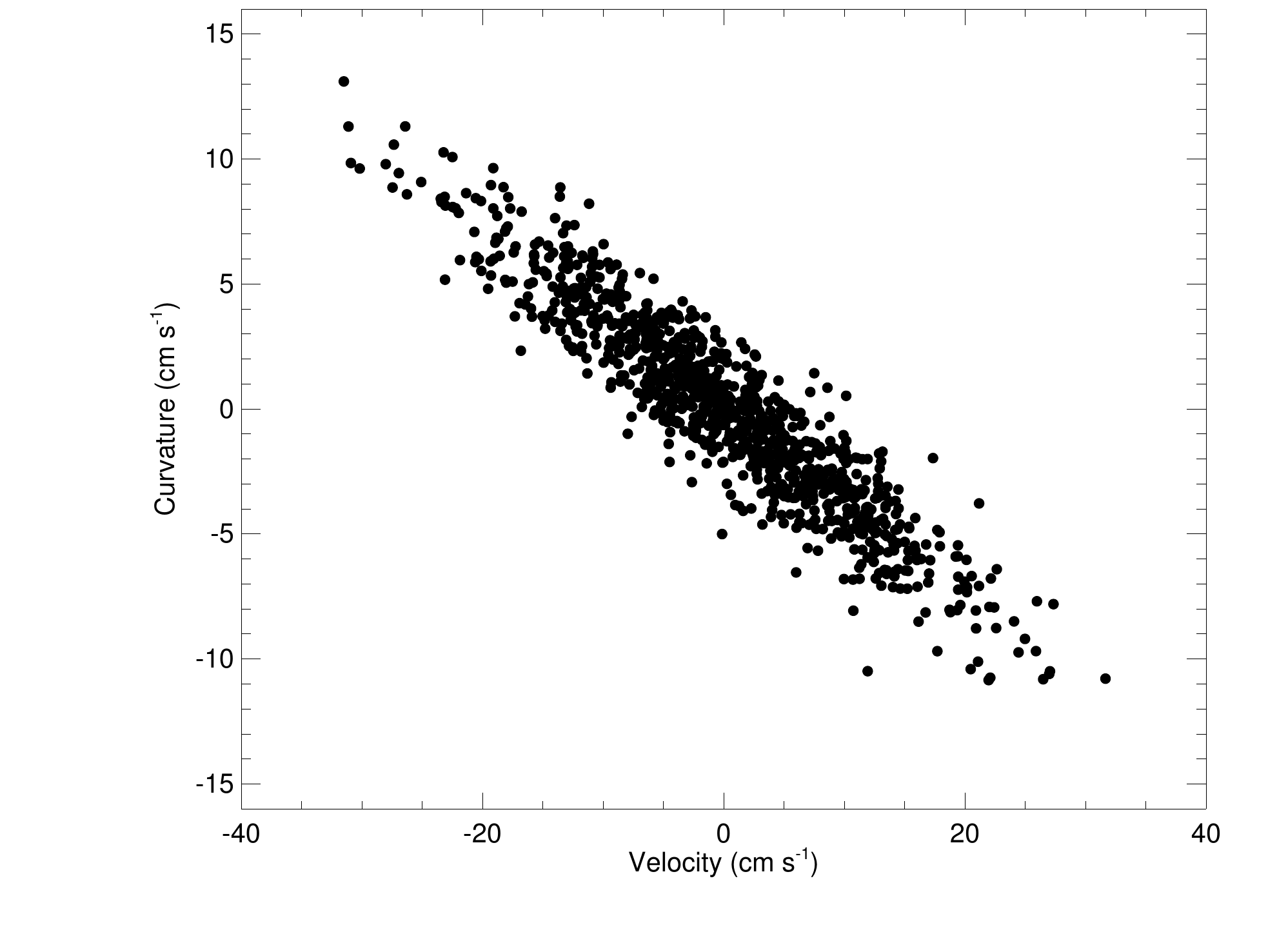} 
\vspace{-15pt}
\caption{Bisector curvature, $C$, versus RV, where the three curvature regions were either defined using standard values in the literature (left) or fine-tuned to the disc-integrated model line bisector shape (right). The mean values for the curvature were subtracted for each, to more easily see their net variation (89.8 and 104.57~m~s$^{-1}$, respectively).}
\label{fig:curve}
\end{figure*}

\begin{figure*}[t]
\centering
\includegraphics[width = 8.5 cm]{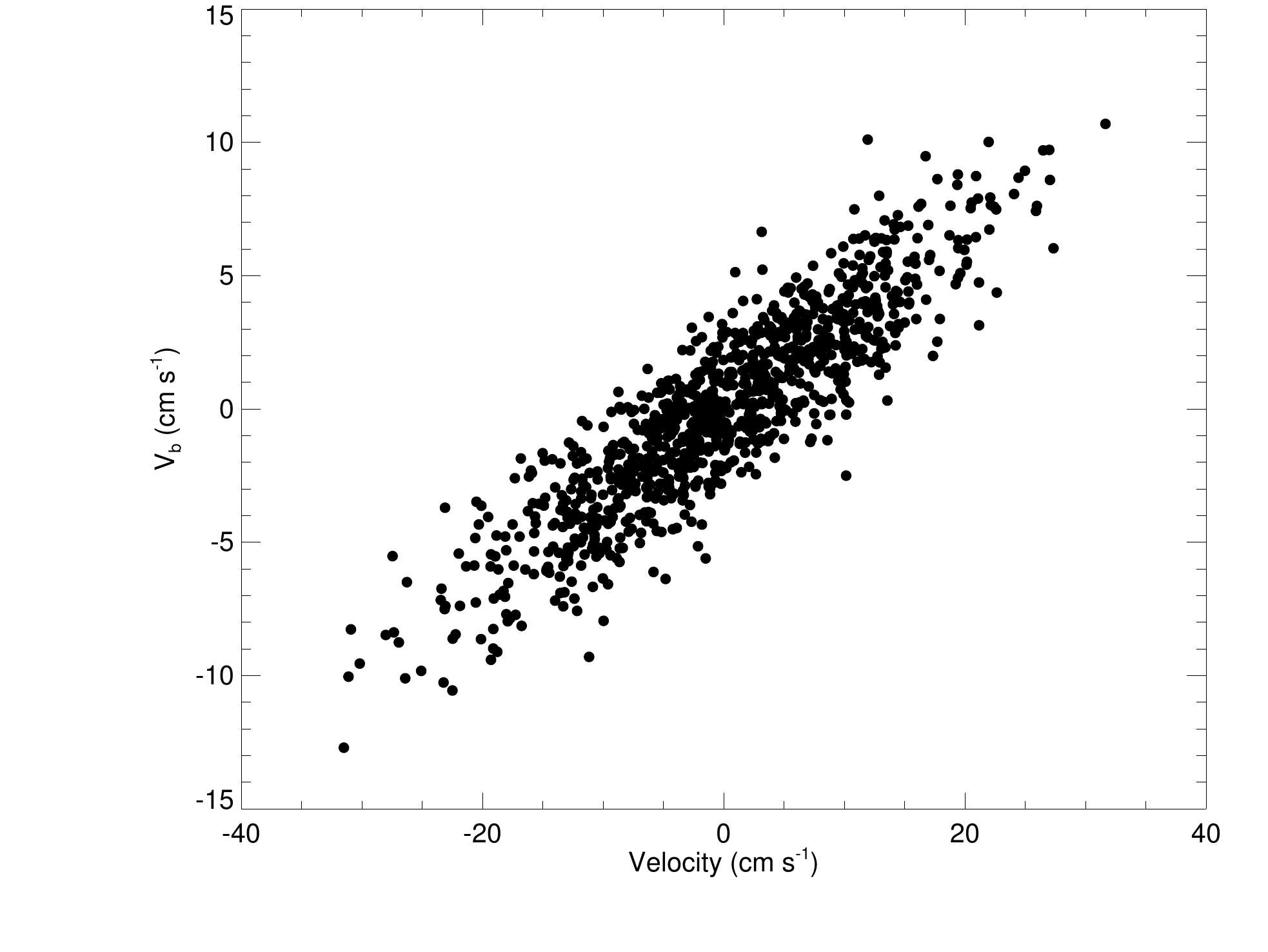} 
\includegraphics[width = 8.5 cm]{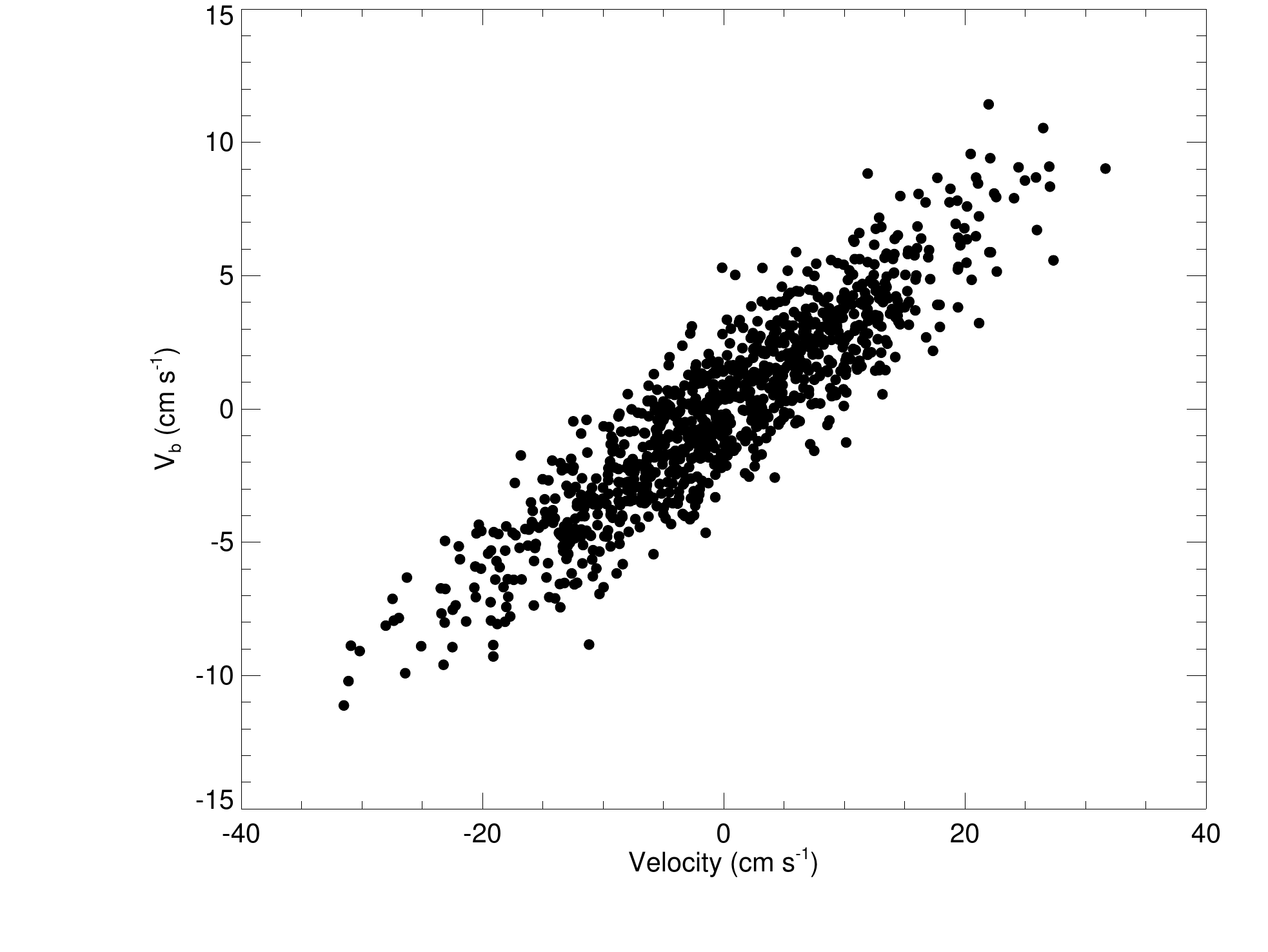} 
\vspace{-15 pt}
\caption{The velocity displacement of the bisector, $V_b$ versus RV, where the bisector regions were defined both using the standard values (left) and the values fine-tuned to the disc-integrated model line bisector (right). The mean values for the $V_b$ were subtracted for each, to more easily see their net variation (-93.47 and -76.58~m~s$^{-1}$, respectively).}
\label{fig:Vb}
\end{figure*}

\begin{figure}[t!]
\centering
\includegraphics[width = 8.5 cm]{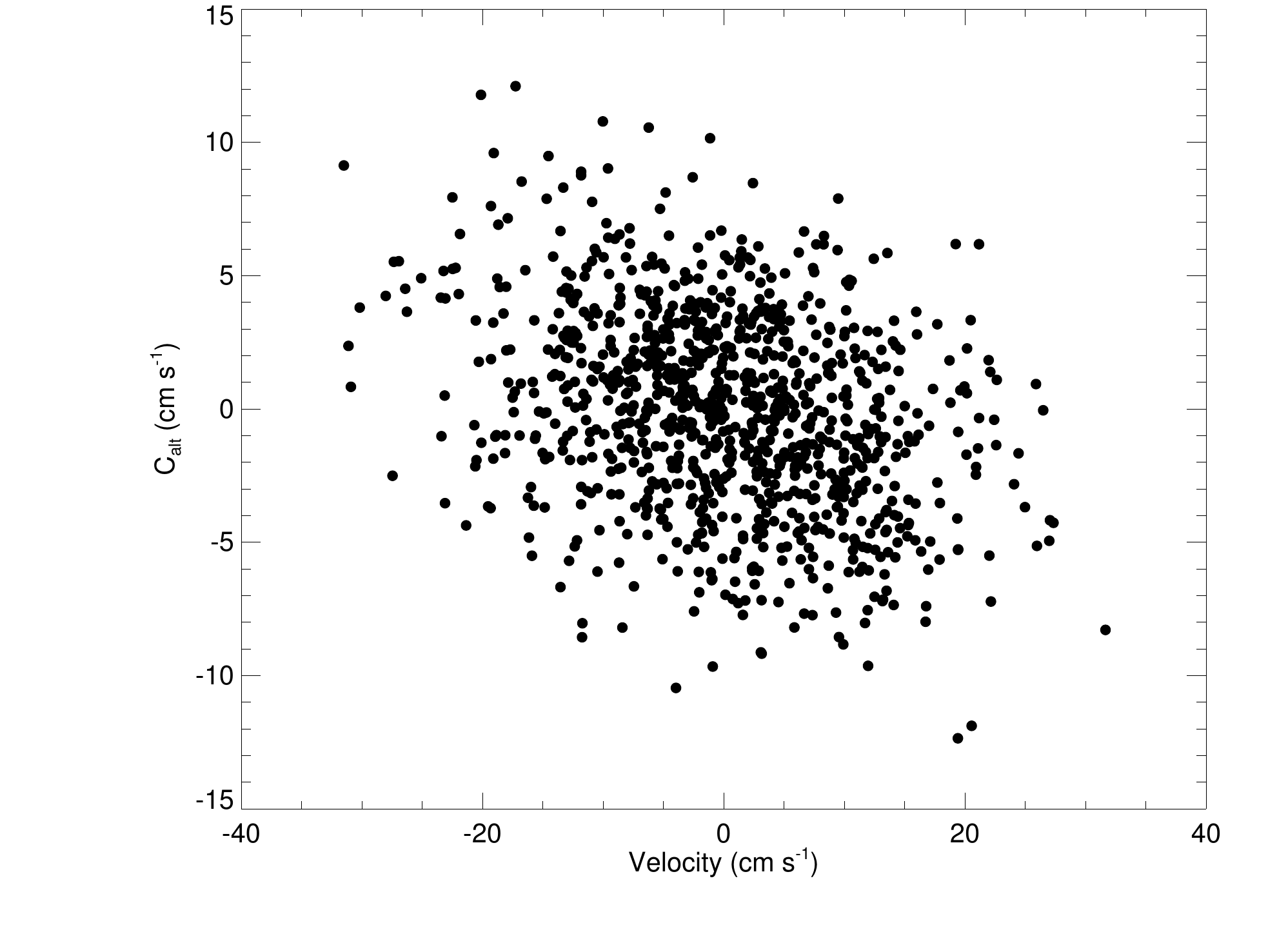} 
\vspace{-25pt}
\caption{An alternative bisector curvature measurement versus RV. The mean curvature, 96.05~m~s$^{-1}$, was subtracted off to more easily see the net variation.}
\label{fig:Calt}
\end{figure}

\begin{figure}[ht!]
\centering
\includegraphics[width = 8.5 cm]{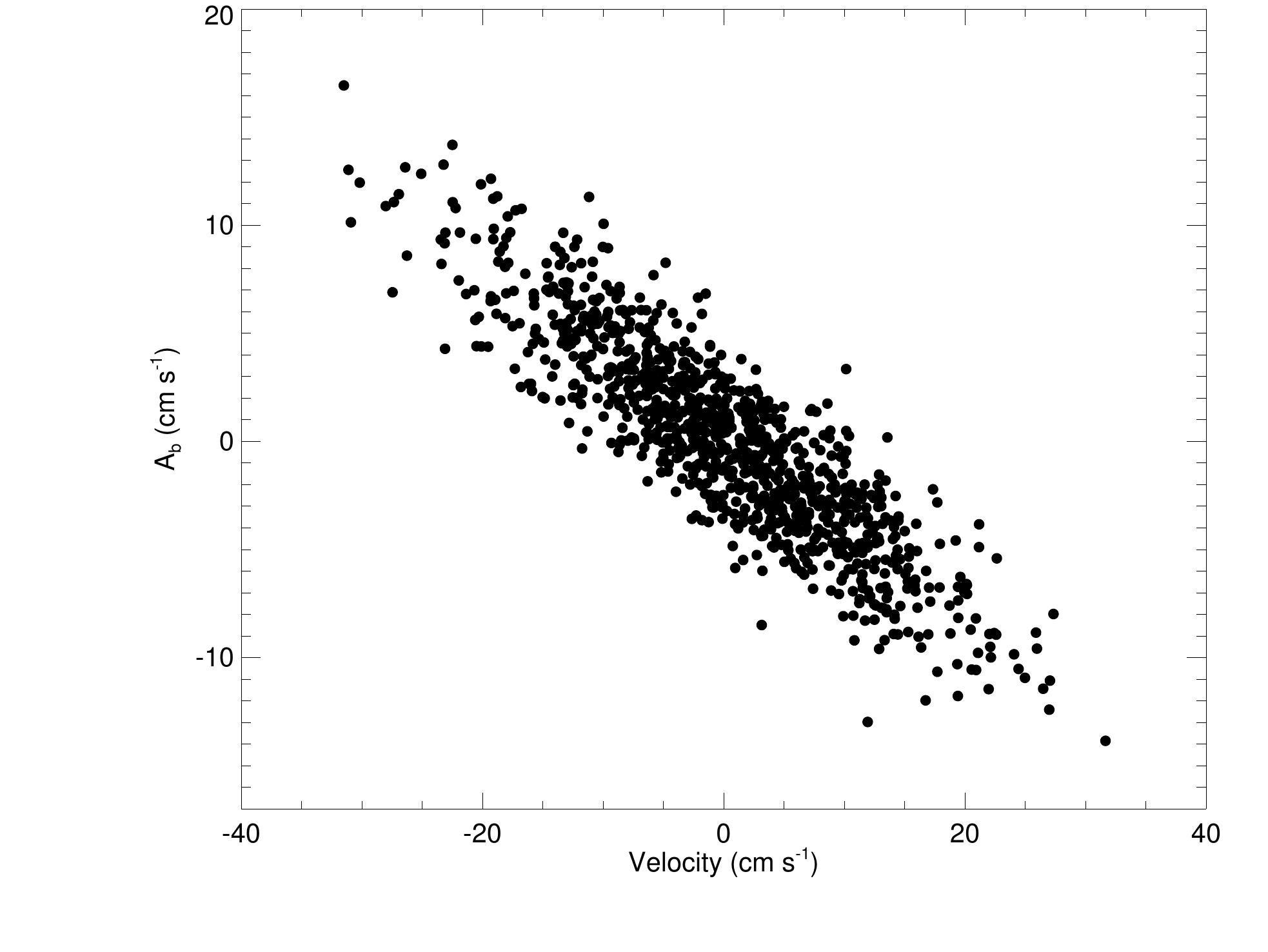} 
\vspace{-25pt}
\caption{The amplitude of the bisector versus RV; the mean amplitude of 125.60~m~s$^{-1}$ was subtracted off to more easily see the net variation.}
\label{fig:Ab}
\end{figure}

\begin{figure*}[t!]
\includegraphics[width = 8.5 cm]{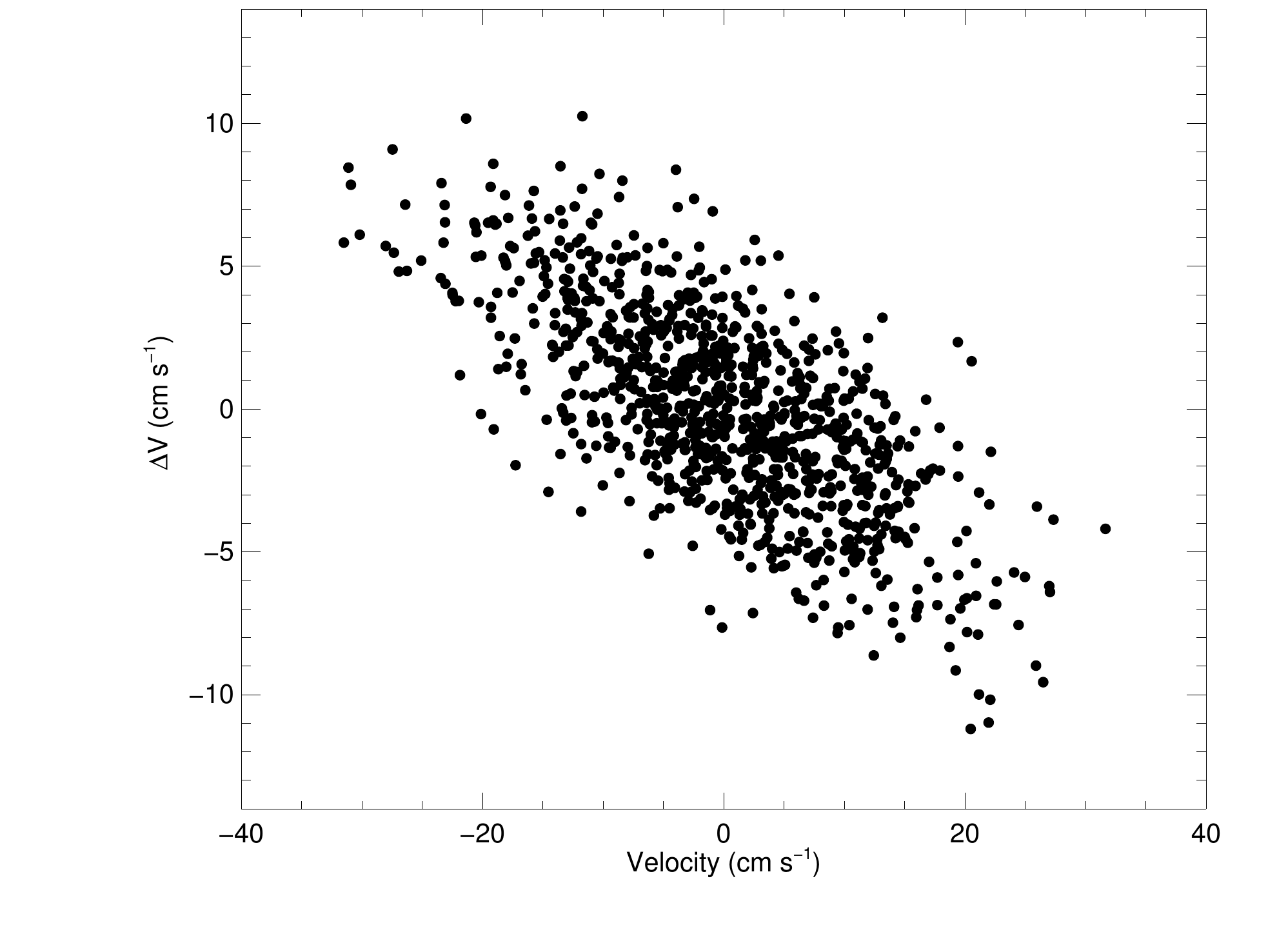} 
\includegraphics[width = 8.5 cm]{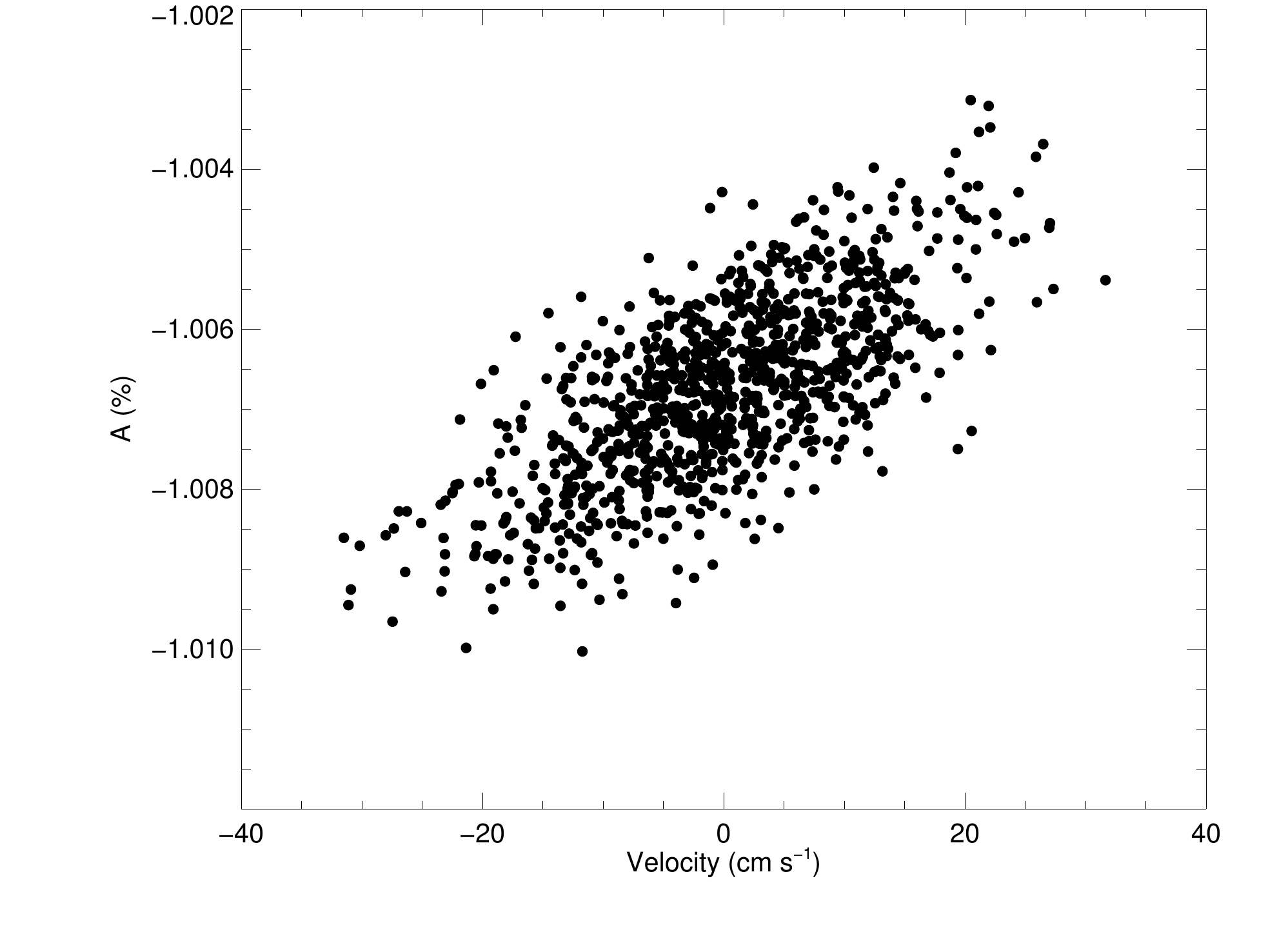} 
\vspace{-15pt}
\caption{Left: the difference in between the RV centroid as determined from bi-Gaussian and (pure) Gaussian fitting, $\Delta V$, versus RV (as determined by cross-correlation); the mean $\Delta V$, 31.45~m~s$^{-1}$, is subtracted to more easily see the net variation. Right: the asymmetry (expressed as a percentage) of the bi-Gaussian fit versus RV.}
\label{fig:biG}
\end{figure*}

\begin{figure}[t]
  \centering
\includegraphics[width = 8.5 cm]{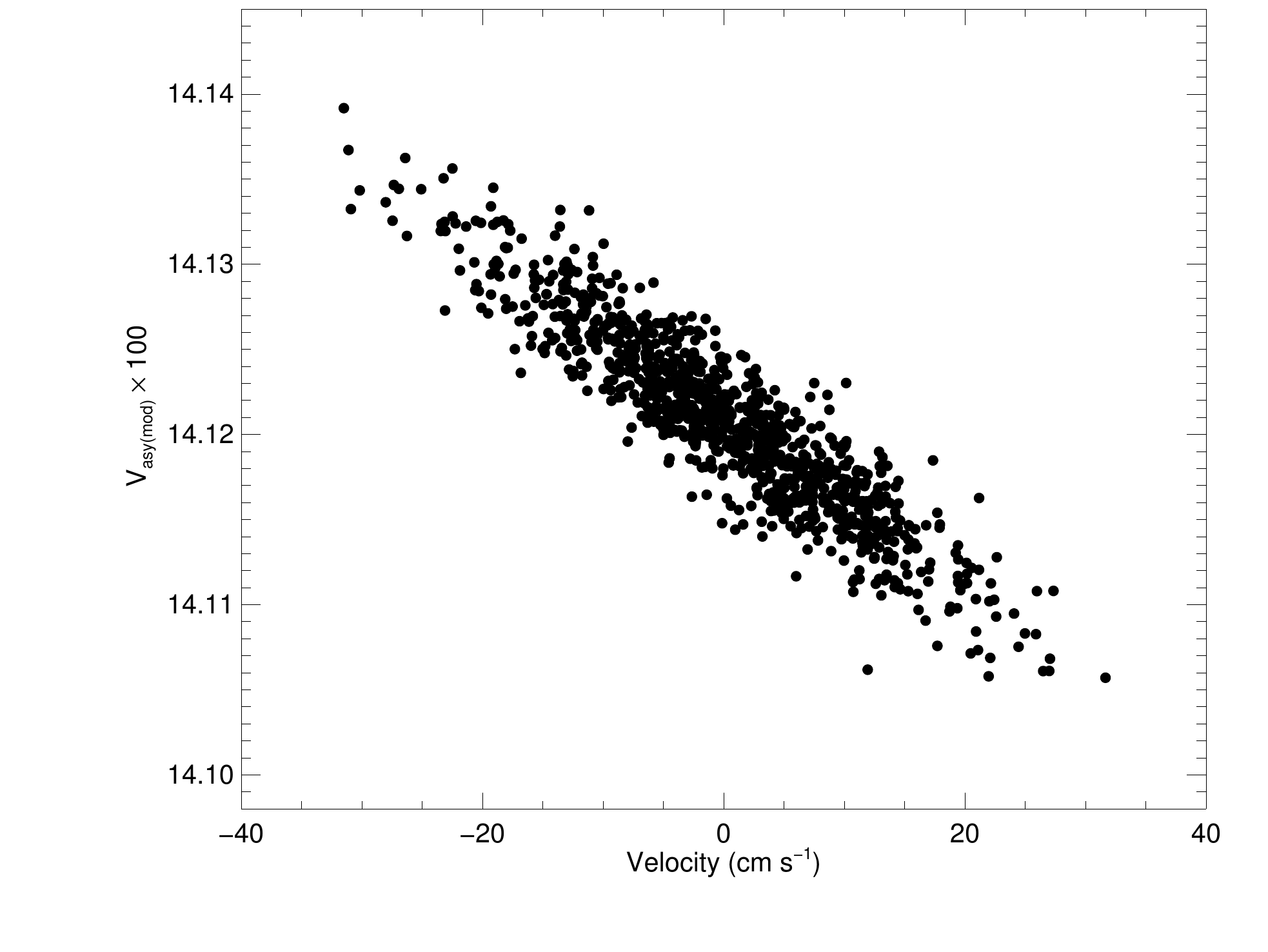} 
\vspace{-25pt}
\caption{The velocity asymmetry, $V_{asy(mod)}$, versus RV, where the $V_{asy(mod)}$ has been multiplied by a factor of 100.}
\label{fig:Vasy}
\end{figure}

\begin{figure}[t]
  \centering
\includegraphics[width = 8.5 cm]{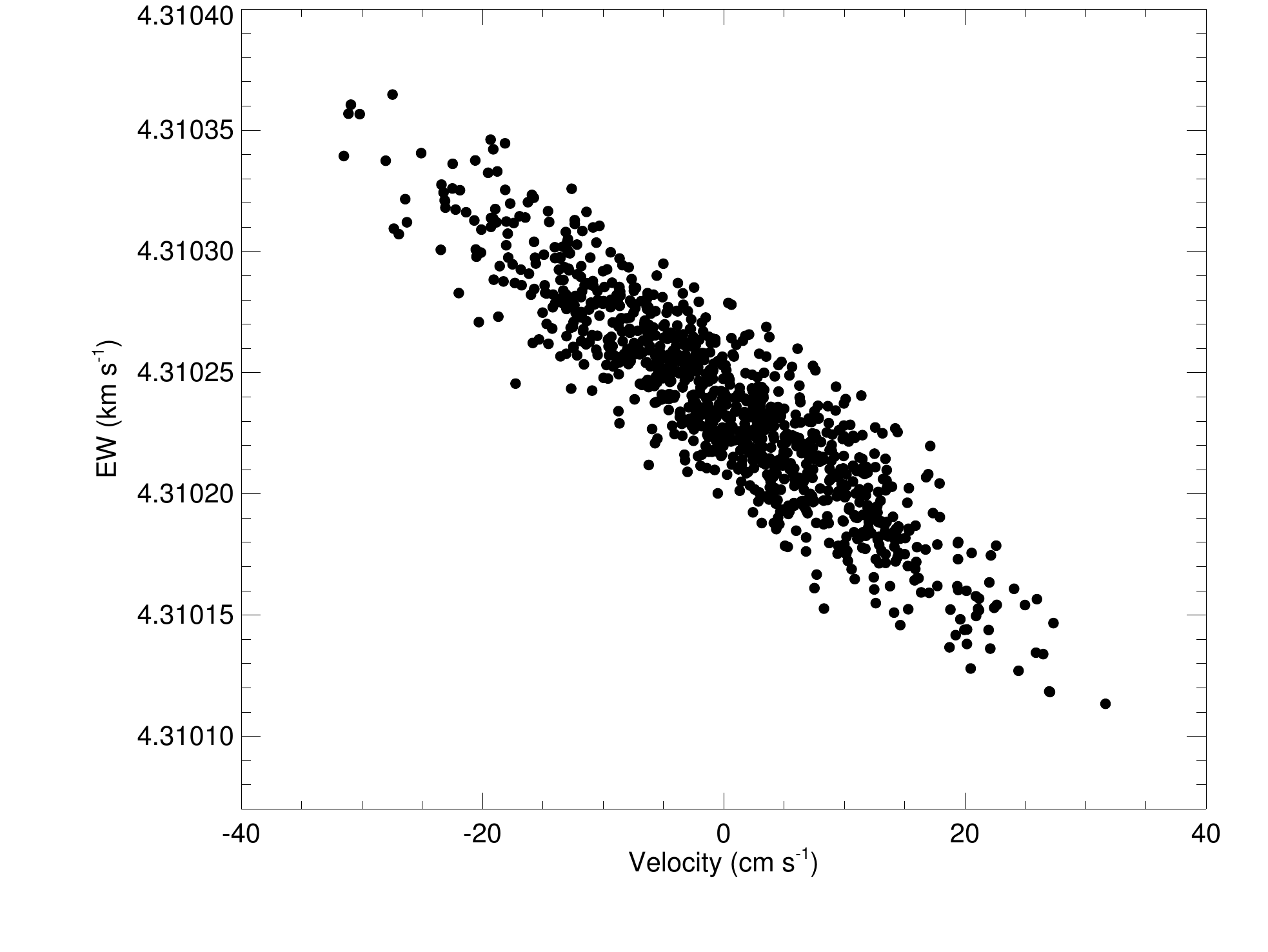} 
\vspace{-25pt}
\caption{The equivalent width (EW) versus RV.}
\label{fig:ew}
\end{figure}

In Section~\ref{sec:corr}, we show how a variety of stellar absorption line profile diagnostics correlate with the corresponding convective-induced RVs from our Sun-as-a-star model observations. The corresponding correlation plots for these diagnostics are shown in Figures~\ref{fig:fwhm}-\ref{fig:ew}. In addition, we provide here the definitions for the bi-Gaussian fitting and the $V_{asy}$ indicator presented in Section~\ref{subsec:full_prof}. The bi-Gaussian fit, following \cite{figueira13} is defined as
\begin{equation}
\label{eq:bigaus1}
\begin{aligned}
bG(RV) = -D \ exp \left ( - \frac{4 \ \mathrm{ln} \ 2(RV - RV_{cen.})^2}{( FWHM \times (1 + A) )^2}  \right ) + Cont.
\\ 
\mathrm{if}~ RV > RV_{cen.} 
\end{aligned}
\end{equation}

\begin{equation}
\label{eq:bigaus2}
\begin{aligned}
 bG(RV) = -D \ exp \left ( - \frac{4 \ \mathrm{ln} \ 2(RV - RV_{cen.})^2}{( FWHM \times (1 - A) )^2}  \right ) + Cont.
 \\ 
 \mathrm{if}~ RV < RV_{cen.} 
\end{aligned}
\end{equation}
with four free parameters: the line depth, $D$, the centre of the bi-Gaussian fit, $RV_{cen.}$, the full width half max ($FWHM$), and the asymmetry, $A$, given as a fraction of the $FWHM$. The RV here is the RV of each point in the line profile and $Cont.$ is the continuum level. Then, from this we can examine the difference in RV as measured from a Bi-Gaussian fit and a pure Gaussian fit (where A = 0): 
\begin{equation}
 \label{eq:deltV}
 \Delta V = RV_{cen.} - RV_{G,cen.}
\end{equation}
The $V_{asy}$ indicator is defined as: 
\begin{equation}
\label{eq:Vasy}
V_{asy(mod)} = \frac{ \sum_{flux} ( W_i(red) - W_i(blue) ) \times \overline{W_i} }{\sum_{flux} \overline{W_i}^2}.
\end{equation}
The weights, $W_i$, were developed by \citeauthor{bouchy01} (2001 -- and references therein) when calculating the fundamental photon noise limit in RV measurements in exoplanet searches. We follow the updates from \cite{figueira15} and \cite{lanza18}:
\begin{equation}
\label{eq:Wi}
W_i = \frac{ 1}{F_0(i)}\ \left (\frac{\partial F_0(i)}{\partial RV(i)} \right)^2 ,
\end{equation}
where $F(i)$ is flux at each point $i$ in the line profile, and $ \partial F_0(i) / \partial RV(i)$ is the slope of the line at each $i$th point; the slope is measured between one flux position, $F(i)$, and the position directly next to it, $F(i+1)$. 

\section{Additional Noise Reductions}
\label{appen:other_noise}
In Section~\ref{sec:add_fact} we discuss the impact on the disc-integrated absorption line profiles from the Sun-as-a-star model observations if the $v \sin i$ is increased from 2 to 10~km~s$^{-1}$. Moreover, how these changes impact the resultant line profile diagnostics and their noise mitigation ability is discussed in Section~\ref{sec:reduct}. Table~\ref{tab:noiseremove_tab1} displays the noise mitigation results for a large variety of line profile diagnostics, at different instrumental resolutions, for model stars with a $v \sin i$ of 2, 4, and 6~km~s$^{-1}$. Here we show these results for model stars with a $v \sin i$ of 8 and 10~km~s$^{-1}$ in Table~\ref{tab:noiseremove_8_10_kms}. See Sections~\ref{sec:add_fact} and \ref{sec:reduct} for more discussion on these results.

The stellar rotation tests performed in Section~\ref{sec:add_fact} assume solid body rotation. However, we know from empirical observations of the Sun and Sun-like stars that stellar surface differential rotation is common. Herein we briefly explore the impact of differential rotation as a function of equatorial velocity. For this, we use the empirical law derived from the Sun: $\Omega = \Omega_{eq} (1 - \alpha \sin^2 \theta)$, where $\Omega$ is the angular rotational velocity, $\Omega_{eq}$ is the value corresponding to the equator, $\theta$ is the stellar latitude, and $\alpha$ is the relative differential rotation rate. In particular,  $\alpha$ is the difference between the equatorial and polar velocities, relative to the equatorial velocity, i.e. $\alpha = (\Omega_{eq} - \Omega_{pole}) / \Omega_{eq}$; on the Sun this is approximately 0.2 (negative $\alpha$ values are considered anti-solar). Naturally, a full investigation of differential rotation would require testing a large range of both solar and anti-solar $\alpha$ values, as well as a full range of stellar inclinations. Such a complete exploration is beyond the scope of the paper. However, to get a feel for the potential impact we created model stars with $\alpha$ = 0.2 and 0.6 -- to match the Sun and to explore a more extreme rotational sheer. We also hold the stellar inclination fixed at 90$^{\rm{o}}$ and model stars with equatorial velocities of 2, 4 and  6~km~s$^{-1}$.

The final noise reduction diagnostics from each of these differentially rotating model stars are shown in Tables~\ref{tab:noiseremove_2_kms_DR}, \ref{tab:noiseremove_4_kms_DR}, and \ref{tab:noiseremove_6_kms_DR}. By comparing these results to those from the model stars with solid body rotation (also shown in Tables~\ref{tab:noiseremove_2_kms_DR} -- \ref{tab:noiseremove_6_kms_DR}.), it is clear that the differential rotation has minimal impact for this range of systems. We expect a larger impact for faster rotating stars, but as rapidly rotating stars are typically younger and more magnetically active, granulation is unlikely to be the dominant noise source for these systems. There may also be larger differences for systems where the stellar rotational axis is highly inclined from the line-of-sight; e.g. a near pole-on system could potentially have a stellar rotational sheer on par with the amplitude of the $v \sin i_{\star}$, in which case it would be necessary to account for the differential rotation if trying to disentangle the convection-induced bisector/line profile changes. 

We note that there were visible differences in the bisector shapes for the model with $v_{eq}$ = 6~km~s$^{-1}$ and $\alpha$ = 0.6 as compared to the equivalent solid body case. As such, if the granulation noise diagnostics for this rotation rate were able to provide tangible noise reduction we may have seen an impact when including this level of differential rotation. Yet, there were also small bisector shape differences for the $v_{eq}$ = 4~km~s$^{-1}$ scenario, and the two diagnostics capable of noise reduction experienced negligible changes when including differential rotation. Consequently, stellar surface differential rotation appears to have a minimal impact on our ability to mitigate granulation-induced RV variability.

\begin{table*}[h]
 \caption{Granulation noise reduction for various diagnostics, alongside their Pearson's R correlation coefficient.}
\centering

\begin{tabular}{c|cc|cc|cc|cc}
    \hline
    $v \sin i$ = 8~km~s$^{-1}$ & & & \multicolumn{2}{c|}{Resolution = 70,000} &  \multicolumn{2}{c|}{Resolution = 140,000}  &  \multicolumn{2}{c}{Resolution = 140,000}   \\
    \hline
     Diagnostic & Reduct. (\%) & R & Reduct. (\%) & R  & Reduct. (\%) & R & Reduct. (\%) & R\\
    \hline
Abs. Depth & -523.34 & -0.16 & -414.24 & -0.19 & -620.55 & -0.14 & -672.25 & -0.13 \\
Norm. Depth & -1070.42 & 0.09 & -873.19 & 0.10 & -670.77 & 0.13 & -649.47 & 0.13 \\
FWHM & -288.11 & -0.25 & -201.05 & -0.32 & -233.99 & -0.29 & -237.97 & -0.28 \\
BIS$_{\rm{std}}$ & -14.45 & -0.66 & -16.28 & -0.65 & -13.97 & -0.66 & -13.66 & -0.66 \\
BIS$_{\rm{best}}$ & -37.94 & -0.59 & -29.39 & -0.61 & -33.46 & -0.60 & -33.91 & -0.60 \\
C$_{\rm{std}}$ & -80.03 & 0.49 & -66.62 & 0.51 & -73.48 & 0.50 & -73.78 & 0.50 \\
C$_{\rm{best}}$ & -41.18 & 0.58 & -28.64 & 0.61 & -36.18 & 0.59 & -35.64 & 0.59 \\
V$_{\rm{b,std}}$ & -68.31 & -0.51 & -34.62 & -0.60 & -49.10 & -0.56 & -54.29 & -0.54 \\
V$_{\rm{b,best}}$ & -61.55 & -0.53 & -29.37 & -0.61 & -41.69 & -0.58 & -46.89 & -0.56 \\
C$_{\rm{alt}}$ & -168.60 & -0.35 & -190.84 & -0.33 & -175.27 & -0.34 & -168.07 & -0.35 \\
A$_{\rm{b}}$ & -101.51 & 0.44 & -707.95 & 0.12 & -74.01 & 0.50 & -83.54 & 0.48 \\
$\Delta$V & -20.38 & 0.64 & -5605.90 & 0.02 & 38.76 & 0.85 & -202.95 & 0.31 \\
A & -20.55 & -0.64 & -17.12 & -0.65 & -18.32 & -0.65 & -18.75 & -0.64 \\
V$_{\rm{asy}}$ & -22.58 & 0.63 & -22.05 & 0.63 & -22.28 & 0.63 & -20.81 & 0.64 \\
EW & -32.87 & -0.60 & -35.70 & -0.59 & -33.81 & -0.60 & -33.73 & -0.60 \\
Brightness & -42.59 & -0.57 & -41.85 & -0.58 & -43.21 & -0.57 & -43.38 & -0.57 \\    
    \hline
    $v \sin i$ = 10~km~s$^{-1}$\\
   \hline
Abs. Depth & -975.82 & -0.09 & -809.11 & -0.11 & -1234.62 & -0.07 & -1348.82 & -0.07 \\
Norm. Depth & -795.23 & 0.11 & -614.56 & 0.14 & -576.67 & 0.15 & -573.34 & 0.15 \\
FWHM & -269.66 & -0.26 & -223.69 & -0.30 & -234.38 & -0.29 & -237.20 & -0.28 \\
BIS$_{\rm{std}}$ & -7.94 & -0.68 & -8.05 & -0.68 & -6.38 & -0.68 & -7.28 & -0.68 \\
BIS$_{\rm{best}}$ & -51.47 & -0.55 & -36.75 & -0.59 & -47.68 & -0.56 & -49.27 & -0.56 \\
C$_{\rm{std}}$ & -99.54 & 0.45 & -80.10 & 0.49 & -95.58 & 0.46 & -95.70 & 0.46 \\
C$_{\rm{best}}$ & -63.39 & 0.52 & -42.48 & 0.57 & -55.35 & 0.54 & -57.98 & 0.53 \\
V$_{\rm{b,std}}$ & -139.24 & -0.39 & -68.06 & -0.51 & -109.81 & -0.43 & -120.27 & -0.41 \\
V$_{\rm{b,best}}$ & -124.32 & -0.41 & -56.08 & -0.54 & -93.63 & -0.46 & -104.04 & -0.44 \\
C$_{\rm{alt}}$ & -136.23 & -0.39 & -158.13 & 0.36 & -137.90 & 0.39 & -150.84 & 0.37 \\
A$_{\rm{b}}$ & -172.03 & 0.35 & -98.47 & 0.45 & -139.02 & 0.39 & -150.26 & 0.37 \\
$\Delta$V & -26.26 & 0.62 & -1915.68 & -0.05 & -1609.51 & 0.06 & -411.69 & 0.19 \\
A & -26.20 & -0.62 & -16.49 & -0.65 & -23.15 & -0.63 & -24.38 & -0.63 \\
V$_{\rm{asy}}$ & -12.01 & 0.67 & -18.59 & 0.64 & -15.00 & 0.66 & -11.70 & 0.67 \\
EW & -60.74 & -0.53 & -71.08 & -0.50 & -61.31 & -0.53 & -60.60 & -0.53 \\
Brightness & -66.34 & -0.52 & -67.73 & -0.51 & -67.28 & -0.51 & -67.16 & -0.51 \\
\end{tabular}
\label{tab:noiseremove_8_10_kms}
\end{table*}  

\begin{table*}[h]
 \caption{Granulation noise reduction for various diagnostics from model observations with $v_{eq}$ = 2~km~s$^{-1}$ and $i_{\star}$ = 90$^{\rm{o}}$, at various levels of differential rotation, alongside their Pearson's R correlation coefficient.}
\centering

\begin{tabular}{c|cc|cc|cc|cc}
    \hline
    $\alpha$ = 0.0 & & & \multicolumn{2}{c|}{Resolution = 70,000} &  \multicolumn{2}{c|}{ Resolution = 140,000}  &  \multicolumn{2}{c}{Resolution = 140,000}   \\
    \hline
     Diagnostic & Reduct. (\%) & R & Reduct. (\%) & R  & Reduct. (\%) & R & Reduct. (\%) & R\\
    \hline
 Abs. Depth & 36.37 & -0.84 & 35.65 & -0.84 & 32.42 & -0.83 & 32.66 & -0.83 \\
Norm. Depth & -13.38 & -0.66 & -20.02 & -0.64 & -28.97 & -0.61 & -28.04 & -0.62 \\
FWHM & -1387.91 & 0.07 & -586.67 & 0.14 & -831.93 & 0.11 & -4468.55 & -0.02 \\
BIS$_{\rm{std}}$ & -42.31 & 0.57 & -81.40 & 0.48 & -48.05 & 0.56 & -44.98 & 0.57 \\
BIS$_{\rm{best}}$ & 61.20 & 0.93 & 11.16 & 0.75 & 52.37 & 0.90 & 57.41 & 0.92 \\
C$_{\rm{std}}$ & -34.29 & -0.60 & 39.12 & -0.85 & -51.39 & -0.55 & -6.30 & -0.69 \\
C$_{\rm{best}}$ & 61.63 & -0.93 & 36.46 & -0.84 & 45.50 & -0.88 & 50.80 & -0.90 \\
V$_{\rm{b,std}}$ & 49.58 & 0.89 & -0.77 & 0.70 & 56.72 & 0.92 & 59.24 & 0.93 \\
V$_{\rm{b,best}}$ & 52.32 & 0.90 & -58.94 & 0.53 & 36.78 & 0.85 & 43.54 & 0.87 \\
C$_{\rm{alt}}$ & -135.77 & -0.39 & -694.33 & 0.12 & -234.10 & -0.29 & -170.33 & -0.35 \\
A$_{\rm{b}}$ & 48.10 & -0.89 & 3.93 & -0.72 & 54.01 & -0.91 & 60.68 & -0.93 \\
$\Delta$V & -0.53 & -0.71 & -66.31 & -0.52 & -25.83 & -0.62 & -15.42 & -0.65 \\
A & -0.63 & 0.70 & -66.35 & 0.52 & -25.81 & 0.62 & -15.44 & 0.65 \\
V$_{\rm{asy}}$ & 57.85 & -0.92 & 49.40 & -0.89 & 36.71 & -0.85 & 36.54 & -0.84 \\
EW & 53.91 & -0.91 & 52.83 & -0.90 & 52.74 & -0.90 & 52.66 & -0.90 \\
Brightness & 52.70 & -0.90 & 54.30 & -0.91 & 53.02 & -0.91 & 52.76 & -0.90 \\   
    \hline
     $\alpha$ = 0.2\\
   \hline
Abs. Depth & 36.36 & -0.84 & 37.01 & -0.85 & 32.63 & -0.83 & 32.86 & -0.83 \\
Norm. Depth & -13.32 & -0.66 & -16.24 & -0.65 & -28.77 & -0.61 & -27.82 & -0.62 \\
FWHM & -1347.94 & 0.07 & -946.72 & 0.10 & -774.74 & 0.11 & -7488.75 & -0.01 \\
BIS$_{\rm{std}}$ & -43.77 & 0.57 & -80.00 & 0.49 & -55.03 & 0.54 & -40.95 & 0.58 \\
BIS$_{\rm{best}}$ & 61.22 & 0.93 & 11.72 & 0.75 & 51.19 & 0.90 & 56.57 & 0.92 \\
C$_{\rm{std}}$ & -32.33 & -0.60 & -49.50 & -0.56 & -47.13 & -0.56 & -53.50 & -0.55 \\
C$_{\rm{best}}$ & 61.78 & -0.93 & 28.90 & -0.81 & 44.13 & -0.87 & 56.67 & -0.92 \\
V$_{\rm{b,std}}$ & 48.90 & 0.89 & 2.57 & 0.72 & 56.01 & 0.92 & 60.81 & 0.93 \\
V$_{\rm{b,best}}$ & 51.37 & 0.90 & -76.40 & 0.49 & 35.16 & 0.84 & 45.56 & 0.88 \\
C$_{\rm{alt}}$ & -131.93 & -0.40 & -983.63 & 0.09 & -225.07 & -0.29 & -139.65 & -0.39 \\
A$_{\rm{b}}$ & 27.90 & -0.81 & 6.98 & -0.73 & 55.58 & -0.91 & 60.12 & -0.93 \\
$\Delta$V & -4.16 & -0.69 & -70.42 & -0.51 & -29.19 & -0.61 & -18.63 & -0.64 \\
A & -4.26 & 0.69 & -70.47 & 0.51 & -29.18 & 0.61 & -18.64 & 0.64 \\
V$_{\rm{asy}}$ & 57.69 & -0.92 & -20.05 & -0.64 & 37.86 & -0.85 & 60.67 & -0.93 \\
EW & 57.23 & -0.92 & 59.08 & -0.93 & 57.77 & -0.92 & 57.50 & -0.92 \\
Brightness & 53.17 & -0.91 & 54.97 & -0.91 & 53.68 & -0.91 & 53.42 & -0.91 \\
   
   \hline
     $\alpha$ = 0.6\\
   \hline
Abs. Depth & 35.22 & -0.84 & 36.29 & -0.84 & 31.73 & -0.83 & 31.87 & -0.83 \\
Norm. Depth & -15.44 & -0.65 & -17.31 & -0.65 & -30.26 & -0.61 & -29.54 & -0.61 \\
FWHM & -1230.14 & 0.07 & -928.12 & 0.10 & -758.77 & 0.12 & -16053.98 & -0.01 \\
BIS$_{\rm{std}}$ & -52.48 & 0.55 & -99.81 & 0.45 & -69.45 & 0.51 & -55.09 & 0.54 \\
BIS$_{\rm{best}}$ & 59.34 & 0.93 & 5.41 & 0.73 & 48.19 & 0.89 & 56.45 & 0.92 \\
C$_{\rm{std}}$ & -33.28 & -0.60 & -9.19 & -0.68 & -47.42 & -0.56 & -42.72 & -0.57 \\
C$_{\rm{best}}$ & 58.33 & -0.92 & 25.07 & -0.80 & 40.25 & -0.86 & 53.62 & -0.91 \\
V$_{\rm{b,std}}$ & 45.95 & 0.88 & -1.83 & 0.70 & 53.86 & 0.91 & 58.74 & 0.92 \\
V$_{\rm{b,best}}$ & 48.81 & 0.89 & -93.99 & 0.46 & 31.72 & 0.83 & 42.68 & 0.87 \\
C$_{\rm{alt}}$ & -139.65 & -0.39 & -1140.95 & 0.08 & -258.50 & -0.27 & -130.25 & -0.40 \\
A$_{\rm{b}}$ & 46.13 & -0.88 & 89.19 & -0.99 & 59.17 & -0.93 & 58.58 & -0.92 \\
$\Delta$V & -9.40 & -0.67 & -82.72 & -0.48 & -36.47 & -0.59 & -25.06 & -0.62 \\
A & -9.47 & 0.67 & -82.78 & 0.48 & -36.43 & 0.59 & -25.06 & 0.62 \\
V$_{\rm{asy}}$ & 55.81 & -0.91 & -43.56 & -0.57 & 35.01 & -0.84 & 59.13 & -0.93 \\
EW & 56.83 & -0.92 & 58.73 & -0.92 & 57.37 & -0.92 & 57.11 & -0.92 \\
Brightness & 52.57 & -0.90 & 54.41 & -0.91 & 53.09 & -0.91 & 52.82 & -0.90 \\

\end{tabular}
\label{tab:noiseremove_2_kms_DR}
\end{table*}

\begin{table*}[h]
 \caption{Granulation noise reduction for various diagnostics from model observations with $v_{eq}$ = 4~km~s$^{-1}$ and $i_{\star}$ = 90$^{\rm{o}}$, at various levels of differential rotation, alongside their Pearson's R correlation coefficient.}
\centering

\begin{tabular}{c|cc|cc|cc|cc}
    \hline
    $\alpha$ = 0.0 & & & \multicolumn{2}{c|}{Resolution = 70,000} &  \multicolumn{2}{c|}{Resolution = 140,000}  &  \multicolumn{2}{c}{Resolution = 140,000}   \\
    \hline
     Diagnostic & Reduct. (\%) & R & Reduct. (\%) & R  & Reduct. (\%) & R & Reduct. (\%) & R\\
    \hline
Abs. Depth & -51.21 & -0.55 & -42.63 & -0.57 & -66.16 & -0.52 & -70.06 & -0.51 \\
Norm. Depth & -258.48 & -0.27 & -305.51 & -0.24 & -466.26 & -0.17 & -493.11 & -0.17 \\
FWHM & -932.73 & -0.10 & -270.23 & -0.26 & -377.25 & -0.21 & -521.71 & -0.16 \\
BIS$_{\rm{std}}$ & -212.15 & -0.31 & -235.44 & -0.29 & -201.93 & -0.31 & -201.09 & -0.32 \\
BIS$_{\rm{best}}$ & -300.44 & -0.24 & -260.68 & -0.27 & -235.40 & -0.29 & -257.31 & -0.27 \\
C$_{\rm{std}}$ & -235.87 & -0.29 & -2655.09 & -0.04 & -213.01 & -0.30 & -213.20 & -0.30 \\
C$_{\rm{best}}$ & -316.02 & 0.23 & -414.35 & 0.19 & -454.37 & 0.18 & -241.24 & 0.28 \\
V$_{\rm{b,std}}$ & -428.24 & -0.19 & -307.87 & -0.24 & -192.87 & -0.32 & -212.02 & -0.31 \\
V$_{\rm{b,best}}$ & -251.70 & -0.27 & -272.93 & -0.26 & -178.59 & -0.34 & -182.96 & -0.33 \\
C$_{\rm{alt}}$ & -167.97 & -0.35 & -404.79 & -0.19 & -189.73 & -0.33 & -182.47 & -0.33 \\
A$_{\rm{b}}$ & -1962.35 & -0.05 & -1641.84 & -0.06 & -741.17 & 0.12 & -648.14 & 0.13 \\
$\Delta$V & -196.16 & 0.32 & -223.28 & 0.30 & -199.24 & 0.32 & -196.04 & 0.32 \\
A & -196.81 & -0.32 & -223.97 & -0.29 & -199.92 & -0.32 & -196.72 & -0.32 \\
V$_{\rm{asy}}$ & -275.98 & 0.26 & -27.84 & 0.62 & -307.18 & 0.24 & -203.44 & 0.31 \\
EW & 25.66 & -0.80 & 28.86 & -0.81 & 26.40 & -0.81 & 25.94 & -0.80 \\
Brightness & 19.88 & -0.78 & 23.50 & -0.79 & 20.67 & -0.78 & 20.17 & -0.78 \\
    \hline
     $\alpha$ = 0.2\\
   \hline
Abs. Depth & -51.07 & -0.55 & -42.69 & -0.57 & -66.09 & -0.52 & -69.96 & -0.51 \\
Norm. Depth & -255.78 & -0.27 & -302.89 & -0.24 & -460.14 & -0.18 & -485.95 & -0.17 \\
FWHM & -974.28 & -0.09 & -274.93 & -0.26 & -384.94 & -0.20 & -534.27 & -0.16 \\
BIS$_{\rm{std}}$ & -205.75 & -0.31 & -235.95 & -0.29 & -204.92 & -0.31 & -202.53 & -0.31 \\
BIS$_{\rm{best}}$ & -296.03 & -0.24 & -249.74 & -0.27 & -230.51 & -0.29 & -252.44 & -0.27 \\
C$_{\rm{std}}$ & -225.41 & -0.29 & -2266.21 & 0.04 & -193.92 & -0.32 & -243.34 & -0.28 \\
C$_{\rm{best}}$ & -307.13 & 0.24 & -313.09 & 0.24 & -413.69 & 0.19 & -282.53 & 0.25 \\
V$_{\rm{b,std}}$ & -450.62 & -0.18 & -311.89 & -0.24 & -194.96 & -0.32 & -236.63 & -0.28 \\
V$_{\rm{b,best}}$ & -255.22 & -0.27 & -280.18 & -0.25 & -179.13 & -0.34 & -187.96 & -0.33 \\
C$_{\rm{alt}}$ & -157.96 & -0.36 & -363.45 & -0.21 & -219.74 & -0.30 & -168.66 & -0.35 \\
A$_{\rm{b}}$ & -1571.27 & -0.06 & -1527.41 & -0.06 & -897.35 & 0.10 & -751.70 & 0.12 \\
$\Delta$V & -192.54 & 0.32 & -217.76 & 0.30 & -195.21 & 0.32 & -192.34 & 0.32 \\
A & -193.16 & -0.32 & -218.42 & -0.30 & -195.85 & -0.32 & -192.98 & -0.32 \\
V$_{\rm{asy}}$ & -280.52 & 0.25 & -25.22 & 0.62 & -290.94 & 0.25 & -249.93 & 0.27 \\
EW & 25.42 & -0.80 & 28.65 & -0.81 & 26.16 & -0.80 & 25.69 & -0.80 \\
Brightness & 19.50 & -0.78 & 23.16 & -0.79 & 20.30 & -0.78 & 19.79 & -0.78 \\   
   \hline
     $\alpha$ = 0.6\\
   \hline
Abs. Depth & -50.20 & -0.55 & -42.64 & -0.57 & -65.23 & -0.52 & -68.91 & -0.51 \\
Norm. Depth & -244.38 & -0.28 & -290.47 & -0.25 & -430.79 & -0.19 & -453.05 & -0.18 \\
FWHM & -1356.79 & -0.07 & -306.02 & -0.24 & -443.25 & -0.18 & -457.30 & -0.18 \\
BIS$_{\rm{std}}$ & -194.96 & -0.32 & -207.54 & -0.31 & -189.30 & -0.33 & -192.36 & -0.32 \\
BIS$_{\rm{best}}$ & -357.93 & -0.21 & -348.25 & -0.22 & -314.65 & -0.23 & -394.98 & -0.20 \\
C$_{\rm{std}}$ & -153.60 & -0.37 & -164.62 & -0.35 & -109.31 & -0.43 & -100.13 & -0.45 \\
C$_{\rm{best}}$ & -394.53 & 0.20 & -297.52 & 0.24 & -363.70 & 0.21 & -376.98 & 0.21 \\
V$_{\rm{b,std}}$ & -707.28 & -0.12 & -315.99 & -0.23 & -271.05 & -0.26 & -299.71 & -0.24 \\
V$_{\rm{b,best}}$ & -323.07 & -0.23 & -251.64 & -0.27 & -200.84 & -0.32 & -204.58 & -0.31 \\
C$_{\rm{alt}}$ & -123.07 & -0.41 & -210.62 & -0.31 & -152.35 & -0.37 & -120.58 & -0.41 \\
A$_{\rm{b}}$ & -938.69 & -0.10 & -775.01 & -0.11 & -12050.25 & -0.01 & -3584.83 & 0.03 \\
$\Delta$V & -193.38 & 0.32 & -209.50 & 0.31 & -192.36 & 0.32 & -190.93 & 0.33 \\
A & -193.90 & -0.32 & -210.08 & -0.31 & -192.91 & -0.32 & -191.47 & -0.32 \\
V$_{\rm{asy}}$ & -340.38 & 0.22 & -56.37 & 0.54 & -359.23 & 0.21 & -263.39 & 0.27 \\
EW & 24.16 & -0.80 & 27.50 & -0.81 & 24.93 & -0.80 & 24.45 & -0.80 \\
Brightness & 17.87 & -0.77 & 21.68 & -0.79 & 18.70 & -0.78 & 18.17 & -0.77 \\
\end{tabular}
\label{tab:noiseremove_4_kms_DR}
\end{table*}  

\begin{table*}[h]
 \caption{Granulation noise reduction for various diagnostics from model observations with $v_{eq}$ = 6~km~s$^{-1}$ and $i_{\star}$ = 90$^{\rm{o}}$, at various levels of differential rotation, alongside their Pearson's R correlation coefficient.}
\centering

\begin{tabular}{c|cc|cc|cc|cc}
    \hline
    $\alpha$ = 0.0 & & & \multicolumn{2}{c|}{Resolution = 70,000} &  \multicolumn{2}{c|}{Resolution = 140,000}  &  \multicolumn{2}{c}{Resolution = 140,000}   \\
    \hline
     Diagnostic & Reduct. (\%) & R & Reduct. (\%) & R  & Reduct. (\%) & R & Reduct. (\%) & R\\
    \hline
Abs. Depth & -210.79 & -0.31 & -180.38 & -0.34 & -262.05 & -0.27 & -280.50 & -0.25 \\
Norm. Depth & -3529.02 & -0.03 & -202717.06 & 0.00 & -1835.29 & 0.05 & -1589.69 & 0.06 \\
FWHM & -375.79 & -0.21 & -193.26 & -0.32 & -244.97 & -0.28 & -60.50 & -0.53 \\
BIS$_{\rm{std}}$ & -40.16 & -0.58 & -42.70 & -0.57 & -38.33 & -0.59 & -39.24 & -0.58 \\
BIS$_{\rm{best}}$ & -38.26 & -0.59 & -41.13 & -0.58 & -36.81 & -0.59 & -37.04 & -0.59 \\
C$_{\rm{std}}$ & -82.91 & 0.48 & -38.43 & 0.59 & -65.99 & 0.52 & -70.29 & 0.51 \\
C$_{\rm{best}}$ & -37.59 & 0.59 & -41.10 & 0.58 & -39.79 & 0.58 & -39.92 & 0.58 \\
V$_{\rm{b,std}}$ & -47.48 & -0.56 & -43.14 & -0.57 & -30.48 & -0.61 & -32.53 & -0.60 \\
V$_{\rm{b,best}}$ & -46.71 & -0.56 & -44.04 & -0.57 & -31.45 & -0.61 & -33.04 & -0.60 \\
C$_{\rm{alt}}$ & -521.42 & -0.16 & -4466.65 & 0.02 & -932.33 & -0.10 & -729.72 & -0.12 \\
A$_{\rm{b}}$ & -170.79 & 0.35 & -42.86 & 0.57 & -43.48 & 0.57 & -75.93 & 0.49 \\
$\Delta$V & -33.02 & 0.60 & -40.75 & 0.58 & -33.90 & 0.60 & -32.86 & 0.60 \\
A & -33.47 & -0.60 & -41.10 & -0.58 & -34.32 & -0.60 & -33.29 & -0.60 \\
V$_{\rm{asy}}$ & -40.52 & 0.58 & -44.78 & 0.57 & -40.70 & 0.58 & -33.70 & 0.60 \\
EW & -6.59 & -0.68 & -4.23 & -0.69 & -6.26 & -0.69 & -6.71 & -0.68 \\
Brightness & -14.76 & -0.66 & -11.20 & -0.67 & -14.37 & -0.66 & -14.90 & -0.66 \\   
    \hline
     $\alpha$ = 0.2\\
   \hline
 Abs. Depth & -205.91 & -0.31 & -178.74 & -0.34 & -257.60 & -0.27 & -274.97 & -0.26 \\
Norm. Depth & -2830.64 & -0.03 & -30521.94 & -0.00 & -2033.33 & 0.05 & -1752.31 & 0.05 \\
FWHM & -393.62 & -0.20 & -197.81 & -0.32 & -252.31 & -0.27 & -280.84 & -0.25 \\
BIS$_{\rm{std}}$ & -39.05 & -0.58 & -42.57 & -0.57 & -38.17 & -0.59 & -38.45 & -0.59 \\
BIS$_{\rm{best}}$ & -39.12 & -0.58 & -41.68 & -0.58 & -36.93 & -0.59 & -38.59 & -0.59 \\
C$_{\rm{std}}$ & -92.98 & 0.46 & -45.54 & 0.57 & -73.75 & 0.50 & -81.98 & 0.48 \\
C$_{\rm{best}}$ & -38.99 & 0.58 & -41.79 & 0.58 & -41.41 & 0.58 & -41.79 & 0.58 \\
V$_{\rm{b,std}}$ & -48.73 & -0.56 & -43.32 & -0.57 & -31.79 & -0.60 & -33.63 & -0.60 \\
V$_{\rm{b,best}}$ & -46.99 & -0.56 & -43.47 & -0.57 & -31.68 & -0.60 & -33.30 & -0.60 \\
C$_{\rm{alt}}$ & -669.80 & -0.13 & -1260.67 & 0.07 & -912.82 & -0.10 & -836.00 & -0.11 \\
A$_{\rm{b}}$ & -180.08 & 0.34 & -53.40 & 0.55 & -53.65 & 0.55 & -79.29 & 0.49 \\
$\Delta$V & -32.69 & 0.60 & -40.16 & 0.58 & -33.48 & 0.60 & -32.48 & 0.60 \\
A & -33.13 & -0.60 & -40.51 & -0.58 & -33.89 & -0.60 & -32.89 & -0.60 \\
V$_{\rm{asy}}$ & -39.92 & 0.58 & -44.80 & 0.57 & -37.48 & 0.59 & -37.73 & 0.59 \\
EW & -7.02 & -0.68 & -4.59 & -0.69 & -6.66 & -0.68 & -7.12 & -0.68 \\
Brightness & -15.42 & -0.65 & -11.77 & -0.67 & -15.01 & -0.66 & -15.55 & -0.65 \\
  
   \hline
     $\alpha$ = 0.6\\
   \hline
Abs. Depth & -187.99 & -0.33 & -165.10 & -0.35 & -227.51 & -0.29 & -239.84 & -0.28 \\
Norm. Depth & -1589.35 & -0.06 & -3101.43 & -0.03 & -7536.76 & 0.01 & -5267.82 & 0.02 \\
FWHM & -508.82 & -0.16 & -236.85 & -0.28 & -319.57 & -0.23 & -370.55 & -0.21 \\
BIS$_{\rm{std}}$ & -42.22 & -0.58 & -46.58 & -0.56 & -40.05 & -0.58 & -41.78 & -0.58 \\
BIS$_{\rm{best}}$ & -58.77 & -0.53 & -61.59 & -0.53 & -51.37 & -0.55 & -53.16 & -0.55 \\
C$_{\rm{std}}$ & -244.47 & 0.28 & -186.22 & 0.33 & -185.68 & 0.33 & -215.08 & 0.30 \\
C$_{\rm{best}}$ & -62.28 & 0.52 & -53.00 & 0.55 & -56.09 & 0.54 & -59.59 & 0.53 \\
V$_{\rm{b,std}}$ & -78.15 & -0.49 & -57.04 & -0.54 & -54.58 & -0.54 & -57.60 & -0.54 \\
V$_{\rm{b,best}}$ & -66.57 & -0.51 & -54.67 & -0.54 & -48.05 & -0.56 & -51.12 & -0.55 \\
C$_{\rm{alt}}$ & -791.74 & -0.11 & -185.36 & 0.33 & -476.29 & -0.17 & -713.14 & -0.12 \\
A$_{\rm{b}}$ & -235.63 & 0.29 & -78.11 & 0.49 & -104.27 & 0.44 & -144.54 & 0.38 \\
$\Delta$V & -39.68 & 0.58 & -45.78 & 0.57 & -605.13 & -0.14 & -38.96 & 0.58 \\
A & -40.04 & -0.58 & -46.09 & -0.56 & -40.15 & -0.58 & -39.31 & -0.58 \\
V$_{\rm{asy}}$ & -53.94 & 0.54 & -65.89 & 0.52 & -57.07 & 0.54 & -54.91 & 0.54 \\
EW & -9.30 & -0.68 & -6.46 & -0.68 & -8.83 & -0.68 & -9.36 & -0.67 \\
Brightness & -18.41 & -0.65 & -14.17 & -0.66 & -17.85 & -0.65 & -18.48 & -0.65 \\
\end{tabular}
\label{tab:noiseremove_6_kms_DR}
\end{table*}

\end{appendix}

\end{document}